\documentclass{aa}
\usepackage{graphicx}
\usepackage[flushleft]{threeparttable}
\usepackage[normalem]{ulem}
\usepackage{subcaption}
\usepackage{amssymb}
\usepackage{natbib}
\usepackage[colorlinks, citecolor=blue, linkcolor=blue, urlcolor=blue]{hyperref}
\usepackage{etoolbox}
\usepackage{tablefootnote}
\usepackage{lipsum} 
\usepackage{booktabs} 
\usepackage[outdir=./]{epstopdf}
\graphicspath{{./}}

\bibpunct{(}{)}{;}{a}{}{,}    

\newcommand{\noun}[1]{\textsc{#1}}
\newcommand{\nr}{{{\sc nr}}}
\newcommand{\agn}{{{\sc agn}}}
\newcommand{\csf}{{{\sc csf}}}

\providecommand{\tabularnewline}{\\}

\begin{document}


\title{High-redshift galaxy groups as seen by Athena/WFI}

\author{Chaoli Zhang\inst{1}\thanks{Member of the International Max Planck Research School (IMPRS) for Astronomy and Astrophysics at the Universities of Bonn and Cologne.}
\and
Miriam E. Ramos-Ceja\inst{1,2}
\and
Florian Pacaud\inst{1}
\and
Thomas H. Reiprich\inst{1}
}

\institute{Argelander-Institut f\"ur Astronomie (AIfA), Universit\"at Bonn, Auf dem H\"ugel 71, 53121 Bonn, Germany\\
(\email{chaoli@astro.uni-bonn.de})
\and
Max-Planck Institut f\"ur extraterrestrische Physik, Postfach 1312, 85741 Garching bei M\"unchen, Germany
}
\date{\today}

  \abstract
   { The first massive galaxy groups in the Universe are predicted to have formed at redshifts well beyond two. Baryonic physics, like stellar and active galactic nuclei (AGN) feedback in this very active epoch, are expected to have left a strong imprint on the thermo-dynamic properties of these early galaxy groups. Therefore, observations of these groups are key to constrain the relative importance of these physical processes. However, current instruments are not sensitive enough to detect them easily and characterize their hot gas content.
   }
   {In this work,
   we quantify the observing power of the Advanced Telescope for High ENergy Astrophysics (Athena), the future large X-ray observatory of the European Space Agency (ESA), for discovering and characterizing early galaxy groups at high redshifts. We also investigate how well Athena will constrain different feedback mechanisms.}
   {We used the SImulation of X-ray TElescopes (SIXTE) simulator to mimic Athena observations, and  a custom-made wavelet-based algorithm to detect galaxy groups and clusters in the redshift range $0.5 \le z \le 4$. We performed extensive X-ray spectral fitting in order to characterize their gas temperature and X-ray luminosity. In the simulations and their analysis, we took into account the main Athena instrumental features: background, vignetting, and point spread function degradation with off-axis angle, as well as all X-ray foreground and background components including a realistic AGN flux distribution. Different physically motivated thermo-dynamical states of galaxy groups were simulated and tested, including central AGN contamination, different scaling relation models (luminosity evolution), and distinct surface brightness profiles. Also, different Athena instrumental setups were tested, including both 15 and 19 mirror rows and the applied optical blocking filter. }
   {In the deep Wide Field Imager (WFI) survey expected to be carried out during part of Athena's first four years (the nominal mission lifetime) more than 10,000 galaxy groups and clusters at $z\ge 0.5$ will be discovered. We find that Athena can detect $\sim20$ high-redshift galaxy groups with masses of $M_{500}\geq 5\times \ensuremath{10^{13}\ \mathrm{\mathrm{M_{\odot}}}}$ and $z\geq2$, and almost half of them will have a gas temperature determined to a precision of $\Delta T/T \le 25$\%. }
   { We demonstrate that high-redshift galaxy groups can be detected very efficiently as extended sources by Athena and that a key parameter determining the total number of such newly discovered sources is the area on the sky surveyed by Athena. We show that these observations have a very good potential to constrain the importance of different feedback processes in the early universe because of Athena's ability not only to find the early groups but also to characterize their hot gas properties at the same time.}

\titlerunning{High-redshift groups as seen by Athena}
\authorrunning{C. Zhang et al.}

\maketitle

%
%


\section{Introduction\label{sec:intr}}

As the largest collapsed structures in the Universe, galaxy groups and clusters are important objects to study the structure formation and evolution of the Universe \citep[e.g.,][]{Voit:2005ab,Vikhlinin09,Borgani:2011aa,Schellenberger:2017ab,Pratt:2019aa}. According to the current cosmological paradigm, they would have formed bottom up, with smaller groups forming first that then grew with time through accretion and mergers to form the rich clusters (mass $\geq\ensuremath{10^{14}\ \mathrm{\mathrm{M_{\odot}}}}$ ). There is no strict criterium to distinguish a galaxy group from a galaxy cluster. In this paper, we refer to systems with masses of $M_{500}\leq\ensuremath{10^{14}\ \mathrm{\mathrm{M_{\odot}}}}$ as groups. Those objects are more numerous than rich clusters. Due to the shallower potential wells in the groups, the physical properties of the intracluster medium (ICM) can be more affected by baryonic physics, including, for example,\ cooling, shock heating, feedback from supernovae (SNe) and active galactic nuclei (AGN), and galactic winds. Also the stellar mass of member galaxies can be similar to (or even higher than) the ICM mass. These factors suggest galaxy groups should be treated as unique objects  in which to study baryonic physics.

The process of galaxy cluster and group formation is well understood in the local Universe. The observed ICM properties do not scale in a self-similar way (i.e., the mass scaling with $L_{\mathrm{X}}\text{--}M$) suggesting the importance of non-gravitational feedback processes \citep[e.g.,][]{Voit:2005ab,Giodini:2013ab,Lovisari:2015aa}. In fact, studies based on observed entropy excess in nearby clusters indicated that feedback processes can transfer significant amounts of energy to the ICM \citep[e.g.,][]{Cavagnolo:2009aa, Pratt:2010aa}. Even though those observations can be explained by theoretical modeling of different feedback mechanisms, it is still unclear which physical processes predominantly shape the hot gas density and temperature profiles of galaxy groups and clusters: stellar feedback from supernova,  feedback from supermassive black holes (SMBHs), or a combination of the two \citep[e.g.,][]{croton2006many,sun2009chandra,short2010evolution,fabjan2010simulating,mccarthy2011gas,Athena132,Ettori:2013ac,Croston:2013aa}.  The difficulty lies in the fact that those non-gravitational feedback processes predict very similar properties at low redshifts; for example, \citet[][]{Truong:2018aa} predicted similar $L_{\mathrm{X}}\text{--}T$ relations at $z<1.0$ for a stellar feedback model and an AGN feedback model (see Fig. \ref{fig:Truong_CSF_LT}). A sample of high-redshift galaxy groups is required to disentangle these scenarios, but only a few detections of low-mass systems are available with current observing facilities \citep[e.g.,][]{Go11,Er13}.

In order to constrain the detailed heating mechanisms involved in the early groups, understanding the thermo-dynamical state and chemical composition of the hot ICM within the range $0.5<z<2.5$ becomes necessary. This is because at higher redshifts, the ICM in the galaxy groups are significantly affected by star formation, galactic nuclear activity, and metal enrichment \citep[e.g.,][]{steidel2010structure,Ettori:2013ac}. The current X-ray observatories, \textit{XMM-Newton} and \textit{Chandra}, cannot facilitate a detailed study of thermo-dynamical physical properties of high-redshift galaxy groups of mass $M_{500}\sim5\times\ensuremath{10^{13}\ \mathrm{\mathrm{M_{\odot}}}}$ at $z>2.0$ owing to the limitations of the measurements of the density and temperature structure, and also to the currently small number of known systems.

The Extended Roentgen Survey with an Imaging Telescope Array (eROSITA) all-sky survey \citep{10.1117/12.856577,Mer12} is expected to detect all galaxy clusters and a large fraction of galaxy groups out to $z\sim1.5$ in the X-ray band \citep{Pi12}. \citeauthor{Pi12} predicted that eROSITA will detect about $10^{5}$ galaxy clusters with masses  $\ge 5\times10^{13}\, h^{-1}\mathrm{\,M_{\odot}}$. In the millimeter band, \citet{Mantz:2019aa} estimated that there will be $\sim50$ galaxy groups with mass  $>1\times10^{14}\,\mathrm{M_{\odot}}$ in the redshift range $2.0<z<2.5$ detected in the future South Pole Telescope (SPT-3G), Cerro Chajnantor Atacama Telescope (CCAT-prime), and Atacama Cosmology Telescope (AdvancedACT) surveys \citep{Benson:2014aa,De-Bernardis:2016aa,henderson2016advanced,Stacey:2018aa}. Thanks to those future surveys, one would expect a reasonable understanding of high-redshift galaxy clusters by 2030. However, the question of the formation and evolution as well as the thermo-dynamical state of galaxy groups with masses $\leq1\times10^{14}\,\mathrm{M_{\odot}}$ at redshift $z\geq2.0$ will likely still be unanswered. To answer this, we need a powerful future X-ray telescope to detect and study the small and faint galaxy groups beyond redshift two. This can be achieved with the next-generation X-ray astronomy observatory Athena (Advanced Telescope for High ENergy Astrophysics).

Athena is a mission proposed to address the `Hot and Energetic Universe' Science Theme. It is the next ESA L-class X-ray observatory with an expected launch in 2031 \citep{Nandra:2013aa}. Athena has a 12 m focal length and an aperture of up to $\sim3.0$ m in diameter. The mirror technology is based on ESA’s silicon pore optics (SPO), which provides a collecting area of up to $2\,\mathrm{m}^2$ at 1 keV and an angular resolution of $\sim5\arcsec$ on-axis \citep{Willingale:2013aa}. The telescope incorporates two observing instruments: an X-ray Integral Field Unit (X-IFU) and a Wide Field Imager (WFI). X-IFU is an X-ray micro-calorimeter spectrometer for high-spectral resolution imaging \citep{Barret:2013aa}. The WFI is a silicon-depleted, p-channel field-effect transistor, active-pixel-sensor camera \citep{Rau:2013aa}. The advanced instruments will give Athena a larger field of view (WFI: $40\arcmin \times 40\arcmin$), an order of magnitude increase in effective area, and a three times better point spread function (PSF) compared to \textit{XMM-Newton}. This large photon grasp will make Athena a unique next-generation, high-energy telescope for discovering the small and faint galaxy groups at high-redshift $z\geq2.0$, which can shed light on the relation between the halo properties and the physical processes involved in their formation and evolution \citep{Athena132,Nandra:2013aa}. 

In this work, we quantify the power of Athena to discover and characterize early galaxy groups, and how well the Athena observations will constrain the importance of different feedback mechanisms. We used the SImulation of X-ray TElescopes \citep[SIXTE,][]{wilms2014athena,Dauser:2019aa} software package to mimic the Athena observations, which takes into account the main Athena instrumental features: background, vignetting, and PSF degradation with off-axis angle. With the  images simulated according to the expected Athena/WFI observations, we can determine the expected number of early galaxy groups at $z\geq2.0$ with  $M_{500}\geq 5\times \ensuremath{10^{13}\ \mathrm{\mathrm{M_{\odot}}}}$ that will be detected. Furthermore, the simulated WFI event files from SIXTE can be utilized for the spectral fitting to determine the temperature precision of the detected groups.

This paper is organized as follows. Section 2 describes the simulation setup. Section 3 details the method of detection and characterization. The results are presented in Sect. 4 and the discussion is given in Sect. 5. We summarize our findings in Sect. 6. Unless otherwise noted, we assume  a $\Lambda$CDM cosmology with $\Omega_{\rm m}=0.27$, $\Omega_{\Lambda}=0.73,$ and $H_{0}=70 \;\rm km \;s^{-1}\; Mpc^{-1}$.


\section{Simulation setup\label{sec:method}}

\subsection{Athena simulator: SIXTE \label{subsec:simulator}}

The Monte Carlo simulations used in this work are realistic Athena/WFI calibrated event lists and images, which were obtained using SIXTE simulator. We used version 2.4.14 of the simulator\footnote{www.sternwarte.uni-erlangen.de/research/sixte/index.php.} together with Athena calibration files version 1.7.0.

\begin{figure}
\begin{centering}
 \includegraphics[trim=10 5 40 35,clip,width=\columnwidth] {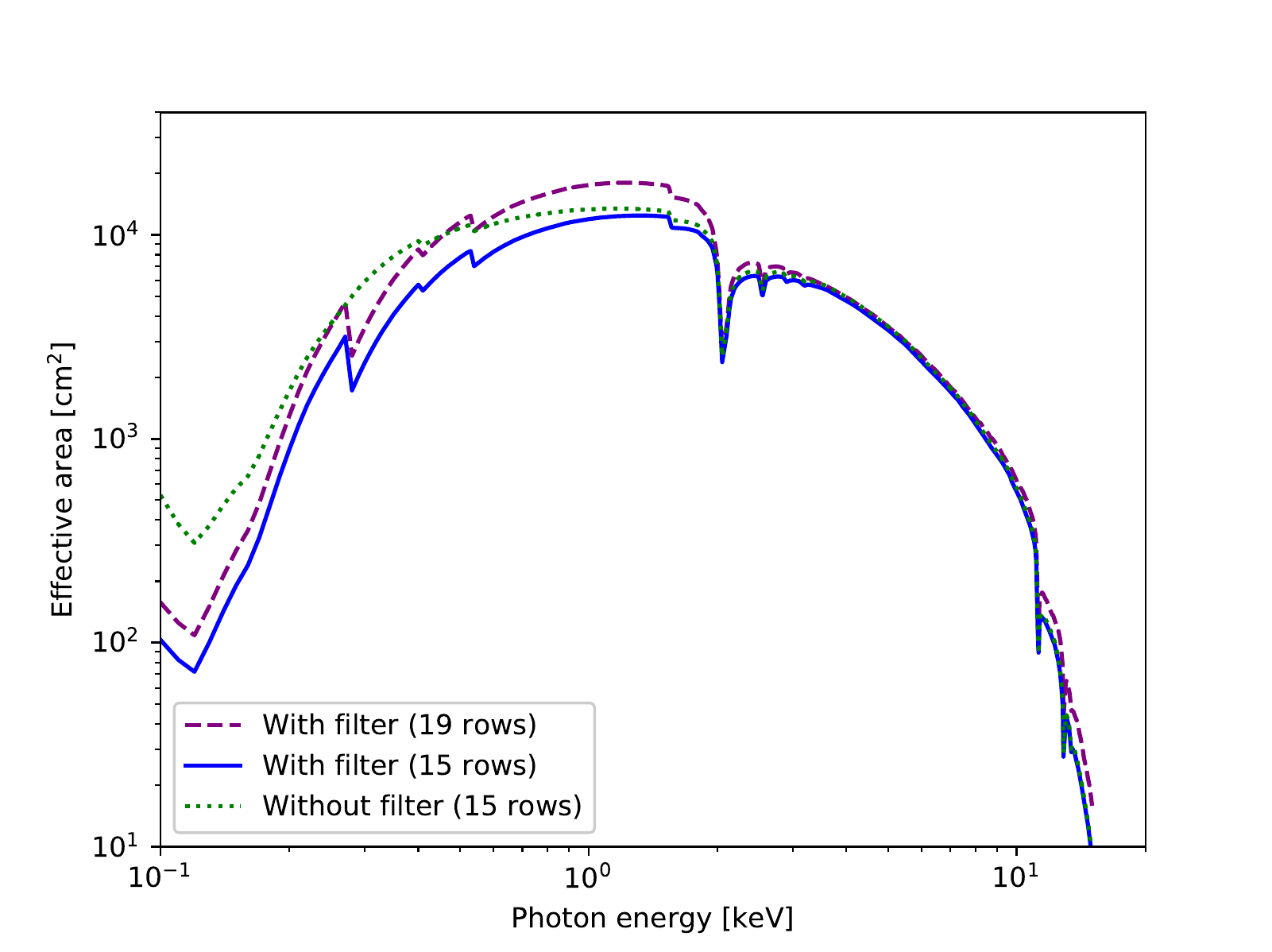} \\
 \includegraphics[trim=10 0 40 30,clip,width=\columnwidth] {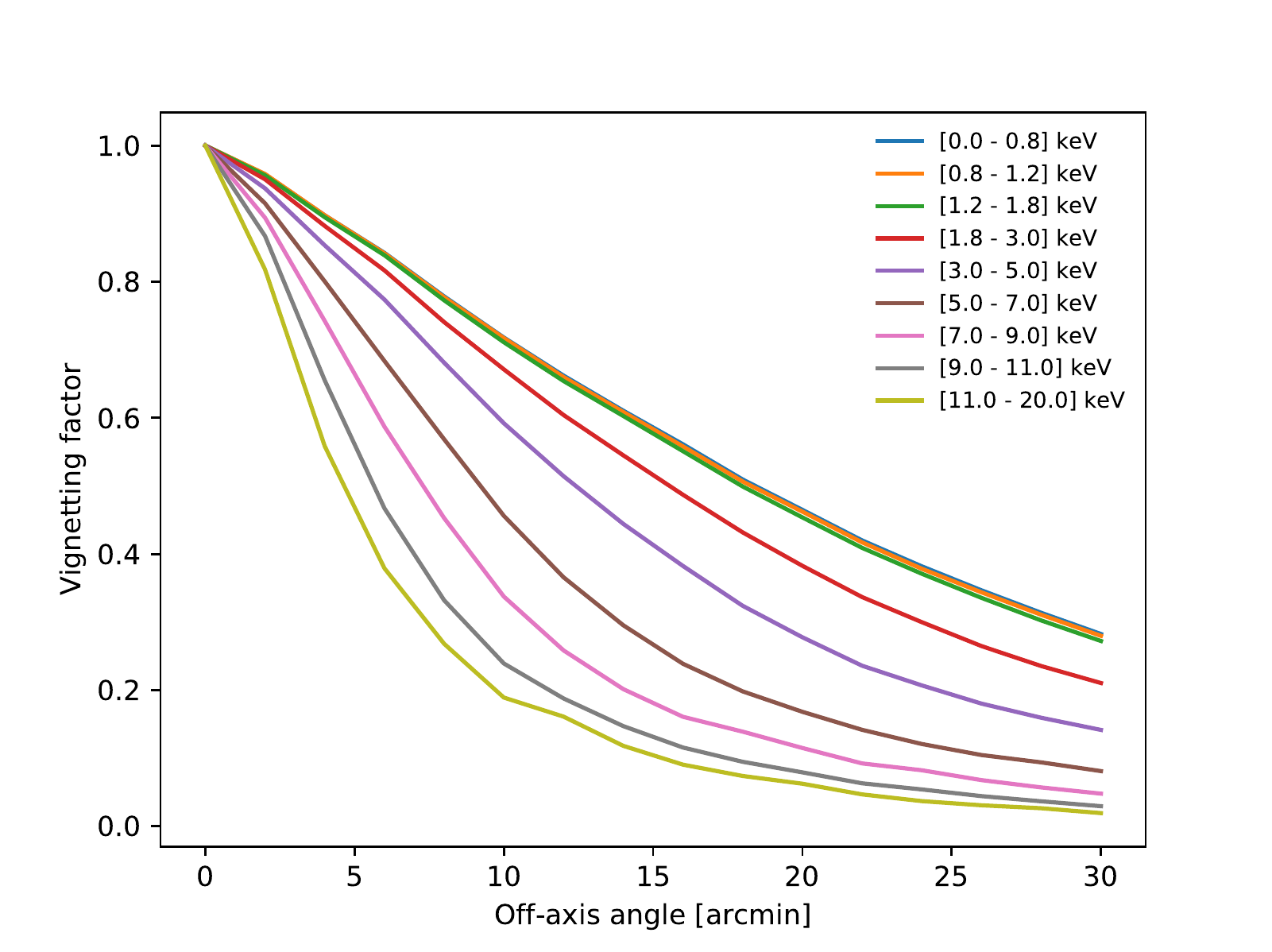}
\par\end{centering}
\caption{Upper panel: Auxiliary Response File (ARF) of Athena with 15 and 19  mirror rows. The blue curve is the ARF of Athena using an optical blocking filter; this can be used to block bright optical and UV sources. The green dotted curve is the ARF of Athena without an optical filter applied. The purple dashed curve is the 19 mirror rows ARF of Athena with an optical filter. Lower panel: Vignetting function at different photon energy, which is determined by collimation imposed by the pore geometry in the SPO modules. These two figures are generated using the Athena calibration files version 1.7.0, which is publicly available online \citep[see][for more details]{Dauser:2019aa}.}
\label{fig:arf_and_vig}
\end{figure}

\begin{figure*}
\begin{centering}
\includegraphics[scale=0.4] {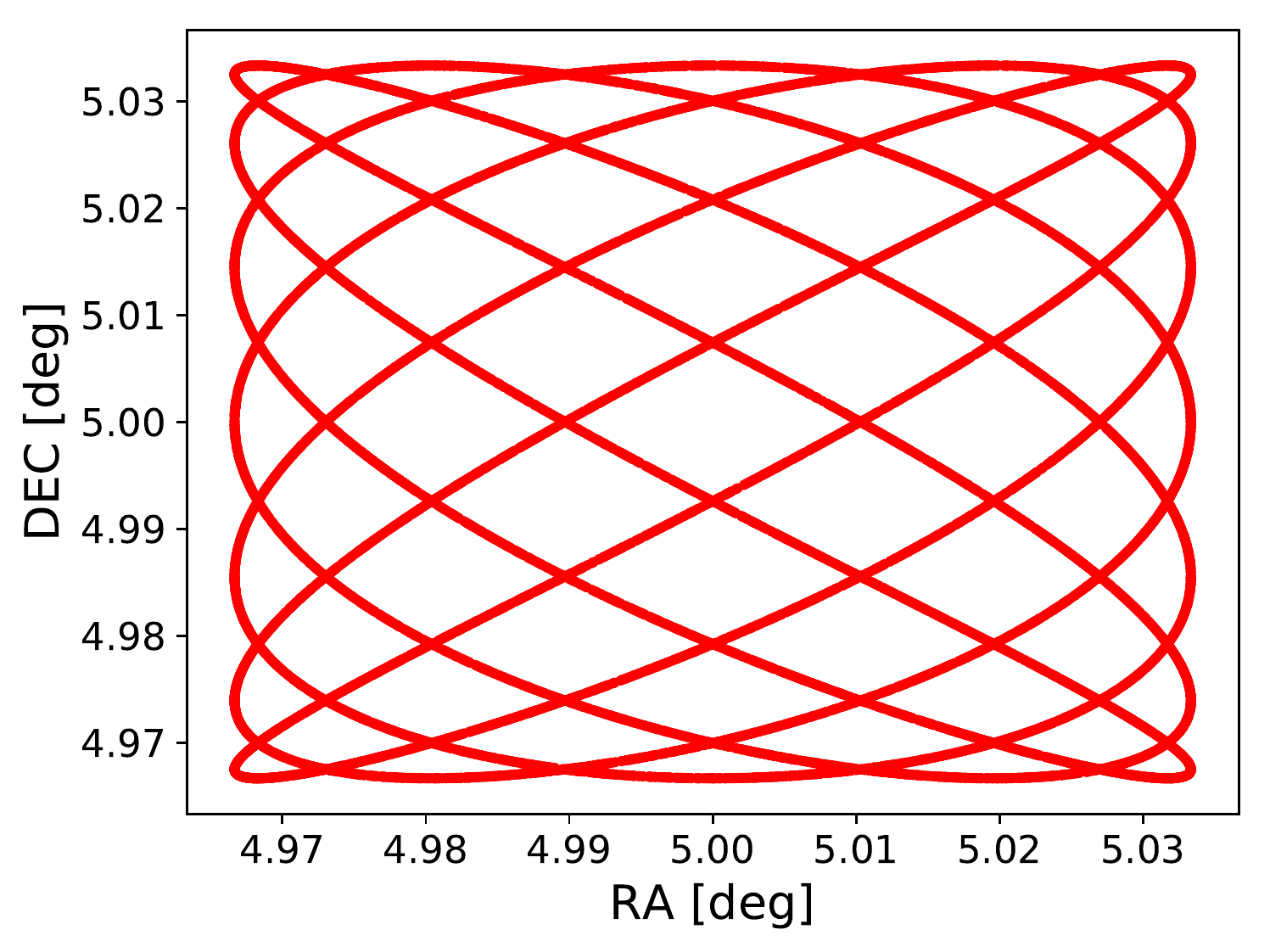}
\includegraphics[scale=0.4] {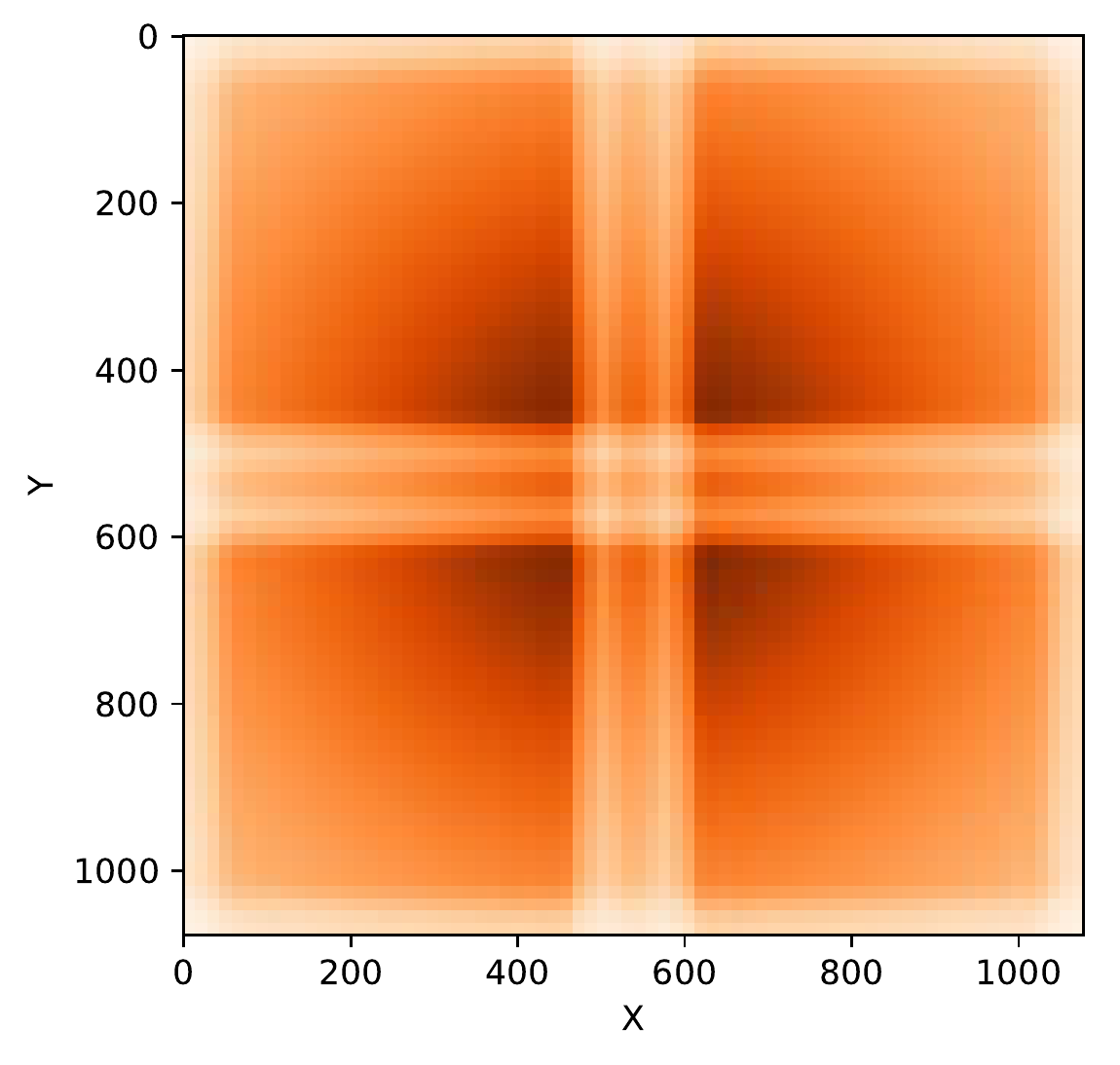}
\includegraphics[scale=0.4] {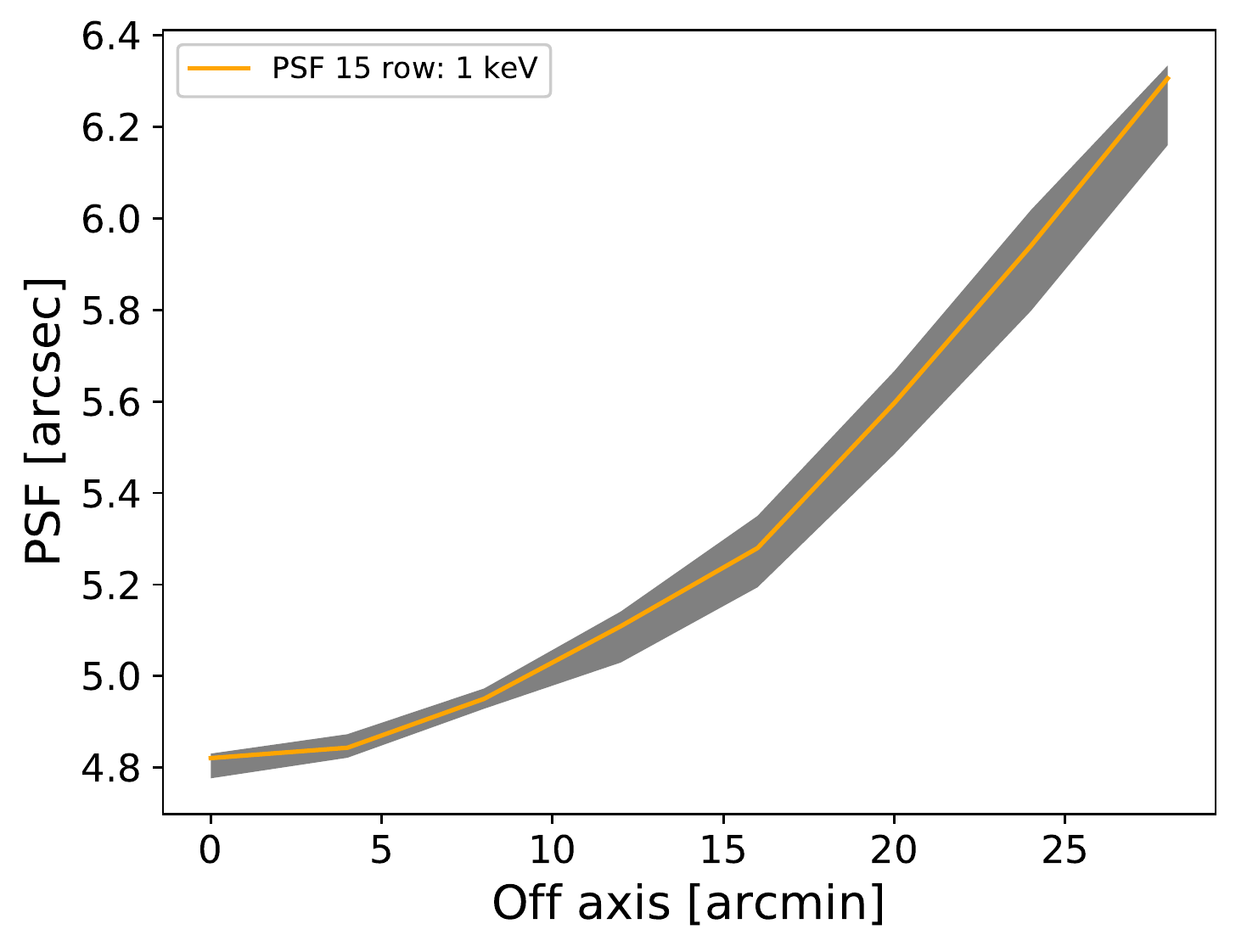}
\par\end{centering}
\caption{Left: Pointing of Athena in RA and Dec as described by the Lissajous pattern for 80 ks exposure time. The X- and Y-axes are given in degrees with the pattern amplitude corresponding to $2\arcmin\times2\arcmin$. Middle: Exposure map for an 80ks exposure with $0.1\arcsec/\mathrm{s}$ dither velocity. Right: Variation of PSF at 1 keV when dithering mode is applied.}
\label{fig:lissajousandpsf}
\end{figure*}

The mirror system of Athena is expected to be composed of $15$~mirror rows. There will be an optical blocking filter to reduce the contamination by optical light. The top panel of Fig.~\ref{fig:arf_and_vig} shows the effect of the filter on the  effective area of Athena. The shape of the mirror system will allow Athena to have a PSF with a value of $\sim 4.8$~arcsec on-axis and $\sim 5.6$~arcsec at $20$~arcmin off-axis angle, both at $1$~keV (see right panel of Fig.~\ref{fig:lissajousandpsf}). For the moment the PSF is described by a Gaussian function. This almost flat PSF across the field of view (FoV) can be achieved thanks to the SPO technology (see Section~\ref{sec:intr}). The Athena PSF used by SIXTE is described as a series of images in steps of $2$~arcmin off-axis angles for two distinct energies at 1~keV and 6~keV. The vignetting is the same and  is tabulated for energy ranges $\Delta E={0.8, 1.2, 1.8, 3.0, 5.0, 7.0, 9.0, 11.0, 20.0}$~keV. At $\sim1$~keV, the effective area is reduced by$~50$\% at $20$~arcmin off-axis angles (see bottom panel of Fig.~\ref{fig:arf_and_vig}).

The detector of the WFI will be composed of four sensors, which will be physically separated by 7~mm (corresponding to an angular element of $120$~arcsec). The Athena mirror system will point to one of the WFI sensors \citep[see][for further details]{Meidinger:2017aa}. We manually changed this default configuration in the internal parameter files of SIXTE, such that the mirror system will point to the center of the WFI's FoV, in the middle of the gaps. To compensate for this insensitive area, Athena will have an observation mode with a dithering pattern. SIXTE can easily accept an attitude file to describe this dithering, and we have used a similar Lissajous-type pattern as in \citet[][]{Dauser:2019aa}; the left panel in Fig.~\ref{fig:lissajousandpsf} shows an example of the Lissajous curve that such an observation mode will follow. The corresponding exposure map is shown in the middle panel of the same figure, and is obtained by running the {\tt exposure\_map} command in SIXTE.

One consequence of the dithering observing mode is a broadening of the Athena PSF at a given position. Since SIXTE provides the PSF size at a given off-axis, we use a linear interpolation to calculate the new PSF size at a new position when the telescope dithers. The PSF variation as a function of the off-axis angle is shown in the right panel of Fig.~\ref{fig:lissajousandpsf}. In general, this PSF variation is $\sim0.2\arcsec$ in size, which is about $4$\% at the edge of the sensor. Given this small effect, we ignore the extra PSF broadening caused by dithering in our work.

Athena is expected to perform a deep survey with the WFI. According to the mock observing plan (MOP) this survey will be similar to the following setup: $103\times84$~ks + $3\times980$~ks + $10\times 840$~ks + $4\times1.4$~Ms, resulting in a total area of 48 $\deg^{2}$. Conservatively, we base our predictions on the assumption of $80$~ks exposure over 48 $\deg^{2}$.
The characteristics described above constitute our baseline Athena setup, which comprises\ a mirror system with 15 rows, an optical filter in place, and observations made with the dithering mode.

\subsection{Input populations}

In the following we describe the main expected components in an X-ray observation of the extra-galactic sky. These are particle and cosmic background, resolved AGNs and, for this particular work, the hot diffuse gas of galaxy groups.

\subsubsection{X-ray background\label{subsec:Background}}

The X-ray background consists of two parts, namely the induced instrumental and the cosmic X-ray background. As the name indicates, the former type of background is produced by the detector, while the latter type consists of the integrated emission from unresolved extra-galactic AGNs, the local hot bubble, and the diffuse Galactic emission.

The instrumental background is assumed to be flat over the FoV, with a value $5\times10^{-3}\ \mathrm{cnt/keV/s/cm^{2}}$ or $6\times10^{-4}\ \mathrm{cnt/keV/s/arcmin^{2}}$ \cite[e.g.,][]{Rau13,lotti2017particle,von2018evaluation} , and is directly implemented within SIXTE. The cosmic X-ray background is described by the following model emission:
\begin{equation}
\underset{\text{(1)}}{\underbrace{\tt{apec}}}+\underset{\text{(2)}}{\underbrace{\tt{wabs}}}\times(\underset{\text{(3)}}{\underbrace{\tt{apec}}}+\underset{\text{(4)}}{\underbrace{\tt{powerlaw}}}).
\label{eq:bkg_model}
\end{equation}
The individual emission components describe (1) the local hot bubble, (2) the galactic absorption, (3) the diffuse galactic emission, and (4) the unresolved AGNs. The parameter values of the model emission are presented in Appendix Table~\ref{tab:Background-Xspec-model}. One minor difference to the Rau (2013) cosmic background emission model is that the {\tt norm} parameter in model 4 in Equation \ref{eq:bkg_model} is $4.0\times10^{-7}$ $\text{pho}/\text{keV}/\text{cm}^{2}/\text{s}$, which corresponds to 60\% resolved point sources. The reason for this change is to match the simulated AGN population, which is explained in next section.

\subsubsection{AGN population\label{subsec:AGN-population}}

The X-ray AGN flux distribution, $S$, and source density, $N$, is usually described by an empirical $\log N-\log S$ relation \citep[e.g.,][]{Lehmer:2012aa,Moretti:2003aa,Gilli:2007aa}. We use the \citet{Lehmer:2012aa} relation to describe the AGN number counts in our simulations. This relation covers a large flux range: $5.1\times10^{-18}-1\times10^{-11}$~ergs~s$^{-1}$~cm$^{-2}$ in the $0.5-2.0$~keV energy band. However, in order to match the $40$\% of the unresolved fraction of AGNs in the cosmic background (see Section~\ref{subsec:Background}) we use the $5\times10^{-17}-1\times10^{-11}\ \text{ergs/s/cm}^{2}$ flux range. This lowest flux limit is chosen because any  sources fainter than $5\times10^{-17}\ \text{ergs/s/cm}^{2}$ would not be detected by Athena with an 80 ks exposure time as shown in Appendix Fig. \ref{fig: LogN-LogS_obs}. Figure~\ref{fig:LogN-LogS-relation} shows the comparison between our sampling method of the $\log N-\log S$ relation and the \citet{Lehmer:2012aa} relation.

The AGN are uniformly distributed in the field, since we do not attempt to model their spatial distribution, and their emission model are described by a {\tt phabs}$\times${\tt powerlaw}. This model has a fixed spectral index value of  $\alpha=1.42$ and a hydrogen column density of $n_{\mathrm{H}}=2\times10^{20}\text{cm}^{2}$.

\subsubsection{Galaxy groups \label{subsec:Galaxy-groups}}

We use the spherically-symmetric $\beta$-model \citep{Cavaliere:1978aa} to describe the surface brightness, $S_\textrm{X}$, of the galaxy groups. The $\beta$-profile is given by
\begin{equation}
 S_\textrm{X}\propto\Bigg[1+\bigg(\frac{r}{r_{\mathrm{c}}}\bigg)^{2}\Bigg]^{-3\beta+1/2},
\end{equation}
where $r_{\mathrm{c}}$ is the core radius of the
galaxy groups with a fixed value of $r_{\mathrm{c}}=0.15\times r_{500}$ \citep[][]{Clerc12}, and $\beta=2/3$. 

We simulate a discrete number of masses,\footnote{We present the group masses in terms of $M_{500}$, which represents the mass within the $r_{500}$ region, where the mean density is 500 times the critical density of the Universe.} $M_{500}=1,~2.5,~4,~5,~7,~10\times\ensuremath{10^{13}\ \mathrm{\mathrm{M_{\odot}}}}$ , at different redshifts, $z=0.5,~1.0,~1.5,~2.0,~2.5,~3.0,~3.5,~4.0$. Their respective temperatures and luminosities are calculated by using the exact cluster mass to temperature, $M-T$, and temperature to luminosity, $T-L$, scaling relations of \citet[][hereafter RE11]{Reichert:2011aa}. Having these quantities, the group emission is modeled by a {\tt phabs}$\times${\tt apec} with the help of the \noun{Xspec} (version 12.9.1u) spectral fitting package \citep{Ar96}. We assume an abundance of $Z=0.3$, which could be an upper limit for all objects at redshift $z\geq3.0$. We also assume a hydrogen column density of $n_{\mathrm{H}}=2\times10^{20}\text{cm}^{2}$. For a galaxy group of mass $5\times\ensuremath{10^{13}\ \mathrm{\mathrm{M_{\odot}}}}$ at $z=2.0$, Athena will typically collect 917 source photons and 2076 background photons within $r_{500}$ in the $0.2-2.0$~keV energy band (on-axis) in 80 ks, which corresponds to a signal-to-noise ratio $\sim17$. With this signal-to-noise ratio, a single $\beta$-model would be very similar to a double $\beta$-model profile within the $r_{500}$ region, because the slight variations in the outer slope of the single $\beta$-profile do not change the photon counts in the outer region for the groups at higher redshifts.

The position of the simulated galaxy groups in the field is not entirely random. We position the groups at different off-axis angles to quantify the effects of vignetting, PSF, and the WFI detector gaps on the detection of galaxy groups. We simulate one galaxy group on-axis and four galaxy groups at the other off-axis angles ($5,~10,~15,~18,~25$~arcmin). Except for the last position, the groups encompassing the same off-axis angle are randomly located. The groups $25$~arcmin away from the center are fixed at the corners of the detector.

\subsection{Baseline model scenario\label{sec:bslnmodeldes}}

On top of the group features mentioned in Section~\ref{subsec:Galaxy-groups}, we include an AGN in the center of the galaxy groups. This is motivated by some studies \citep[e.g.,][]{Er13,Gobat:2011aa} that have shown that at higher redshifts galaxy groups tend to have AGN in their central parts. The presence of a central AGN in central galaxies in galaxy groups may have a significant effect on the thermo-dynamics of the ICM, as the ICM can be feedback-heated via the outburst of AGNs \cite[e.g.,][]{Croston:2013aa}. In particular, \citet{Biffi:2018aa} studied the effect of central AGN contamination for detecting the galaxy clusters at $z\sim1-1.5$. They found the X-ray emission from AGN is roughly a factor of five less in photon counts than from the whole ICM, with significant scatter. We take this into account by assigning a 20\% flux to the AGN as compared to the corresponding galaxy group. The spectrum of this central AGN is calculated with \noun{Xspec} using {\tt phabs$\times$pow}, with $n_{\mathrm{H}}=2\times10^{20}\text{cm}^{2}$ and $\alpha=1.42$.

As mentioned in Section~\ref{sec:intr}, very few galaxy groups of masses $M_{500}=5\times10^{13}\,\mathrm{\mathrm{M_{\odot}}}$ at $z\geq2.0$ will have been discovered by $\sim2030$. Our aim is to quantify the  capabilities of Athena to detect and, importantly, characterize them.  Therefore, we create simulations with three different exposure times: 50, 80, and 130~ks. For each group mass, redshift, and exposure time, we simulated 20 Monte Carlo realizations. Since our simulations involved a large number of different setup parameters, we summarize all the possibilities in Table~\ref{tab:The-parameters-for-GGs}. The baseline model is indicated as parameters marked in bold in the table, which consists of the features of the galaxy groups and the Athena telescope setup. The galaxy group features include a mass $5\times10^{13}\,\mathrm{\mathrm{M_{\odot}}}$ with a central AGN that has 20\% flux of the group, the core radius of the $\beta$-model is $r_{\mathrm{c}}=\text{15\%\ensuremath{\times}}r_{500}$ , and the temperature and luminosity of the group are calculated using the RE11 scaling relation. The Athena setup consists of observations with an exposure time of $t_{\mathrm{exp}}=80$~ks, and the telescope has an optical filter applied and 15 mirror rows.

The SIXTE simulator generates Athena calibrated event files in the $0.1-15$~keV energy band  with a pixel size of $2.2$~arcsec. However, we restrict the image analysis to the $0.2-2\ \mathrm{keV}$ energy band. This energy band is where we expect to detect most of the galaxy group emission, due to the redshifted bremsstrahlung exponential cut-off, and where Athena has a larger effective area. An example of a simulated image for the baseline model scenario is shown in the left panel of Fig. \ref{fig: SIXTE_FILTER_IM}.

\begin{table*}
\begin{centering}
\begin{tabular}{cc}
\hline 
\hline
Parameter & Value\tabularnewline
\hline
$t_{\mathrm{exp}}${[}ks{]} & \textbf{80}, 50, 130\tabularnewline
Optical filter & \textbf{with}, without\tabularnewline
Central AGN & \textbf{yes}, no\tabularnewline
Number of mirrow rows & \textbf{15}, 19\tabularnewline
$r_{\mathrm{c}}=\text{\#\ensuremath{\times}}r_{500}$ & \textbf{15\%}, 30\%, 45\%\tabularnewline
Scaling relation & \textbf{RE11}, TR18 \nr, \csf, \agn\tabularnewline
Mass $[\mathrm{M_{\odot}}]$ & \textbf{5}, 1, 2.5, 4, 7, $10\times10^{13}$\tabularnewline
Redshift & 0.5, 1.0, 1.5, 2.0, 2.5, 3.0, 3.5, 4.0\tabularnewline
Off-axis $[\arcmin]$ & 0, 5, 10 , 15 , 18 , 25\tabularnewline
Number of simulated groups per off-axis & 4\tabularnewline
Number of realization per redshift & 20\tabularnewline
\end{tabular}
\par\end{centering}
\caption{Parameters used in the simulations. The bold parameter indicated in the first column is the baseline model setup. TR18 represents the scaling relations obtained from \citet[][]{Truong:2018aa} (see Section~\ref{subs:TR18_relation} for further details). \label{tab:The-parameters-for-GGs} }
\end{table*}

\begin{figure}
\begin{centering}
\includegraphics[trim=5 0 35 30,clip,width=\columnwidth]{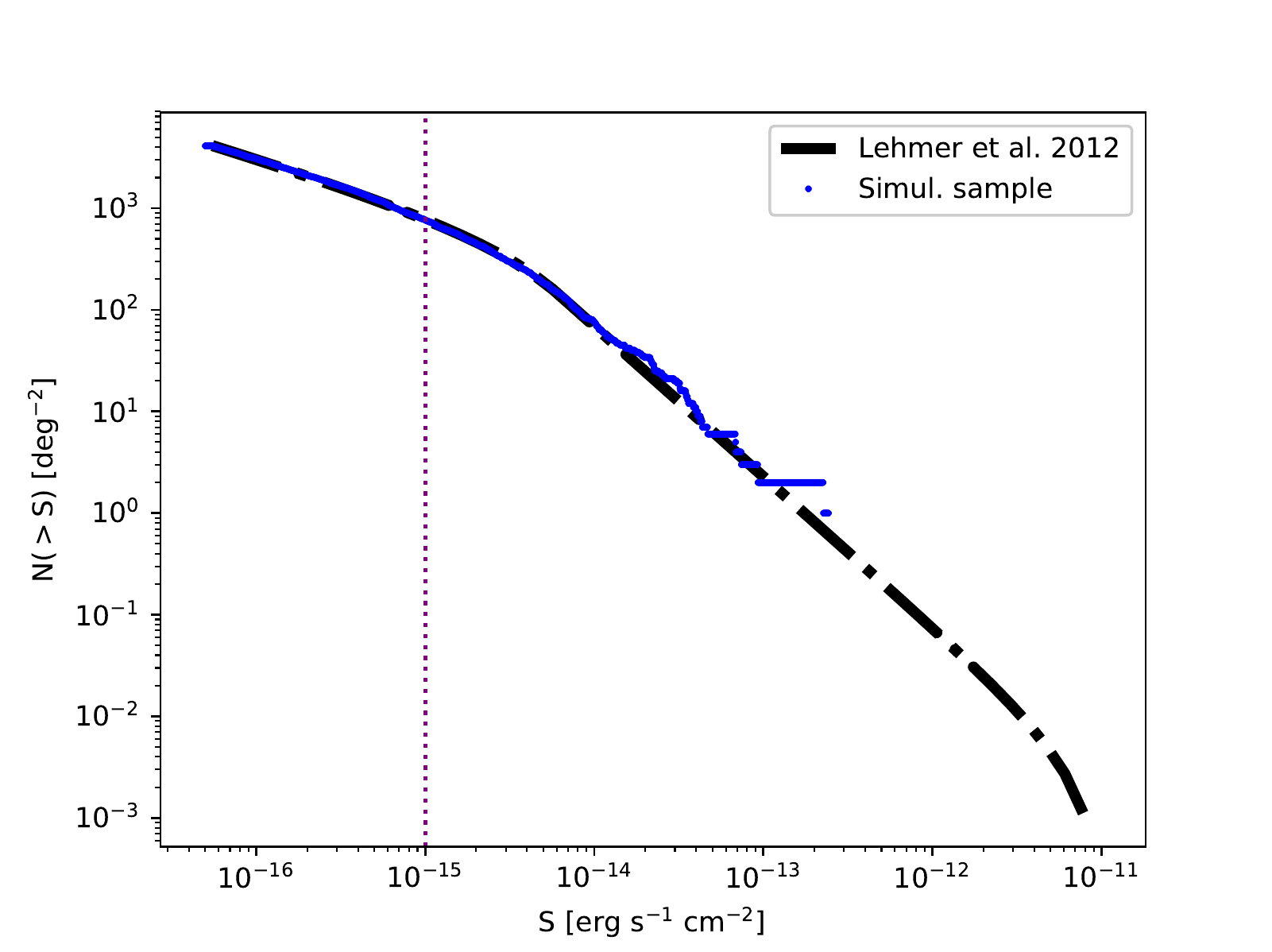}
\par\end{centering}
\caption{Cumulative number counts
vs. flux for the AGNs in the $0.5-2.0$~keV energy band. The black curve is from \citet{Lehmer:2012aa};
the blue dotted curve represents one realization of an AGN population
used for the simulation. The purple dotted line indicates the AGN flux cut ($\geq1\times10^{-15}\ \text{ergs/s/cm}^{2}$)
for the 90\% encircled energy fraction
when removing the AGN contamination at the stage of the spectral fitting (Section~\ref{subsubsec:spec_fit}).}
\label{fig:LogN-LogS-relation}
\end{figure}


\section{Methodology}

\subsection{Source detection and characterization \label{subsec:Detection_and_charc}}

The observed X-ray objects can be broadly divided into two categories: point-like and extended sources. The former are mainly composed of AGNs, whose angular size is smaller than the telescope PSF. The latter have an extended X-ray emission, such as galaxy groups and clusters, whose angular size extends from a few tens of arcseconds to degrees.

Despite the fact that X-ray extended sources are larger than the PSF size, the detection of these sources, especially the ones at high redshifts, is challenging. This is because these X-ray objects often emit very few photons distributed over a large area. Therefore, their signal will likely be completely submerged into the background. Also, contamination by AGN emission can make it hard to determine the extended nature of the underlying group emission. Furthermore, a few faint AGN lying close together in projection can mimic a single extended source, resulting in false detections.

To detect the high-redshift galaxy groups as extended sources, we use comprehensive source detection and characterization algorithms. The method is the same as the one applied by \citet{Xu:2018aa} when searching for very extended galaxy groups in R\"ontgensatellit (ROSAT) All-Sky Survey data, and it is similar to the one implemented by \citet{Pacaud:2006aa} and \citet{Faccioli:2018aa}. The algorithm comprises three steps: (i) wavelet filtering of the photon image, (ii) source detection by \noun{SExtractor,} and (iii) maximum likelihood fitting procedure. In the following, a brief description of each step is presented.

\begin{figure}
\begin{centering}
\includegraphics[trim=0 0 35 25,clip,width=\columnwidth]{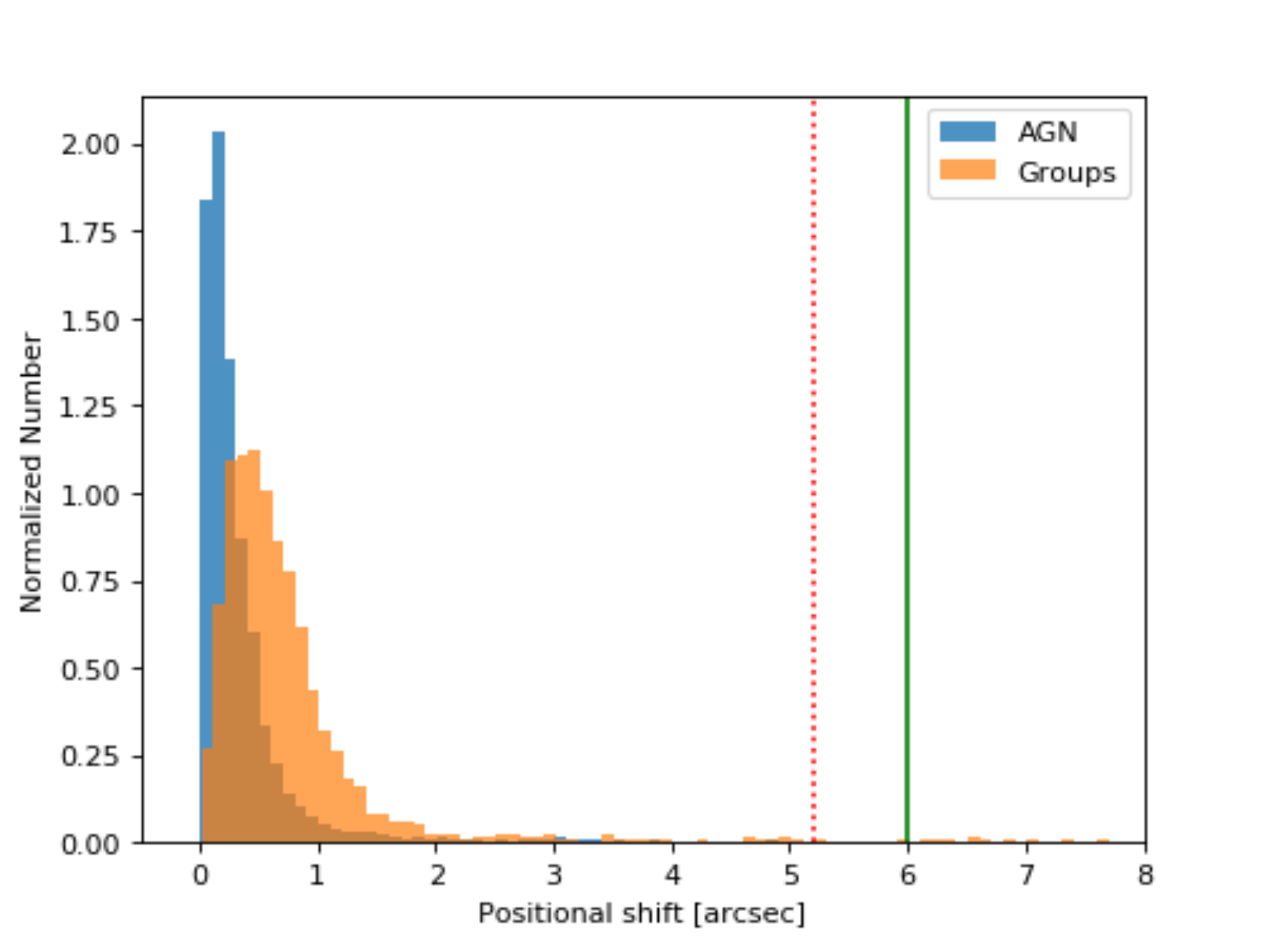}
\par\end{centering}
\caption{\label{fig:Histograms-positional_acc}Histograms show the positional
accuracy on AGN (blue) and galaxy groups (orange). The position shift
is defined as the distance between the input positions of the simulated
sources and the positions of their corresponding closest detected
sources. The red vertical dotted line indicates the Athena FOV-averaged PSF $5.2\arcsec$,
and the green solid line at $6.0\arcsec$ is the adopted cross-matching radius for the detected sources.}
\end{figure}

The simulated Athena images are filtered using the \noun{erwavelet} software, which uses the \`a trous transform with a cubic B-spline wavelet and has a rigorous treatment of the Poisson noise while accounting for a varying exposure time. This allows us to remove the insignificant features directly in the wavelet space using a thresholding algorithm \citep{starck_murtagh_bijaoui_1998,starck1998structure}. The result is a smooth and denoised image.

\begin{figure*}
\begin{centering}
\includegraphics[scale=0.2]{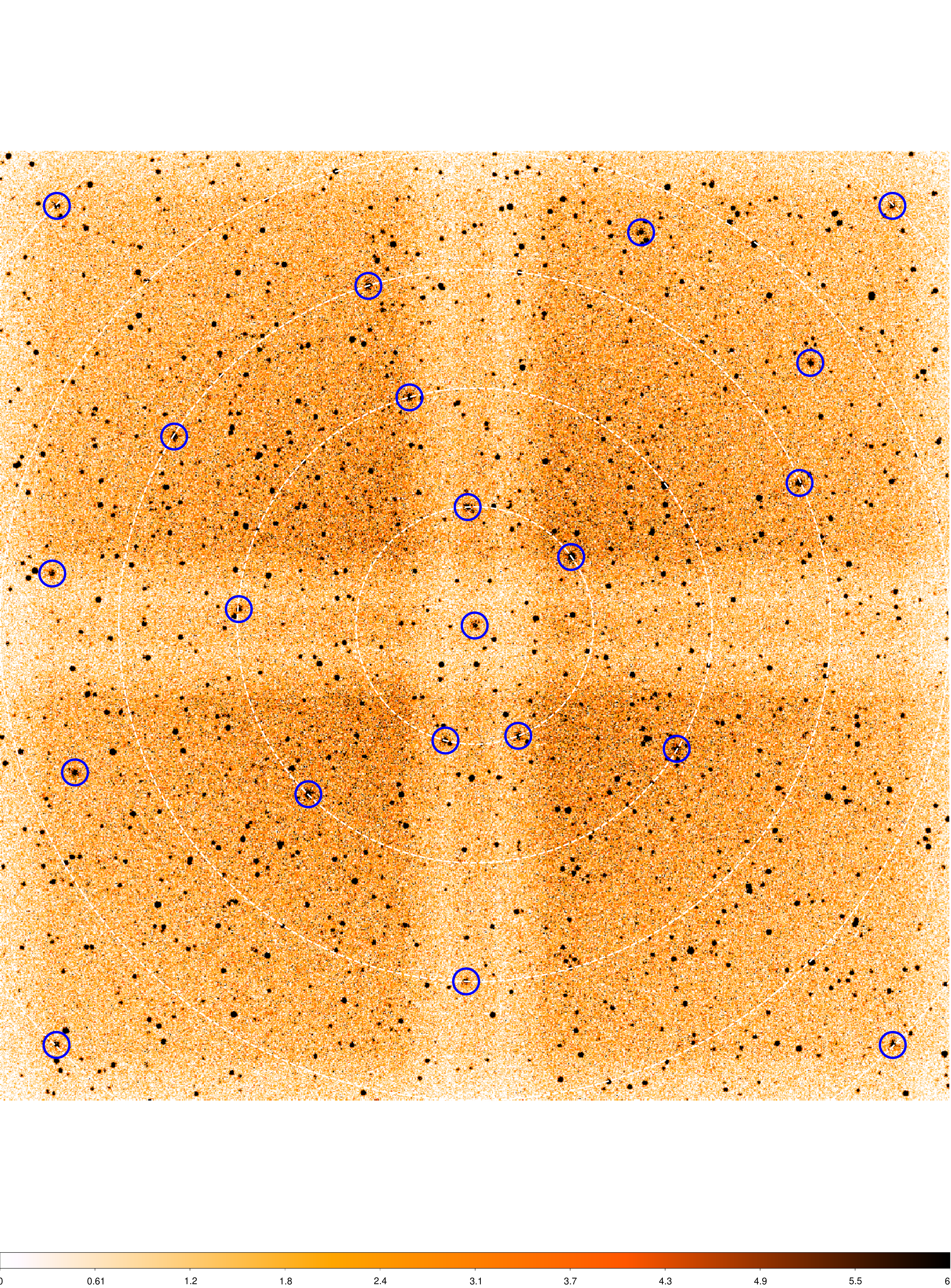}
\includegraphics[scale=0.2]{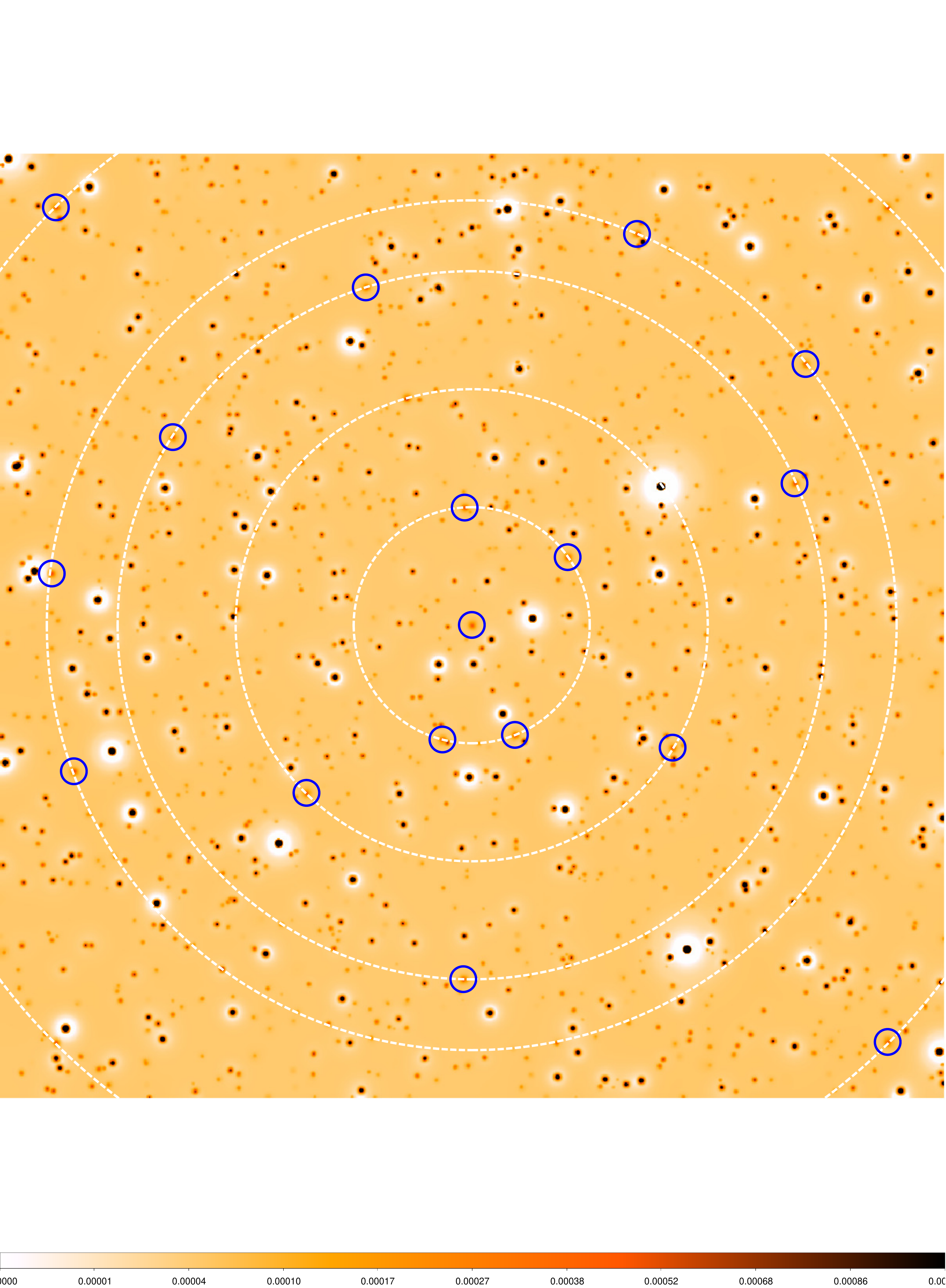}
\par\end{centering}
\caption{Left: Example of the SIXTE simulated image of an 80 ks Athena/WFI observation. Cosmic X-ray background, Galactic foreground, and particle background are included as well as telescope vignetting and PSF
degradation with an off-axis angle. The blue circles on the left correspond to the size of $r_{500}$ of the input groups, which are located at $z=2.0$ with $M_{500}=5\times10^{13}\,\mathrm{M_{\odot}}$. Right: Wavelet image of this SIXTE simulated image. The blue circles represent the detected sources after \noun{SExtractor} and the maximum likelihood fitting detection algorithm are run; three sources are undetected by our algorithm.}
\label{fig: SIXTE_FILTER_IM}
\end{figure*}

Source detection on filtered images is then performed by the \noun{SExtractor} software \citep{Bertin:1996aa}, which produces a preliminary source list. The sources identified by \noun{SExtractor} are then analyzed by a maximum likelihood fitting procedure. This algorithm uses some of the \noun{SExtractor} output parameters as an input for source characterization. The fitting code determines which surface brightness model maximizes the probability of generating the observed source photon distribution. Two models are fitted (1) a point-like model, and (2) a point-like plus a $\beta$-profile model. Both models are convolved with the Athena PSF, and the algorithm takes into account the Athena instrumental effects including the exposure time. Further details can be found in \citet[]{Pacaud:2006aa} and \citet{Xu:2018aa}. 

As shown in \citet[]{Pacaud:2006aa} and \citet{Xu:2018aa}, we also use the output parameters from the maximum likelihood fitting method to look for criteria for the identification of high-redshift groups as extended X-ray sources. We found that the best criteria are given by the {\tt extension likelihood} and the {\tt extent} of the source. The extension-likelihood represents the significance of a source being described by an extended source model, which compares the Poisson likelihoods of the point-source model and extended source model \citep[see Eq. 2 in][for further details]{Xu:2018aa}, and the extent parameterizes the core radius of the detected source.

An example of a wavelet filtered image, together with the sources detected as extended, is shown in the right panel of Fig.~\ref{fig: SIXTE_FILTER_IM}. We note the good reconstruction of the filtered image when applying the exposure map correction.

\subsection{Analysis techniques  \label{subsec:Analysis techniques}}

In the following we describe the main analysis techniques used in our work. These are cross matching radius determination, selection criterion used for the extended sources, detection probability projection over the WFI detector, mass functions to calculate expected number of galaxy groups and spectral fitting.

\subsubsection{Positional accuracy}

The input source positions can be displaced due to source extent, finite photon statistics, and the convolution of the simulated sources with the PSF. We estimate this positional shift by measuring the distance between the input positions of the sources and the positions of their corresponding closest detected sources. We investigate this by simulating images with only AGN and background or only groups\footnote{The simulations are based on group masses of $5\times\ensuremath{10^{13}\ \mathrm{\mathrm{M_{\odot}}}}$ at redshifts $z=0.5,~1.0,~1.5,~2.0,~2.5,~3.0,~3.5,~4.0$.} and background. The results are shown in Fig.~\ref{fig:Histograms-positional_acc}. About $99.7$\% of the detected sources lie within $6.0\arcsec$ of their input position. Therefore, we use a value of $6.0\arcsec$ as a correlation radius when matching input and detected sources.

\subsubsection{Source contamination\label{srcconta}}

One important ingredient in the process of determining the detection efficiency of X-ray sources is the ability to measure the source contamination. For example, nearby point-like sources or point-like sources in the vicinity of extended ones can result in a mis-classification of the detected point-like source as extended \citep[see][for other examples]{valtchanov2001comparison}.

\begin{figure}
\begin{centering}
\includegraphics[trim=25 0 25 20,clip,width=\columnwidth]{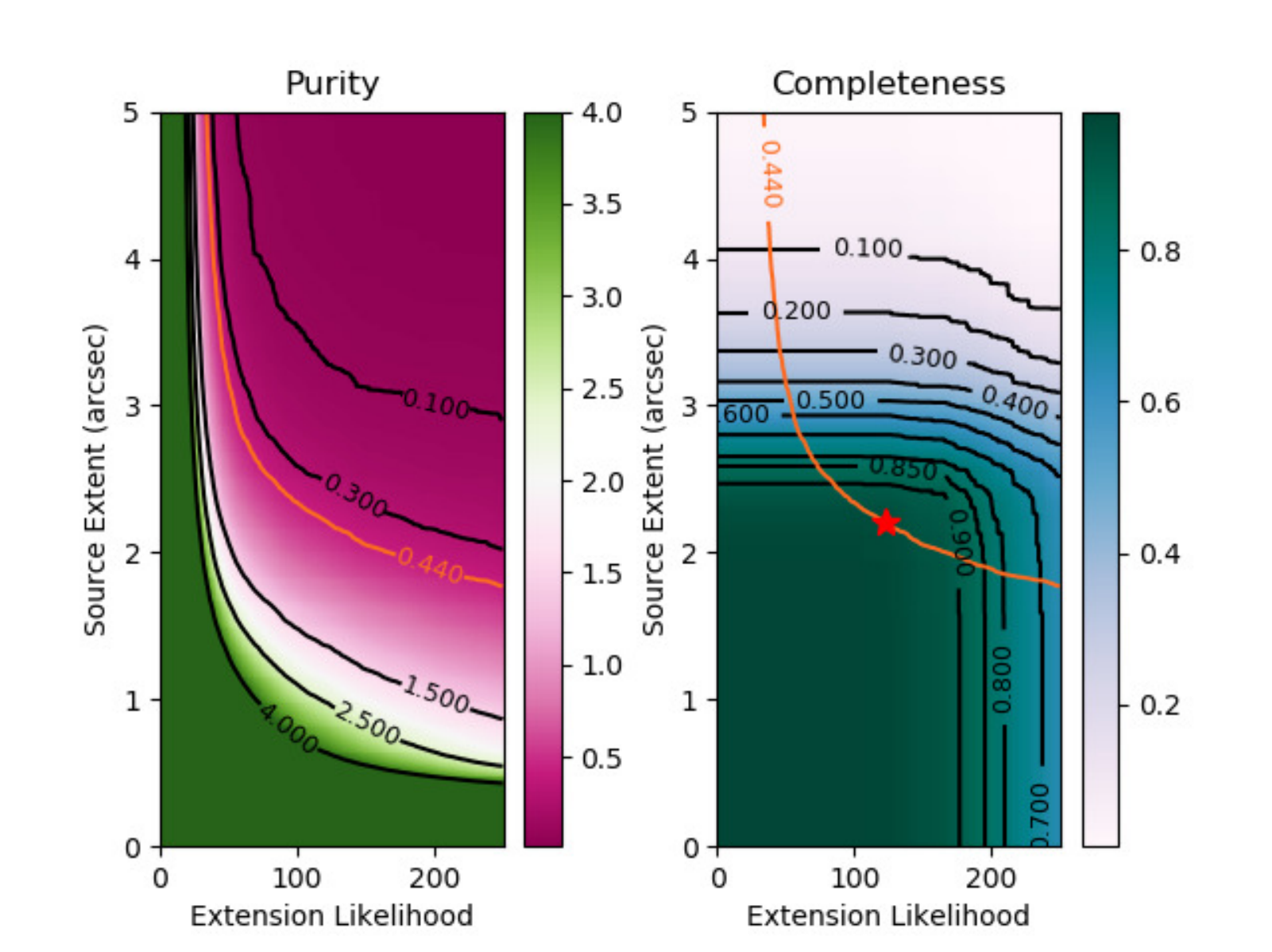}
\par\end{centering}
\caption{\label{fig:AGN_POP}
Purity and completeness contours to identify the optimal location of  {\tt extent}-{\tt extension-likelihood} plane for the baseline 80 ks simulation.
Left: Black purity contour lines are used to identify the source contamination levels, the orange line represents a 0.44 sources per image contamination rate, which corresponds to one source per $\mathrm{deg}^{2}$ false detection rate. Right: Completeness contour is then used to find the location to maximize the detection probability of galaxy groups. The orange line is the same as in the left panel, and the orange star represents the selection criteria to distinguish between point-like and extended sources, at $\sim93\%$ completeness level.}
\end{figure}

\begin{figure}
\begin{centering}
\includegraphics[trim=70 20 120 75,clip,width=\columnwidth]{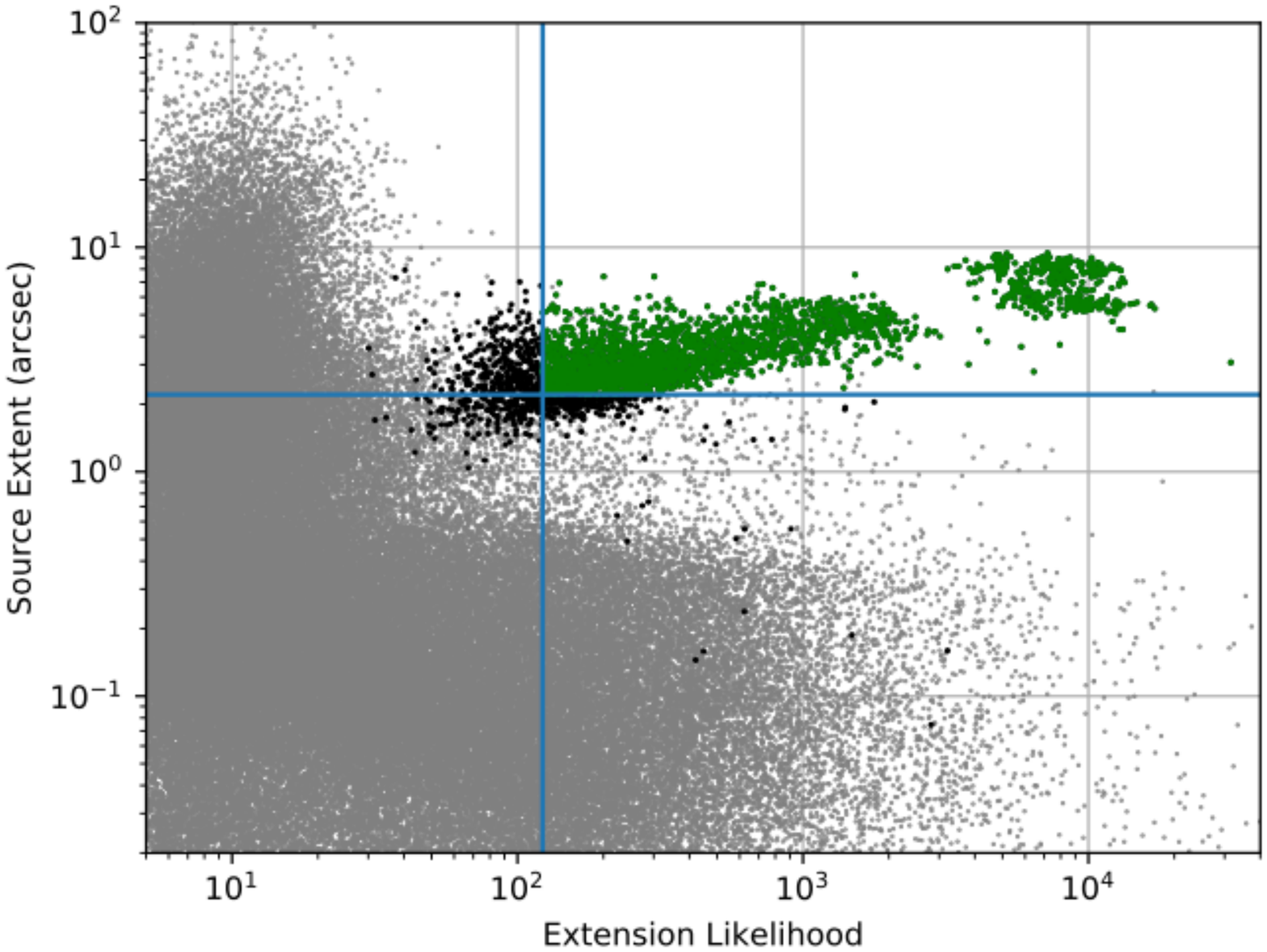}
\par\end{centering}
\caption{\label{fig:ext_ml_extent} Best-fitting values in the {\tt extent}-{\tt extension-likelihood} plane. Point-like sources (AGNs) are displayed as gray points, the simulated extended sources (galaxy groups) are marked as black, and the green dots represent the detected galaxy groups. The solid blue lines at {\tt extension-likelihood} $=104$ and $\mathtt{extent}=2.3$
determine the selection criteria for extended-like sources with a $1/\mathrm{deg}^{2}$ false detection rate.}
\end{figure}

We set the criterion to be one source per $\mathrm{deg}^{2}$ as the false detection rate, which corresponds to 0.44 source per image\footnote{The FoV of Athena is $40\arcmin\times40\arcmin$, therefore one Athena observation covers the sky area of size $40/60\,\mathrm{deg}\times40/60\,\mathrm{deg}\sim0.44\,\mathrm{deg}^{2}$.}. This selection criterion is based on two considerations: first, minimizing the number of false detections because follow-up observations (e.g., using Athena/X-IFU) on false detections are expensive; and second, maximizing the detection efficiency. For example, the MOP for the Athena/WFI survey has 103 pointings with 84 ks exposure time, which can give $\sim45$ false detections in total. Athena is expecting to discover over 1500 galaxy groups above redshift 0.5 with $M_{500}\geq5\times\ensuremath{10^{13}\ \mathrm{\mathrm{M_{\odot}}}}$ (as indicated in Table~\ref{tab:LT_num_table_Baseline}), so the 45 false detections correspond to only a 3\% contamination level. This criterion helps to find the place in the {\tt extent}-{\tt extension-likelihood} plane that allows us to distinguish between point-like and extended sources. This is achieved by simulating the baseline model setup by increasing the number of realizations per redshift from the original 20 to 200 (ten times larger than the default setup, see Table~\ref{tab:The-parameters-for-GGs}), running the source detection and characterization algorithms on them, and scanning the output parameter space. The false detections within $r_{500}$ of the galaxy groups are excluded, because typically if a point-like source is located inside $r_{500}$ of the groups, it will be detected as an extended source. The orange line in the purity contour in the left panel of Fig.~\ref{fig:AGN_POP} shows the location of one false detection per deg$^{2}$ in the {\tt extent}-{\tt extension-likelihood} plane. The purity contours at levels ${[0.1- 4.0]}$ indicate lines of constant false detection rate in the plane. This same orange line is over-plotted in the completeness contour {\tt extent}-{\tt extension-likelihood} plane in the right panel of Fig.~\ref{fig:AGN_POP}. The completeness contours refer to the probability (between 0 and 1) of detecting the galaxy groups. Looking for the maximum point where the detection efficiency is the largest, we found {\tt extension-likelihood} $>104$ and $\mathtt{extent}>2.3$. These limits are used as our standard selection criteria for identifying galaxy groups as extended sources. Figure \ref{fig:ext_ml_extent} shows one example of the {\tt extent}-{\tt extension-likelihood} plane. The gray data points are the detected AGNs while the green dots are the detected galaxy groups. The distinction between point-like and extended sources is visible. These classification criteria can effectively distinguish the galaxy groups from contaminating sources.

\subsubsection{Projected detection probability \label{subsect:proj_det_prob}}

As mentioned in Section~\ref{subsec:simulator} the Athena simulations that we perform are in a dithering observing mode to compensate for the gap between the sensors. Therefore the Athena exposure is not uniform across the FoV. We take this variation into account when quantifying the group detection probability, $P(z,M,\theta)$. This detection efficiency is a function of the redshift ($z$), mass ($M$), and off-axis position ($\theta$) of the detected groups. To obtain the final group detection efficiency, we average $P(z,M,\theta)$ over the FoV,
\begin{equation}
P(z, M)=\frac{\intop_{0}^{20\sqrt{2}'}\theta\times\alpha(\theta)\times P(z,\theta, M)\ d\theta}{\intop_{0}^{20\sqrt{2}'}\theta\times\alpha(\theta)\ d\theta} \label{eq:total_DP}
,\end{equation}
where $\alpha(\theta)$ is the weighting function that measures the arc length of the circle centered at the FoV enclosed by the detector, which is defined as

\begin{equation}
\alpha(\theta)=\begin{cases}
2\pi & \theta<20\arcmin\\
2\pi-8\arccos\left(\frac{20\arcmin}{\theta}\right) & 20\arcmin\leq\theta\leq20\sqrt{2}\arcmin.
\end{cases}
\end{equation}

We see that at the edge of the detector $\theta=20\arcmin$, $\alpha(\theta)$
reduces to $2\pi$, and at the edge of the corner of the detector
$\theta=20\sqrt{2}\arcmin$, $\alpha(\theta)$ is equal to zero. The integral at the denominator is a normalization constant.

\subsubsection{Mass function \label{subsect:MF}}

The number of galaxy groups expected to be detected by Athena can be estimated through the cluster mass function. The cluster mass function gives the theoretical expectation of the number density, $N$, of
halos of a given mass and at a given redshift,
\begin{equation}
N(>M,\ >z)=\int_{M_{\mathrm{lim}}}^{\infty}\int_{z}^{\infty}\frac{dn}{d\ln(M)}\times V_{z} \times \Omega_{\text{s}} \times P_{(z,M)}\ dz\ dM.
\label{eq:MF}
\end{equation}
Here, $\frac{dn}{d\ln(M)}$ is the cumulative mass function for which we use the \citet[][hereafter tinker08]{Tinker:2008aa} fitting function. We also test for the mass functions obtained from \citet[][hereafter watson13]{watson2013halo} and \citet[][hereafter despali16]{despali2016universality}; we give the results in Table. \ref{tab:no_groups_table_summary}. We obtain $\frac{dn}{d\ln(M)}$ using the COLOSSUS python package \citep{Diemer:2018aa}. The fraction of the survey area is represented by $\Omega_{\text{s}}$   and $V_{z}$ is the differential co-moving volume at redshift $z$. The $P_{(z,M)}$ is the projected detection probability in Eq.~(\ref{eq:total_DP}). Since we have discrete values of $M$ and $z$ , we use a bi-linear interpolation for $P_{(z,M)}$. More details are given in Appendix Fig. \ref{fig: Det_2D_interpolation}.

\subsubsection{Spectral fitting\label{subsubsec:spec_fit}}

We use the {\tt makespec} command from SIXTE to extract spectra of the detected galaxy groups. The group's spectrum is extracted from a region of area $\pi r_{500}^{2}$ and centered around its detected position. A respective background spectrum is extracted from an annulus of width $r_{500} - 2\times r_{500}$ around the group. The X-ray emission for the galaxy group is strictly within the $r_{500}$ region so that the annulus for the background subtraction from $r_{500}$ to $2\times r_{500}$ contains no photons from the group. It is worth mentioning that we tested a scenario with the X-ray emission of the galaxy group extended to $3\times r_{500}$. This extended emission of the groups does not introduce a bias due to the sparse photons spread out across the vast outskirts region. We remove the AGN that fall in the background region within a certain area covered by their encircled energy fraction (EEF); if the AGN has a flux lower than $1\times10^{-15}\ \text{ergs/s/cm}^{2}$ then $90\%$ of the EEF is removed, if the flux is larger then $99\%$ of the EEF is deleted.

For each spectrum, we generate an ancillary response file (ARF) by multiplying the default ARF (on-axis) by a factor that takes into account the energy-dependent vignetting at a given off-axis angle. As mentioned in Section~\ref{subsec:simulator}, our Athena simulations employ an observation mode with a dithering pattern, which can result in slight ARF variations. For the spectral fitting, we do not take this dithering effect into account, and use the ARF at a given off-axis angle for the respective galaxy group. The redistribution matrix file (RMF) is also used for each spectrum, obtained from default Athena calibration files (version 1.7.0) as part of the SIXTE simulator.

We use \noun{Xspec} to perform the spectral fitting. The particle and cosmic X-ray background are assumed to be stable and without spectral or spatial variation, and to have the same properties as in the source area. With these assumptions, we directly subtract the background spectrum from the source one. Moreover, we use Cash statistics \citep{Cash:1979aa} to estimate the best-fit value for the fitting model parameters and the {\tt error} command to determine $68\%$ confidence intervals.

We modeled the group thermal emission with an absorbed {\tt apec} model with \noun{Xspec} in the $0.2-3$~keV energy band. To properly interpret the fitting results from \noun{Xspec}, we follow the methodology of \citet{Borm:2014aa} to define the precision, accuracy, and accuracy of the uncertainty of the group's temperature. These quantities are defined as follows.
 \begin{enumerate}
 \item Precision is the spread of the fitted results given by
 \begin{equation}
 \frac{\Delta T}{<T_{\mathrm{fit}}>}=\frac{T_{\mathrm{1\sigma+}}-T_{\mathrm{1\sigma-}}}{2\times<T_{\mathrm{fit}}>_{\mathrm{median}}}.\label{eq:precision}
 \end{equation}
 Here $[T_{\mathrm{1\sigma-}},T_{\mathrm{1\sigma+}}]$ encloses the  true 68\% confidence range from the distribution of the fit values.
 \item The accuracy of the temperature is interpreted as the bias on the best-fit cluster temperature,
 \begin{equation}
 \frac{<T_{\mathrm{fit}}>_{\mathrm{median}}}{T_{\mathrm{input}}}.\label{eq:accuray_temp}
 \end{equation}
 \item The accuracy of the error estimation gives the deviation between the median uncertainty obtained from the {\tt error} command and the uncertainty obtained from the distribution of the fitted results: 
 \begin{equation}
 \frac{<\Delta T_{\mathrm{error}}>}{\Delta T}=\frac{<T_{\mathrm{error+}}-T_{\mathrm{error-}}>_{\mathrm{median}}}{T_{\mathrm{1\sigma+}}-T_{\mathrm{1\sigma-}}}.\label{eq:accuray_error}
 \end{equation}
 \end{enumerate}


\section{Results}

\subsection{Plausible high-redshift physical scenarios \label{sec:res:Detection probability}}

As mentioned in Section~\ref{sec:intr}, there is not much information about and few observations of galaxy groups at and beyond $z=2$. Therefore little is known about the physical processes and thermo-dynamical state of the collapsed structures at $z>2$. In the following, the results of different physically motivated scenarios are presented, and Table~\ref{tab:The-parameters-for-GGs} summarizes the simulation setups for different scenarios. We do not modify the cuts in the {\tt extent}-{\tt extension likelihood} plane for the different scenarios.

\begin{figure}
\begin{centering}
\includegraphics[trim=90 25 20 40,clip,width=\columnwidth]{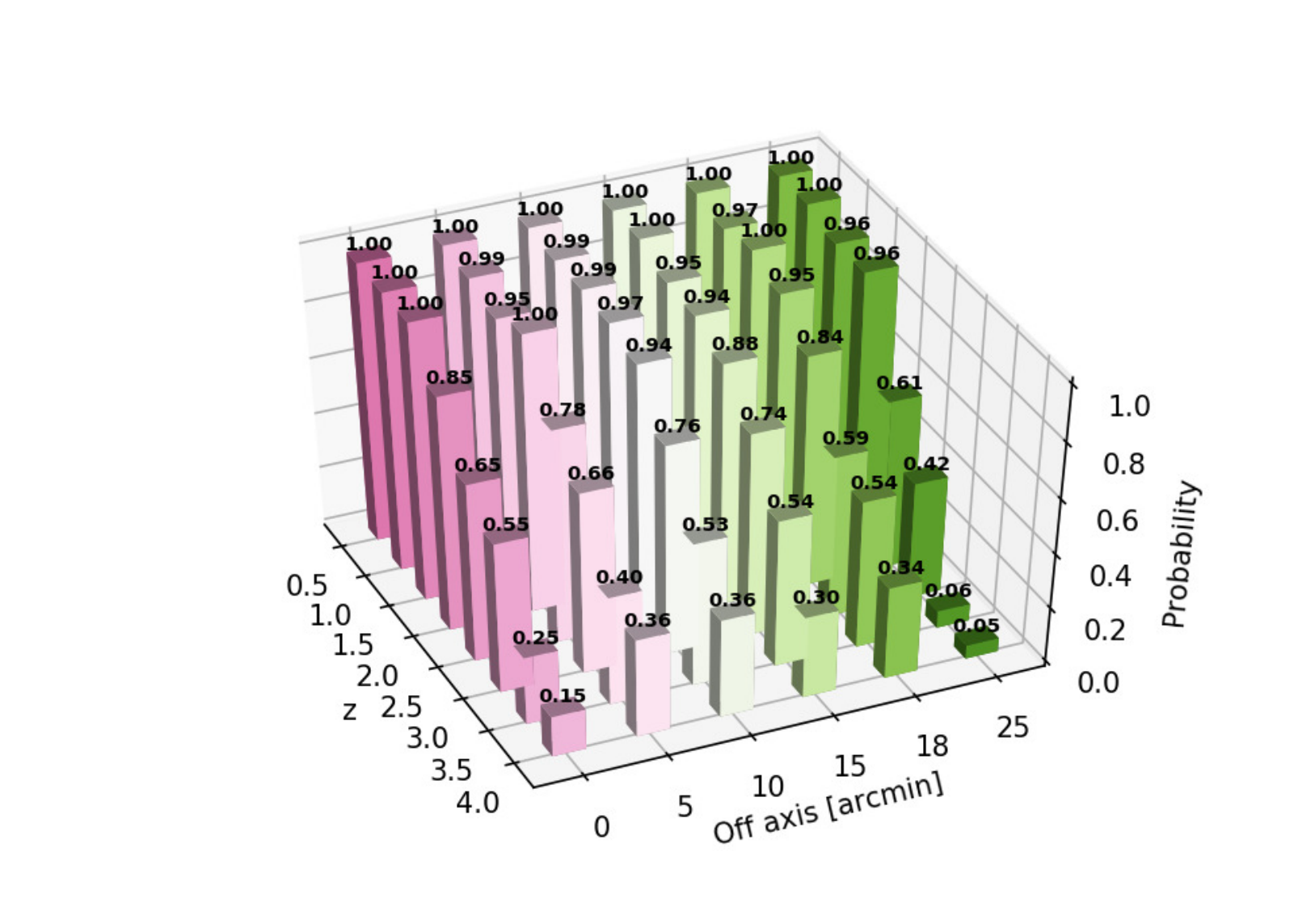}
\par\end{centering}
\caption{\label{fig:probability_3d} Three-dimensional detection probability
of $5\times10^{13}\mathrm{\mathrm{M_{\odot}}}$ galaxy groups with 80 ks exposure time. The x- and y-axes indicate  the redshift ($z$) and  off-axis angle $[\mathrm{arcmin}]$, respectively.
The z-axis represents the detection probability. A detection probability equal to one means that all simulated galaxy groups are detected as extended sources.}
\end{figure}

\begin{figure*}[]
\begin{centering}
\includegraphics[trim=30 90 80 130,clip,width=\textwidth]{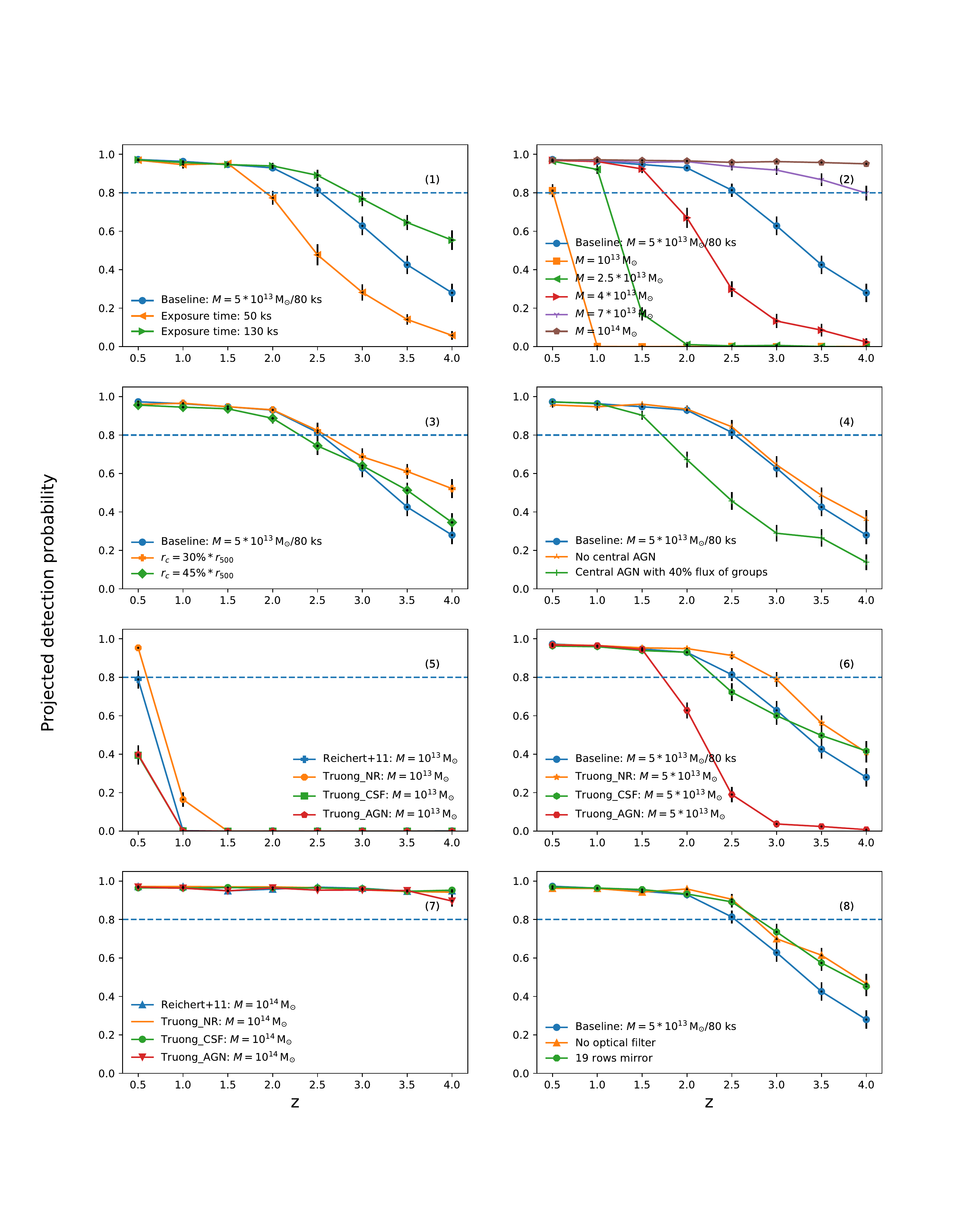}
\par\end{centering}
\caption{\label{fig:probability_1d} Projected detection probability averaged over the Athena FoV, which is a weighted probability from on-axis to the edge of the detector as described in Section~\ref{subsect:proj_det_prob}. The dashed blue horizontal line represents the projected detection probability at 0.8. Panel 1: Baseline model with exposure times of 50 ks, 80 ks, and 130 ks. Panel 2: Baseline model for different masses. Panel 3: Assuming different surface brightness profile: $r_{c} = 30\%\times r_{500}$ and $r_{c} = 45\%\times r_{500}$. Panel 4: Assuming there is no AGN in the center of galaxy groups, and assuming 40\% flux of the groups to the central AGN. Panel 5: Modified scaling relations of mass $10^{13}\,\mathrm{M_{\odot}}$. Panel 6: Modified scaling relations of mass $5\times10^{13}\,\mathrm{M_{\odot}}$. Panel 7: Modified scaling relations of mass $10^{14}\,\mathrm{M_{\odot}}$. Panel 8: Different Athena configurations used: without applying optical filter and with 19 mirror rows. The modified  scaling relations are based on \citet{Reichert:2011aa} and \citet{Truong:2018aa}.
}
\end{figure*}

\subsubsection{Three-dimensional detection probability for the baseline model \label{sec:res:baseline}}

The simulations created with our baseline model scenario described in Section~\ref{sec:bslnmodeldes} are processed with the source detection and characterization algorithms (see Section~\ref{subsec:Detection_and_charc}) for an exposure time of 80 ks. Once we applied the source criteria presented in Section~\ref{srcconta}, we obtained the group detection probability shown in Fig.~\ref{fig:probability_3d}. In this 3D histogram, the $x$-axis and $y$-axis represent the redshift and the off-axis angles, respectively, while the $z$-axis shows the group detection probability. A detection probability equal to $1$ means that all simulated galaxy groups are detected as extended sources.

The 3D detection probability shows that (i) at low redshift $z\leq1.5$, the detection probability remains high at any off-axis angle; (ii) the on-axis detection probability is lower than $5\arcmin$ off-axis owing to the shallower exposure time in the center, and (iii) at higher redshift ($z\geq2.0$), the detection probability starts to drop at an intermediate off-axis angle $\geq15\arcmin$. This is due to two combined effects. First, the flux of the galaxy groups at high-redshift drops and also their intrinsic sizes decrease\footnote{Based on scaling relations from  \cite{Reichert:2011aa}, a galaxy group with mass $5\times\ensuremath{10^{13}\ \mathrm{\mathrm{M_{\odot}}}}$ in the $0.2-2$~keV energy band has flux $1.69\times10^{-14}\ \text{ergs/s/cm}^{2}$ with $r_{500}=76.9\arcsec$ at $z=0.5$, whereas at $z=2.0$, a galaxy group with the same mass has flux $1.71\times10^{-15}\ \text{ergs/s/cm}^{2}$ with $r_{500}=32.5\arcsec$, and at $z=4.0$, the flux reduces to $9.58\times10^{-16}\ \text{ergs/s/cm}^{2}$ with $r_{500}=23.9\arcsec$.}. Second, due the increase in vignetting and PSF size with off-axis angle, sources result in fewer detected source photons and extended sources become harder to differentiate from point sources. Overall, Athena is able to detect the galaxy groups with high detection efficiency for redshifts up to 2.0 at any part of the WFI detector. Beyond redshift 2.5 and at intermediate off-axis angles the detection probability starts to decrease.

No galaxy groups with $M_{500}=5\times\ensuremath{10^{13}\ \mathrm{\mathrm{M_{\odot}}}}$ at $z\geq4.0$ and groups of mass $M_{500}=1\times\ensuremath{10^{14}\ \mathrm{\mathrm{M_{\odot}}}}$ at $z\geq3.0$ are expected to exist given our current knowledge (see first panel of Fig. \ref{fig:Mass_function_plot} in Section \ref{sec:hmf}). The determination of the detection probability is merely shown here to demonstrate that, should unexpectedly an early population of such groups exist (the so-called pink elephants), Athena would possibly detect them.

\subsubsection{Projected detection probability for the baseline model \label{sec:res:projected_baseline}}

Panel 1 in Figure. \ref{fig:probability_1d} compares the projected detection probability (see Section~\ref{subsect:proj_det_prob}) of our baseline model with those of simulations with two other exposure times, 50 and 130 ks. In general, galaxy groups can be detected as extended sources with more than 80\% probability up to redshift 2.5 as long as the exposure time is $\geq80$~ks. This probability is reduced to $\sim50\%$ for the 50~ks simulations.


\subsubsection{Galaxy groups with different masses}

As described in Section~\ref{subsec:Galaxy-groups}, we also simulated galaxy groups with masses of $M_{500}=1,~2.5,~4,~7,~10\times\ensuremath{10^{13}\ \mathrm{\mathrm{M_{\odot}}}}$ in the baseline scenario. We aim to check the masses and redshifts up to which Athena will be capable of detecting galaxy groups, and of estimating the number of expected groups (see Section~\ref{sec:hmf}).
The projected detection probability for different group masses is shown in panel 2 of Fig.~\ref{fig:probability_1d}. The corresponding {\tt extent}-{\tt extension-likelihood} planes for the different masses are shown in Appendix Fig.~\ref{fig: Extension_Li_EXT}.

We can see that Athena will be able to detect low-mass groups ($M_{500}<2.5\times\ensuremath{10^{13}\ \mathrm{\mathrm{M_{\odot}}}}$) only below $z<1$. For higher masses, $M_{500}\geq4\times\ensuremath{10^{13}\ \mathrm{\mathrm{M_{\odot}}}}$, Athena will detect such groups with a probability $\geq70\%$ up to $z\leq2.0$. The most massive groups with mass $1\times\ensuremath{10^{14}\ \mathrm{\mathrm{M_{\odot}}}}$ will be detected by Athena at any redshift.

\subsubsection{Surface brightness profiles with different core radii}

Active galactic nucleus feedback, star formation, and supernova explosions are well-known astrophysical processes that release a significant amount of energy into the surrounding interstellar and intra-cluster media. These feedback processes can redistribute the ICM into galaxy groups and clusters, causing a less concentrated surface brightness \citep[e.g.,][]{Eckert12}.
Also, some nearby galaxy groups with very flat surface brightness distributions were discovered recently that were missed in previous surveys \citep{Xu:2018aa}.
In consequence, galaxy groups no longer appear self-similar and their detection probability might be affected. 

To see this effect, as a first approximation we simulate a flatter surface brightness in galaxy groups. For this, we increased the core radius value from $0.15\times r_{500}$ to $0.30\times r_{500}$ and $0.45\times r_{500}$. We emphasize that the normalization of the surface brightness profiles is readjusted in order to maintain the total luminosity. The comparison of the projected detection probability for these scenarios is displayed in panel 3 of Fig.~\ref{fig:probability_1d}. We found that the flatter profile with $0.30\times r_{500}$ shown in orange has a higher detection probability than the baseline model, especially at $z\geq2.5,$ with a 66\%\ increase at $z=4.0$; while for the case of $0.45\times r_{500}$, the detection probability reduces at higher redshift (only 15\%\ increase compared to the baseline model), as shown in green. We also discovered that the detected groups' samples shown in green for $0.45\times r_{500}$ are more scattered and more extended than $0.30\times r_{500}$ in {\tt extent}-{\tt extension-likelihood} planes, as shown in Appendix Fig. \ref{fig: Extension_Li_EXT}. This result demonstrates the detection probability of high-redshift galaxy groups can be sensitive to the $r_{c}$ parameter, and we will discuss this in Section~\ref{Discussion}.

\subsubsection{No central AGN}

The presence of a central AGN in the galaxy groups can play an essential role in the detection of high-redshift groups, as we already discussed in Section~\ref{sec:bslnmodeldes}. To see if this could be a limiting factor, we simulated a scenario without a central AGN in the galaxy groups. The results are shown in panel 4 of Fig. \ref{fig:probability_1d}. There is a small increase in the detection probability, with a 5\%\ increase at $z=2.5$ and a 23\% increase at $z=4.0$ when central AGNs in the galaxy groups are removed. This slight increment in detection probability is due to the fact that the maximum likelihood fitting algorithm tends to give a higher value in extension likelihood, since there is no central "peak" within $r_{500}$. The absence of a central peak in the galaxy group can lead the detected source being more easily characterized as an extended source; as a consequence, the algorithm accepts the detection as an extended source, and thus gives a higher detection probability. We also perform a test where the emissivity of the central AGN is increased to have 40\% flux of the galaxy groups (20\% is assumed for the baseline model, see Section \ref{sec:bslnmodeldes}; the results are displayed as a green curve in panel 4 of Fig. \ref{fig:probability_1d}). We found there is an overall decrease in the detection probability at $z\geq1.5$ as compared to the baseline model, meaning there is a 27\% decrease at $z=2.0$, and a 54\% decrease at $z=3.0$ and at $z=4.0$.

\subsubsection{Luminosity evolution: Different scaling relations\label{subs:TR18_relation}}

Up to now, we have taken the group luminosity from the $L_{\mathrm{X}}\text{--}T$ relation measured by \citet[][]{Reichert:2011aa} (see Section~\ref{subsec:Galaxy-groups} for further details). RE11 used galaxy groups and clusters from $z=0$ to $z=1.46$ to derive their $L_{\mathrm{X}}\text{--}T$ relation. In general, this relation exhibits a flatter redshift evolution than the self-similar evolution. Since there are no comprehensive observational studies with higher redshift systems ($z\geq2.0$) that have derived an $L_{\mathrm{X}}\text{--}T$ relation, we have assumed that the scaling relation from RE11 holds for systems beyond $z\geq1.5$ for our baseline model.

In order to test quantitatively how well Athena will be able to rule out physical feedback models at high redshift, we select a different set of simulated $L_{\mathrm{X}}\text{--}T$ scaling relations from \citet[][hereafter TR18]{Truong:2018aa}. In their work, they analyzed three sets of hydrodynamical simulations that actually can reproduce a wide range of cluster properties, and they pushed the study of the evolution of the scaling relations to $z=2.0$. There are three different ICM descriptions in TR18: {\nr}  (non-radiative); {\csf}  (cooling, star formation and stellar feedback); and {\agn} (the same as the {\csf} scenario plus AGN feedback). The corresponding parameters are given in Table~\ref{tab:Truong_SC_App}. Those three models are good candidates for Athena to test its capability in differentiating various heating feedback models, especially at higher redshift regions, $z>1.5$. The TR18 relations predict, however, quite high X-ray luminosities compared to the extrapolation from observations. Without any change, all simulated groups at any redshift would be very easily detected, as a result. This seemed too optimistic to us. In our simulations, therefore, we made a change to bring TR18 scaling relations closer to the observed \cite{Reichert:2011aa} scaling relation. We scaled the TR18 {\agn} $L_{\mathrm{X}}\text{--}T$ scaling relation in luminosity at each redshift bin to match the RE11 luminosity at 1 keV. This luminosity shift in the {\agn} model is then applied to TR18 {\nr} and TR18 {\csf} models. The luminosity correcting factor is 0.122 at $z=2.0$. This luminosity shift only changed the constant ($C_{LT}$) in the $L_{\mathrm{X}}\text{--}T$ relation, while keeping the slope ($\beta$) the same. Since the slope is fixed, so the shape of different models in TR18 is strictly preserved as well as their \emph{relative} normalizations.

The impact of different scaling relations on the detection probability of high-redshift groups is displayed in panels 5 -- 7 of Fig. \ref{fig:probability_1d}. Panel 5 shows the detection probability for the groups with masses of $M_{500}=1.0\times\ensuremath{10^{13}\ \mathrm{\mathrm{M_{\odot}}}}$, which indicated that Athena is unlikely to detect galaxy groups with such small mass systems except at low redshift 0.5. Panel 6 shows the result for masses of $M_{500}=5.0\times\ensuremath{10^{13}\ \mathrm{\mathrm{M_{\odot}}}}$. We see that the detection probability of the TR18 {\csf} model is close to the baseline model, whereas in the TR18 {\agn} model, the detection probability drops significantly at $z\sim2.5$, due to the smaller\footnote{The smaller predicted luminosity for the TR18 {\agn} relation is because we firstly use $\ensuremath{M-T_{\mathrm{sl}}}$ to calculate the temperature of the galaxy groups, and then use $\ensuremath{L_{\mathrm{X}}-T_{\mathrm{sl}}}$ to obtain the luminosity. After that, we scale the TR18 {\agn} relation such that it has the \emph{same} luminosity as the RE11 at 1 keV.} predicted luminosity by TR18. The TR18 {\nr} is the brightest among the three models and therefore predicts the highest detection probability, $\sim80\%$ at $z<3.0$. Panel 7 shows the detection probability for masses of $M_{500}=1.0\times\ensuremath{10^{14}\ \mathrm{\mathrm{M_{\odot}}}}$, which shows that Athena will detect those massive objects at any redshifts with $\sim100\%$ probability.

\subsection{Different Athena configurations}

The final setup, payload, and design of Athena is still under development. In the following, different telescope configurations are tested in order to measure their impact on the detection of distant groups. 

\subsubsection{Optical filter\label{subsec:optical_filter}}

The WFI camera, based on arrays of Depleted field effect transistors (DEPFET) active pixel sensors, is sensitive to optical and UV photons. In general, an optical blocking filter is needed to avoid biasing the reconstructed X-ray photon energies in fields with bright optical and UV sources
\citep{Barbera:2015aa}. Therefore, the filter is included as our baseline.
However, if we assume the WFI survey fields are chosen to avoid the presence of optically bright sources, then we may be able to avoid usage of the optical blocking filter, which also reduces the effective area of Athena for soft X-ray photons (Fig.~\ref{fig:arf_and_vig}). To quantify the effect on the detection probability of the early groups, we simulated a scenario of Athena observation without applying this optical filter, and the result is shown in Fig. \ref{fig:probability_1d} panel 8. We see that the probability of detection increases as compared to the baseline model, especially at the high-redshift end, with an 11\%\ increase at $z = 2.5$ and a 67\%\ increase at $z = 4.0$. This increase in detection probability can be explained by the fact that Athena without the optical filter has a larger effective area than otherwise, in particular in the $0.2-2$~keV energy band (where most of our galaxy groups emission originated), as shown in Fig. \ref{fig:arf_and_vig}. 

\subsubsection{Number of mirror rows}

The optics technology based on silicon pore optics (SPO) consists of a set of high-quality Si plates, which are stacked together between small ribs \citep{Willingale:2013aa}. The final number of stacks of plates is not yet determined, although the baseline assumes 15 rows, which we also adopt in this work as a baseline. The final number of rows is subject to mass and cost constraints; it is conceivable that a small number of rows might be added. Therefore, we simulated a scenario of 19 mirror rows for comparison, which was a previous Athena baseline. The result is displayed in Fig.  \ref{fig:probability_1d} panel 8. There is an increase in detection probability with a 9.5\%\ increase at $z = 2.5$ and a 62\%\ increase at $z = 4.0$ for  the 19 rows configuration. We see that the 19 rows would increase the photon reception capability (see Fig. \ref{fig:arf_and_vig}), which would naturally increase the detection probability. This result shows that the effective area of Athena is quite an important factor for the detection of high-redshift galaxy groups. 

\begin{figure}[!ht]
\begin{centering}
\includegraphics[trim=10 80 40 120,clip,width=\columnwidth]{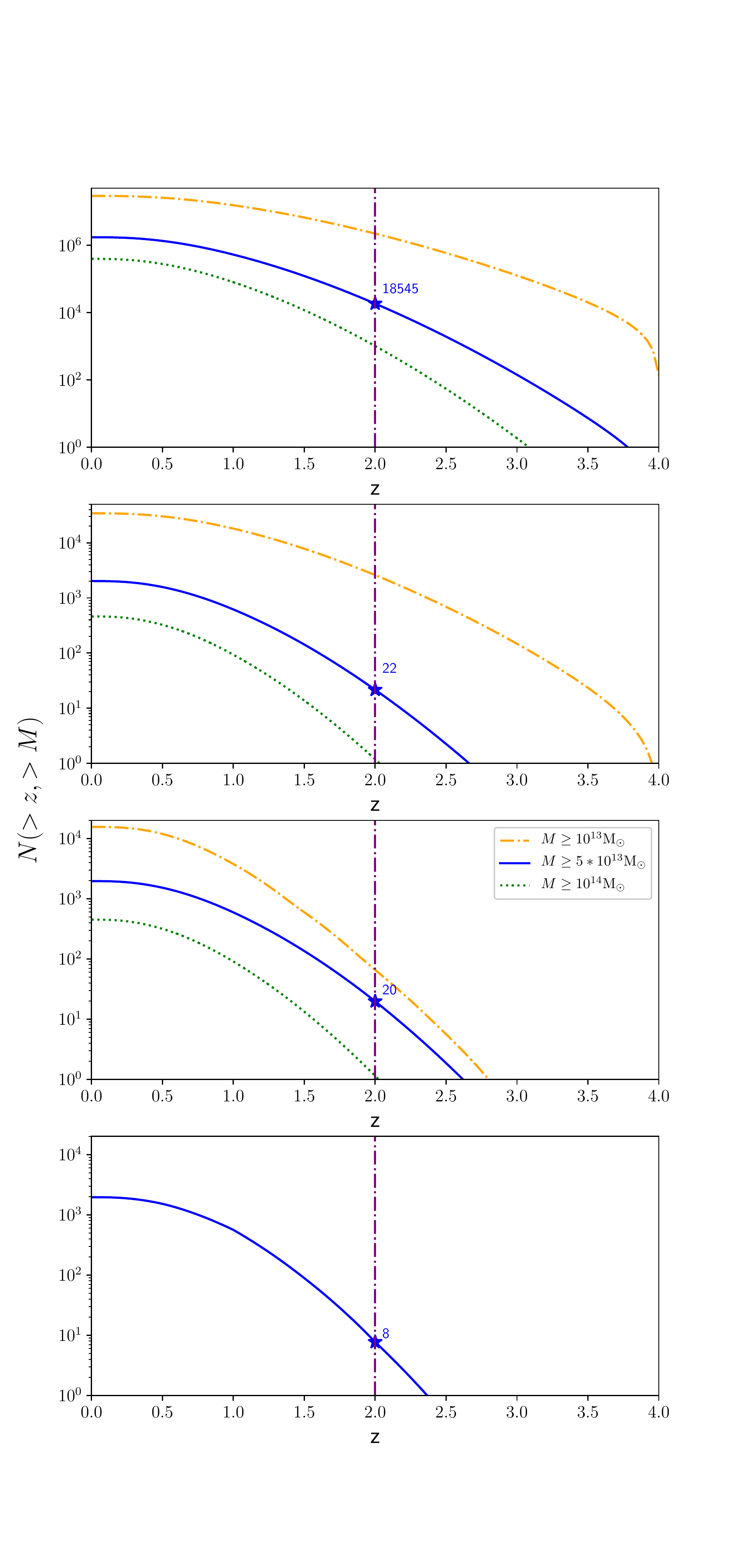}
\par\end{centering}
\caption{\label{fig:Mass_function_plot} Total number of galaxy groups of masses $M_{500}=10^{13}\,\mathrm{M_{\odot}}$, $5\times10^{13}\,\mathrm{M_{\odot}}$ , and $10^{14}\,\mathrm{M_{\odot}}$ as a function of redshift. Planck cosmology is assumed. First panel: Number of galaxy groups in the entire sky. Second panel: Number of galaxy groups in the Athena WFI survey area assumed for the nominal mission lifetime of four years (48 $\text{deg}^{2}$). Third panel: Same as the second panel, but taking into account the Athena detection efficiency for the RE11 baseline simulation in 80 ks exposure time. Fourth panel: Expected number of galaxy groups with temperature measurements with a precision $\Delta T/T\le 25\%$.}
\end{figure}

\subsection{ Expected number of galaxy groups detected by Athena\label{sec:hmf}}

\begin{table*}
\begin{centering}
\begin{tabular}{ccccccccc}
\hline
\hline
 Panel & Mass $[\mathrm{M_{\odot}}]$ & Mass function & $z\geq0$ & $z\geq0.5$ & $z\geq1.0$ & $z\geq1.5$ & $z\geq2.0$ & $z\geq2.5$\tabularnewline
\midrule 
 &  & tinker08 & 29416719  & 25968096 & 15546565 & 6715285 & 2229519 & 592860\tabularnewline
\cmidrule{3-9} 
 & $\geq\ensuremath{10^{13}}$ & watson13 & 29253536  & 25695123 & 15622655 & 6983156 & 2436322 & 682620\tabularnewline
\cmidrule{3-9} 
 &  & despali16 & 28430787 & 25107575 & 15124462 & 6561228 & 2203987 & 597631\tabularnewline
\cmidrule{2-9} 
 &  & tinker08 & 1738760 & 1354850 & 531448 & 123060 & \textbf{18545} & 1941\tabularnewline
\cmidrule{3-9} 
1st  & $\geq\ensuremath{5\times10^{13}}$ & watson13 & 2098401 & 1616700 & 633586 & 146831 & 22284 & 2336\tabularnewline
\cmidrule{3-9} 
 &  & despali16 & 1583691 & 1233233 & 489433 & 114486 & 17665 & 1921\tabularnewline
\cmidrule{2-9} 
 &  & tinker08 & 396586 & 279505 & 80316 & 11895 & 1019 & 54\tabularnewline
\cmidrule{3-9} 
 & $\geq\ensuremath{10^{14}}$ & watson13 & 524014 & 363217 & 102711 & 14826 & 1237 & 63\tabularnewline
\cmidrule{3-9} 
 &  & despali16 & 354015  & 249096 & 72717 & 10943 & 969 & 54\tabularnewline
\midrule 
2nd & $\geq\ensuremath{5\times10^{13}}$ & tinker08 & 1960  & 1522 & 593 & 135 & \textbf{20} & 1\tabularnewline
\midrule 
3rd & $\geq\ensuremath{5\times10^{13}}$ & tinker08 & 1960 & 1522 & 562 & 89 & \textbf{8} & 0\tabularnewline
\end{tabular}
\par\end{centering}
\caption{Comparison for the number of galaxy groups for different mass functions: tinker08 \citep[default model,][]{Tinker:2008aa}, watson13 \citep{watson2013halo}, and despali16 \citep{despali2016universality}. Planck18 cosmology is used for the calculation \citep{2018arXiv180706209P}. First panel: Number of galaxy groups in the whole sky (first panel in Fig. \ref{fig:Mass_function_plot}). Second panel: Number of galaxy groups expected to be discovered by Athena for a nominal mission lifetime of four years with sky coverage of 48 $\text{deg}^{2}$ (third panel in Fig. \ref{fig:Mass_function_plot}). The results are based on the RE11 baseline simulation. Third panel:  Expected number of galaxy groups with $\Delta T/T\le 25\%$ in the temperature measurements (fourth panel in Fig. \ref{fig:Mass_function_plot}). The three bold numbers in the table are indicated as blue stars in Fig. \ref{fig:Mass_function_plot}. \label{tab:no_groups_table_summary}}
\end{table*}

In this section, we show step-by-step how we estimate the expected number of galaxy groups that Athena will be able to discover and characterize. For this, we use the methodology described in Section~\ref{subsect:MF}.

The total number of galaxy groups in the whole sky as a function of redshift is displayed in the first panel of Fig.~\ref{fig:Mass_function_plot}. We calculate this number for Planck cosmology\footnote{We performed a test with WMAP~9 cosmology, and the results are similar: Planck18 predicts 18545 galaxy clusters with masses $\geq5\times10^{13}\,\mathrm{M_{\odot}}$ at $z\geq2.0$, while WMAP~9 predicts 18232 galaxy groups.} \citep[$\Omega_{\rm m}=0.3111$, $\sigma_{8}=0.8102$ and $H_{0}=67.66 \;\rm km \;s^{-1}\; Mpc^{-1}$,][hereafter Planck18]{2018arXiv180706209P}. The total numbers of galaxy groups are calculated for masses greater than $10^{13}\,\mathrm{M_{\odot}}$ (dash-dotted line), $5\times10^{13}\,\mathrm{M_{\odot}}$ (solid line), and $10^{14}\,\mathrm{M_{\odot}}$ (dotted line), which are shown in the figure. Since different mass functions have been proposed in the last ten years, we therefore compare the number of galaxy groups in the whole sky for different mass functions. We use tinker08 \citep{Tinker:2008aa}, watson13 \citep{watson2013halo}, and despali16 \citep{despali2016universality}, obtained from COLOSSUS python package (see Section~\ref{subsect:MF}) for the comparison. The reason is they have the same mass definition with spherical over-density (SO) and include a redshift dependence in the package. The results are displayed in first panel of Table \ref{tab:no_groups_table_summary}. We found that the three mass functions predict similar number of groups with masses $\geq5\times10^{13}\,\mathrm{M_{\odot}}$ at $z\geq2.0$ in the whole sky, ranging from 17665 (despali16) to 22284 (watson13), and tinder08 predicts 18545, which is between the models of watson13 and despali16.

In a second step, we include the planned Athena survey area ($\Omega_{\text{survey}}=48\,\text{\ensuremath{\deg}}^{2}$, see Section~\ref{subsect:MF}) in the calculation of the expected number of halos above a given redshift. The results are shown in the second panel of Fig. \ref{fig:Mass_function_plot}. As expected, the number of galaxy groups and clusters reduces significantly by almost 99.8\% (from 18545 to 22) at $z=2.0$ due to the fact that the survey area covers only a tiny fraction of the whole sky.

Finally, by taking into account the group detection efficiency of Athena (baseline model, see Section~\ref{sec:bslnmodeldes} and Section~\ref{sec:res:projected_baseline}), by integrating the $P_{(z,M)}$ in Eq.~(\ref{eq:MF}), we obtain the expected number of galaxy groups to be detected by Athena. The results are displayed both in the third panel of Fig.~\ref{fig:Mass_function_plot} and in the second panel in Table \ref{tab:no_groups_table_summary}. We found that for the baseline model (see Section~\ref{sec:bslnmodeldes}), Athena will be able to detect 20 galaxy groups at $z\geq2.0$ with $M_{500}\geq5\times 10^{13}\,\mathrm{M_{\odot}}$. From the figure, we see that it is unlikely that Athena will detect galaxy groups of mass $M_{500}\geq10^{14}\,\mathrm{M_{\odot}}$ beyond redshift $2.0$ because of the limited sky area covered. By comparing the first three panels of Fig. \ref{fig:Mass_function_plot}, we found that (i) the survey area is the limiting factor for Athena's detection of high-redshift galaxy groups, (ii) Athena has great detection efficiency in discovering the systems, where the expected number of groups only reduces by $\sim10\%$ (i.e., from 22 groups to 20).


\subsection{Temperature determination and $L_{\mathrm{X}}\text{--}T$ scaling relation }

One of the science objectives of Athena is to determine the physical processes that dominate the injection of non-gravitational energy, which is the energy released that is not due to the gravitational collapse of the gas, into the ICM. These non-gravitational processes include, for example, the feedback produced by the outflows of supernova, or heating by jets from the central AGN \cite[e.g.,][]{Ettori:2013ac,rau2016athena}. In this section, the temperature determination of simulated high-redshift groups is presented. A simple test is performed to evaluate if Athena is able to differentiate the distinct physically motivated scenarios for the heating mechanisms.

\subsubsection{Spectral fitting results for different off-axes and redshifts \label{subs:specral_fitting_plot}} 

 \begin{figure*}
 \begin{centering}
 \includegraphics[trim=20 40 80 80,clip,width=0.85\textwidth]{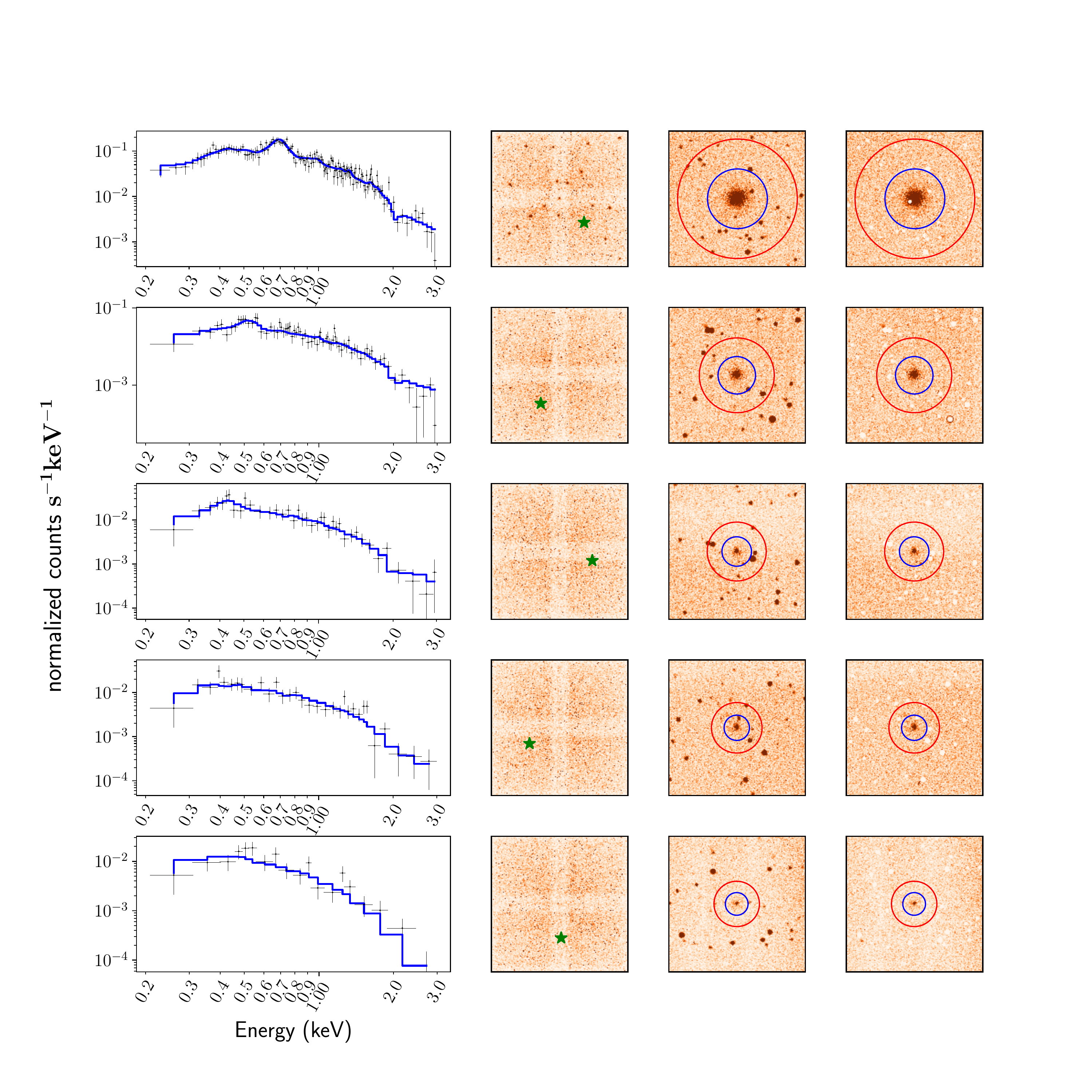}
 \par\end{centering}
 \caption{\label{fig:spec_IM_diff_z} Simulated background-subtracted WFI spectrum of a  $5\times10^{13}\,\mathrm{M_{\odot}}$ group at $z=0.5,~1.0,~1.5,~2.0,~2.5$ (from top to bottom), at an off-axis angle of $10\arcmin$, and an exposure time of  80 ks. First panel (from the left): Galaxy group spectra with the background subtracted. Second panel:  Extracted location of galaxy group in Athena FoV. Third panel: zoom-in view of the galaxy groups (box size $6\arcmin\times6\arcmin$); the blue circle represents $r_{500}$; the red circle is $2.0\times r_{500}$; the annulus is used for background subtraction. Fourth panel: Detected AGNs are excluded with a PSF corresponding to a 90\%  encircled energy fraction (99\% for the bright ones outside $r_{500}$).}
 \end{figure*}

\begin{figure*}
\begin{centering}
\includegraphics[trim=90 0 90 60,clip,width=\textwidth]{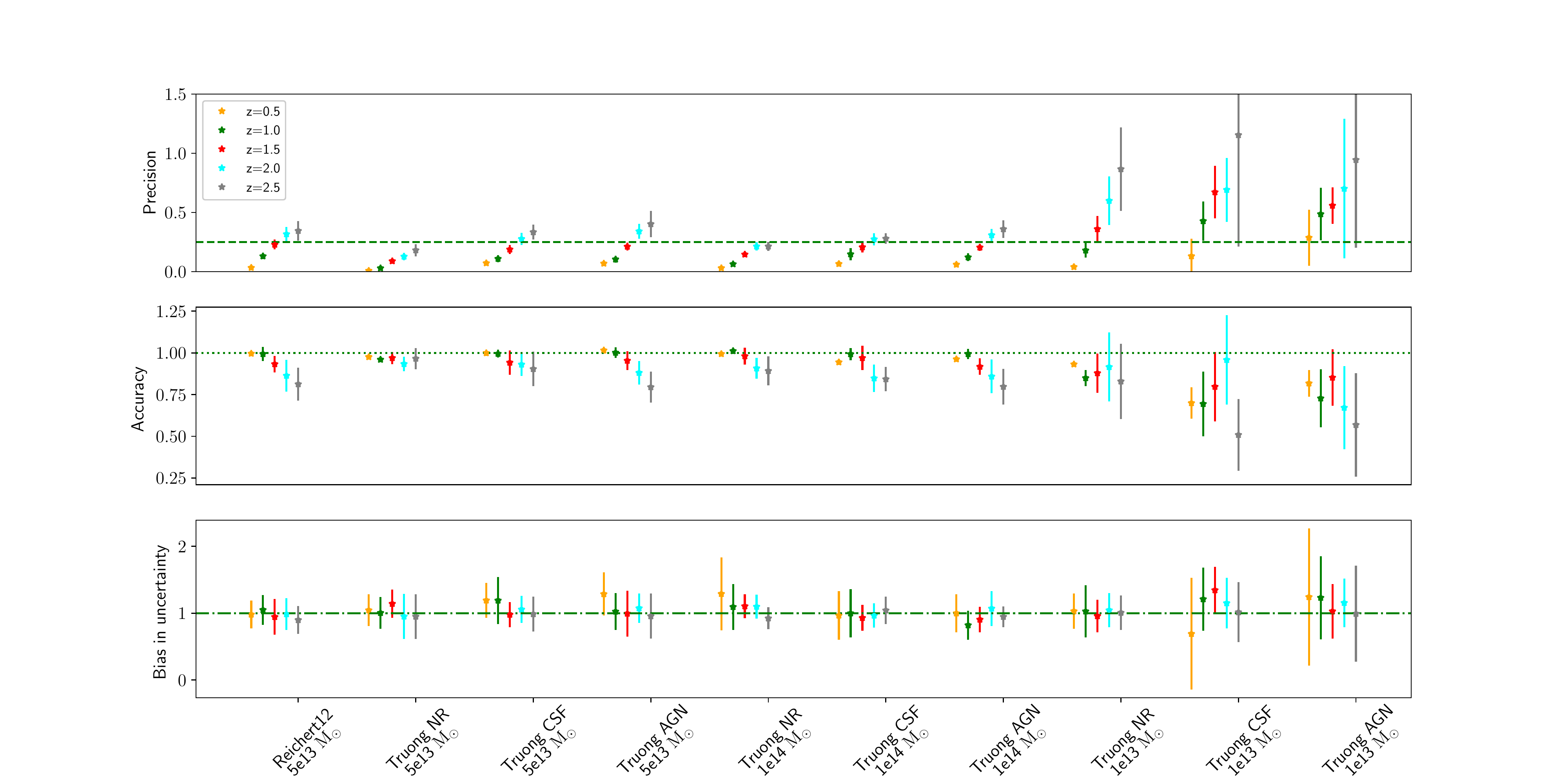}
\par\end{centering}
\caption{\label{fig:Temp_stat_fig} Temperature statistics for the simulations  by RE11 and TR18 (Truong {\nr}, Truong {\csf} and Truong {\agn}) in 80 ks exposure time, with masses of $M_{500}=1\times 10^{13}\,\mathrm{M_{\odot}}$, $5\times10^{13}\,\mathrm{M_{\odot}}$ and $10^{14}\,\mathrm{M_{\odot}}$. Top: Median precision of the temperature fit. The green dashed line indicates 25\% precision. Middle:  Accuracy of the temperature measurements. Bottom: Bias in uncertainty. The precision, accuracy, and bias in uncertainty are defined in Section~\ref{subsubsec:spec_fit}.}
\end{figure*}

The spectra of galaxy groups are extracted and fitted using \noun{Xspec} as described in Section~\ref{subsubsec:spec_fit}. The spectral fitting is applied on those galaxy groups that have been detected as extended sources in our baseline simulations with an exposure time of 80~ks, where the groups have a mass value of $M_{500}=5\times10^{13}\,\mathrm{M_{\odot}}$ (see Section~\ref{sec:res:baseline}). Given that we assume these high-redshift groups are contaminated by AGN emission, we fit the background-subtracted spectrum of the group with an {\tt apec+pow} model, where {\tt pow} describes the emission of the central AGN (or, in fact, the excess sum of emission above background from all AGNs that have not been detected as such and are included in the spectral extraction region). A fixed spectral index value of $\alpha=1.42$ is used for the fitting in our work, while in an actual analysis this parameter may be left free to vary within some reasonable range. Assuming that we perfectly know this value is optimistic. However, the detection and first characterization of these high-redshift groups is likely to trigger further follow-up observations, for  example, with the X-IFU, which will allow us to determine the temperature  with much higher precision even than determined here. Figure~\ref{fig:spec_IM_diff_z} shows one example of background-subtracted WFI spectra of galaxy groups located at a fixed off-axis angle of $10\arcmin$, at $z=0.5,~1.0,~1.5,~2.0,~2.5$ in the 0.2- 3~keV energy band. A similar plot with spectra of groups located at distinct off-axis angles, $\theta=5,~10,~15,~18,~25$~arc\-min, but at redshift $z=2.0$ is discussed in Appendix \ref{spec_off_axis_fix_z} (see Fig. \ref{fig:spectra_z_diff_off_axis}). In Fig.~\ref{fig:spec_IM_diff_z} the spectrum at redshift $2.5$ shows very few counts because the targeted group is located at the gap of the detector with shallower exposure time. No spectrum is shown beyond redshift $2.5$, given that Athena will not be able to detect any galaxy groups at $z\geq3.0$ (see in Fig. \ref{fig:Mass_function_plot}).

\subsubsection{Gas temperature statistics\label{gas_temp}}

In order to obtain statistics on the determined temperature, we perform the spectral analysis for the simulations RE11 and TR18 (Truong {\nr}, Truong {\csf,} and Truong {\agn} with group masses of $M_{500}=1,~5,~10\times\ensuremath{10^{13}\ \mathrm{\mathrm{M_{\odot}}}}$) with an exposure time of 80 ks. We filtered out bad fits that have their 68\% confidence interval calculated by the {\tt error}-command not including the best-fit value. Then the temperature precision, temperature accuracy, and accuracy of error estimation, as described in Section~\ref{subsubsec:spec_fit}, are calculated for each scenario, and the results are displayed in Fig. \ref{fig:Temp_stat_fig}.

For the baseline RE11 model, the temperature can be measured to a (median) precision better than 25\% up to $z=1.5$, but it drops to $30-35$\% for higher redshifts. For the TR18 {\nr}, {\csf,} and {\agn} scenarios, we found: (i) the temperature precision of the TR18 {\nr} can achieve better than $25\%$ at all redshifts due to its high predicted luminosity; (ii) for the TR18 {\agn} the temperature precision can only achieve up to $\sim25\%$ at lower redshift ranges with $z\leq1.5$ (at higher redshift ranges with $z\geq2.0$, the precision is only $\sim37-40\%$); (iii) the TR18 {\csf} model is very close to the RE11 model, and this is because the two scenarios have a similar $L_{\mathrm{X}}\text{--}T$ scaling relation. For higher masses of $M_{500}=10^{14}\,\mathrm{M_{\odot}}$, the temperature precision is, in general, better than the one for the baseline model; whereas for lower masses of $M_{500}=10^{13}\,\mathrm{M_{\odot}}$, the temperature measurements can still achieve 25\% for the low redshift range with $z=0.5$.

 The accuracy of the temperature measurement gradually decreases from unity at $z=0.5$ to about $0.8$ at $z=2.5$ for all scenarios, as shown in the middle panel of Fig. \ref{fig:Temp_stat_fig}. We see that the accuracy for the lower mass (with $M_{500}=10^{13}\,\mathrm{M_{\odot}}$) has a large bias, which indicates Athena cannot measure the temperature accurately enough using WFI observations; high spectral resolution X-IFU observations may be needed for such low-mass objects. Finally the bias in uncertainty is similar for all simulated scenarios and scattered around unity (as shown in the bottom panel of Fig. \ref{fig:Temp_stat_fig}).

\subsubsection{Comparison between TR18 models through the $L_{\mathrm{X}}\text{--}T$ relation and the number of groups expected by TR18 models \label{L_T_relation}}

\begin{table*}[!ht]
\begin{centering}
\begin{tabular}{ccccccc}
\hline 
\hline
 & $z$ & %
\begin{tabular}{c}
0.5\tabularnewline
\hline 
$\int_{0.5}^{1}dz$\tabularnewline
\end{tabular} & %
\begin{tabular}{c}
1.0\tabularnewline
\hline 
$\int_{1.0}^{1.5}dz$\tabularnewline
\end{tabular} & %
\begin{tabular}{c}
1.5\tabularnewline
\hline 
$\int_{1.5}^{2.0}dz$\tabularnewline
\end{tabular} & %
\begin{tabular}{c}
2.0\tabularnewline
\hline 
$\int_{2.0}^{2.5}dz$\tabularnewline
\end{tabular} & %
\begin{tabular}{c}
2.5\tabularnewline
\hline 
$\int_{2.5}^{\infty}dz$\tabularnewline
\end{tabular}\tabularnewline
\hline 
 & %
\begin{tabular}{c}
$M=\ensuremath{10^{13}\,\mathrm{M_{\odot}}}$\tabularnewline
$\int_{10^{13}\,\mathrm{M_{\odot}}}^{5\times10^{13}\,\mathrm{M_{\odot}}}dM$\tabularnewline
\end{tabular} & 7191 & 2716 & 415 & 42 & 4\tabularnewline
\cline{2-7} 
RE11 & %
\begin{tabular}{c}
$M=\ensuremath{5\times10^{13}\,\mathrm{M_{\odot}}}$\tabularnewline
$\int_{5\times10^{13}\,\mathrm{M_{\odot}}}^{10^{14}\,\mathrm{M_{\odot}}}dM$\tabularnewline
\end{tabular} & 704 & 380 & 104 & 17 & 2\tabularnewline
\cline{2-7} 
 & %
\begin{tabular}{c}
$M=\ensuremath{10^{14}\,\mathrm{M_{\odot}}}.$\tabularnewline
$\int_{10^{14}\,\mathrm{M_{\odot}}}^{\infty\,\mathrm{M_{\odot}}}dM$\tabularnewline
\end{tabular} & 225 & 77 & 12 & 1 & 0\tabularnewline
\hline 
 & %
\begin{tabular}{c}
$M=\ensuremath{10^{13}\,\mathrm{M_{\odot}}}$\tabularnewline
$\int_{10^{13}\,\mathrm{M_{\odot}}}^{5\times10^{13}\,\mathrm{M_{\odot}}}dM$\tabularnewline
\end{tabular} & %
\begin{tabular}{c}
7052\tabularnewline
3972\tabularnewline
4003\tabularnewline
\end{tabular} & %
\begin{tabular}{c}
2525\tabularnewline
1811\tabularnewline
1822\tabularnewline
\end{tabular} & %
\begin{tabular}{c}
812\tabularnewline
798\tabularnewline
694\tabularnewline
\end{tabular} & %
\begin{tabular}{c}
249\tabularnewline
226\tabularnewline
121\tabularnewline
\end{tabular} & %
\begin{tabular}{c}
64\tabularnewline
50\tabularnewline
8\tabularnewline
\end{tabular}\tabularnewline
\cline{2-7} 
TR18 & %
\begin{tabular}{c}
$M=\ensuremath{5\times10^{13}\,\mathrm{M_{\odot}}}$\tabularnewline
$\int_{5\times10^{13}\,\mathrm{M_{\odot}}}^{10^{14}\,\mathrm{M_{\odot}}}dM$\tabularnewline
\end{tabular} & %
\begin{tabular}{c}
704\tabularnewline
699\tabularnewline
702\tabularnewline
\end{tabular} & %
\begin{tabular}{c}
381\tabularnewline
378\tabularnewline
379\tabularnewline
\end{tabular} & %
\begin{tabular}{c}
104\tabularnewline
103\tabularnewline
94\tabularnewline
\end{tabular} & %
\begin{tabular}{c}
17\tabularnewline
16\tabularnewline
11\tabularnewline
\end{tabular} & %
\begin{tabular}{c}
2\tabularnewline
2\tabularnewline
1\tabularnewline
\end{tabular}\tabularnewline
\cline{2-7} 
 & %
\begin{tabular}{c}
$M=\ensuremath{10^{14}\,\mathrm{M_{\odot}}}.$\tabularnewline
$\int_{10^{14}\,\mathrm{M_{\odot}}}^{\infty\,\mathrm{M_{\odot}}}dM$\tabularnewline
\end{tabular} & %
\begin{tabular}{c}
226\tabularnewline
224\tabularnewline
224\tabularnewline
\end{tabular} & %
\begin{tabular}{c}
77\tabularnewline
77\tabularnewline
76\tabularnewline
\end{tabular} & %
\begin{tabular}{c}
12\tabularnewline
12\tabularnewline
12\tabularnewline
\end{tabular} & %
\begin{tabular}{c}
1\tabularnewline
1\tabularnewline
1\tabularnewline
\end{tabular} & %
\begin{tabular}{c}
0\tabularnewline
0\tabularnewline
0\tabularnewline
\end{tabular}\tabularnewline
\end{tabular}
\par\end{centering}
\caption{Expected number of galaxy groups to be detected by Athena at each redshift and mass bin based on the RE11 baseline simulation and simulations of TR18 {\nr} (top value), {\csf} (middle value), and {\agn} (bottom value). The numbers are calculated by integrating Eq.~(\ref{eq:MF}) with the limits of the integration shown in the integrals. Planck18 cosmology is assumed for the calculation \citep{2018arXiv180706209P}. \label{tab:LT_num_table_Baseline}}
\end{table*}

The $L_{\mathrm{X}}\text{--}T$ relation is one of the most studied X-ray scaling relations because both temperature and luminosity can easily be measured from X-ray data. We know that the current observed $L_{\mathrm{X}}\text{--}T$ relation deviates away from the self-similar prediction, which indicates that possible non-gravitational processes -- feedback from AGN or supernovae -- happened in the early formation era of galaxy groups \citep[e.g.,][]{Giodini:2013ab}. Moreover, different feedback models can result in different $L_{\mathrm{X}}\text{--}T$ relation predictions and observations of high-redshift galaxy groups are essential because the predictions by the non-gravitational feedback processes only differ strongly at high-redshifts. Athena will be ideal for studying this due to its unprecedented sensitivity and high efficiency in discovering high-redshift groups.

In an attempt to test if Athena/WFI can differentiate between different feedback models, we used the simulation of TR18 {\csf} as the observed data with three simulated masses, $M_{500}=10^{13}\,\mathrm{M_{\odot}}$, $5\times10^{13}\,\mathrm{M_{\odot}}$ , and $10^{14}\,\mathrm{M_{\odot}}$. We note that for the smallest mass bin of $M_{500}=10^{13}\,\mathrm{M_{\odot}}$, since the detection probability is zero at redshifts $z=~1.0,~1.5,~2.0,~2.5$ as shown in green in panel 5 of Fig. \ref{fig:probability_1d}, we excluded the data points for this mass bin at redshift $z\geq1.0$. The final numbers at each redshift and mass bin are calculated by integrating the Eq.~(\ref{eq:MF}) as described in Section~\ref{subsect:MF}. Table~\ref{tab:LT_num_table_Baseline} summarizes the expected numbers of galaxy groups to be detected by Athena for the baseline model RE11 and TR18 models of {\nr} (top value), {\csf} (middle value), and {\agn} (bottom). From this, we found that the TR18 {\csf} is expecting to detect 19 galaxy groups at $z\geq2.0$ with $M_{500}\geq5\times 10^{13}\,\mathrm{M_{\odot}}$, and TR18 {\nr} can detect 20 groups, while TR18 {\agn} can only detect 13. We emphasize the numbers calculated in the TR18 table are based on bi-linear interpolation of only three masses ($M_{500}=1,~5,~10\times\ensuremath{10^{13}\ \mathrm{\mathrm{M_{\odot}}}}$). This sparse interpolation in the lower mass range might lead to an imperfect number estimation when performing the integration from mass $1\times\ensuremath{10^{13}\ \mathrm{\mathrm{M_{\odot}}}}$ to $5\times\ensuremath{10^{13}\ \mathrm{\mathrm{M_{\odot}}}}$ , but this will not affect the main conclusion we draw for this section.

\begin{figure*}[!ht]
\begin{centering}
\includegraphics[trim=90 20 90 30,clip,width=\textwidth]{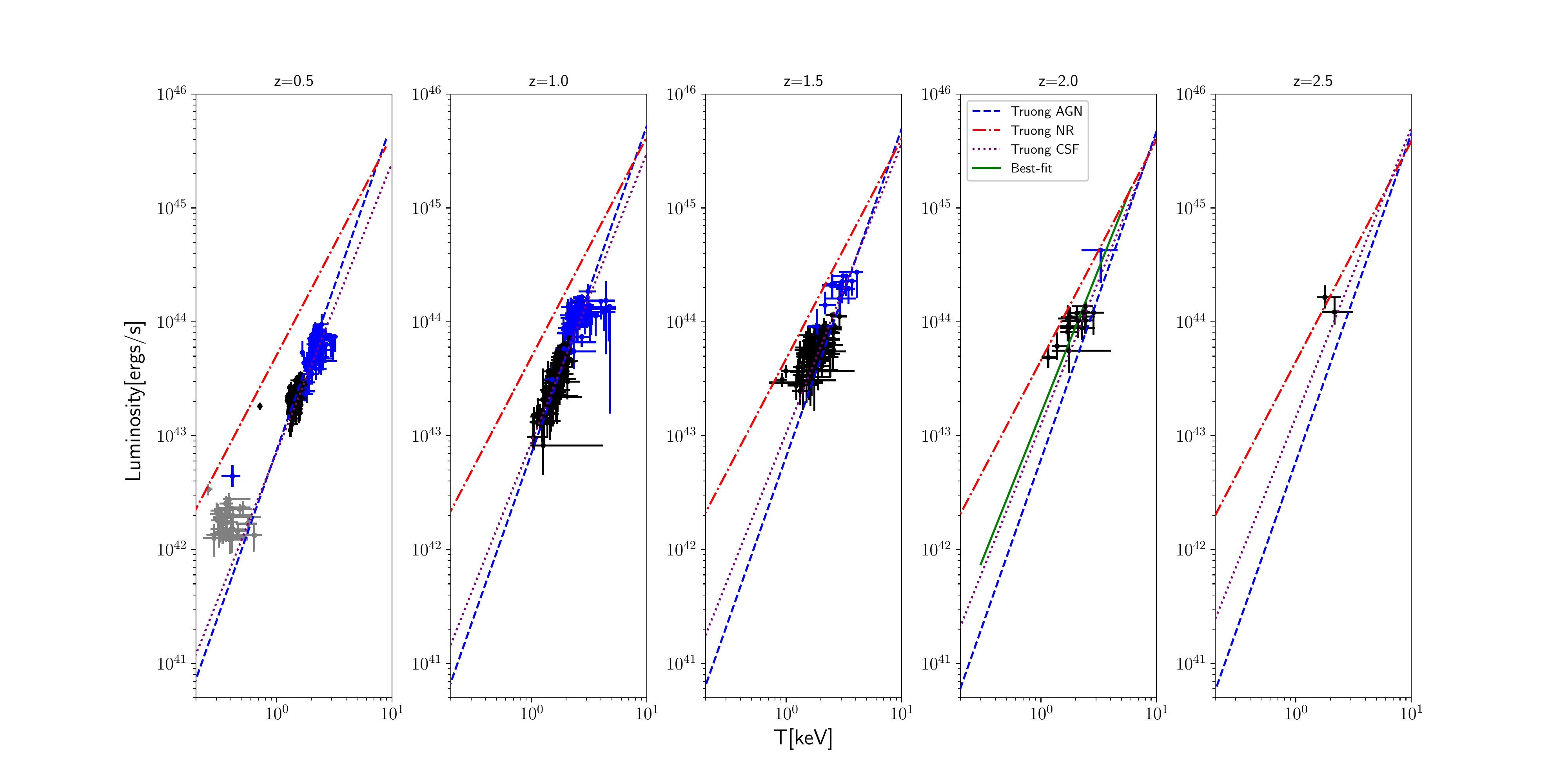}
\par\end{centering}
\caption{\label{fig:Truong_CSF_LT} Observed $L_{\mathrm{X}}\text{--}T$ relation based on TR18 {\csf} simulation. The TR18 {\nr},  {\csf,}  and {\agn} models are shown in red (dash-dotted), purple (dotted), and blue (dashed) lines. The gray, black, and blue data points represent the simulations with masses of $M_{500}=10^{13}\,\mathrm{M_{\odot}}$, $5\times10^{13}\,\mathrm{M_{\odot}}$ , and $10^{14}\,\mathrm{M_{\odot}}$ respectively. The number of data points are the expected number of groups to be detected by Athena for the TR18 {\csf} model (which is given in the middle values in Table~\ref{tab:LT_num_table_Baseline}). No data points are shown for the mass of $M_{500}=10^{13}\,\mathrm{M_{\odot}}$ beyond $z\geq1.0$ due to the zero detection probability predicted in panel 5 of Fig. \ref{fig:probability_1d}.  The solid green line is the linear fit result with a fixed slope from the TR18 {\csf} model.}
\end{figure*}

The final recovered $L_{\mathrm{X}}\text{--}T$ relation of the TR18 {\csf} model is shown in Fig. \ref{fig:Truong_CSF_LT}. It is evident that the TR18 {\nr} model shown as the red dash-dotted line does not pass through the observed data at the lower redshift range from 0.5 to 1.5, and the observed data strongly prefer the model of TR18 {\csf} and TR18 {\agn}. However, the model prediction of TR18 {\csf} and TR18 {\agn} is too similar at lower redshift. It is not sufficient to differentiate which model is better based on the low redshift observed data. For this, we performed a simple linear fit with a fixed slope (with TR18 {\csf} $\beta$ as the input value ) to the observed $L_{\mathrm{X}}\text{--}T$ relation at redshift $z=2.0$, and obtained the normalization of the TR18 {\csf} model $\log_{10}(C\times E(2.0)^{\gamma})=3.303\pm0.536$ with a relative uncertainty of 16\%. The relative offset between the model normalizations for all three TR18 models can be defined as 
\begin{equation}
\delta_{\mathrm{CSF-X}}(z)=\frac{\log\left(C_{\mathrm{CSF}}E(z)^{\gamma_{\mathrm{CSF}}}\right)-\log\left(C_{\mathrm{X}}E(z)^{\gamma_{\mathrm{X}}}\right)}{\mathrm{\log}\left(C_{\mathrm{CSF}}E(z)^{\gamma_{\mathrm{CSF}}}\right)}\,.
\end{equation}
We found the normalization offsets between the models to be  $\delta_{\mathrm{CSF-NR}}(2.0)\simeq43\%$ and $\delta_{\mathrm{CSF-AGN}}(2.0)\simeq23\%$. This result demonstrates that the galaxy groups detected at higher redshifts can distinguish the TR18 {\csf} and TR18 {\agn} in the sense that the relative uncertainties from observed data (16\%) are much smaller than the model offset predictions (43\% and 23\%).

\subsection{Expected number of detected galaxy groups with $\Delta T/T\leq25\%$:}

\begin{figure}[!ht]
\begin{centering}
\includegraphics[trim=15 20 40 50,clip,width=\columnwidth]{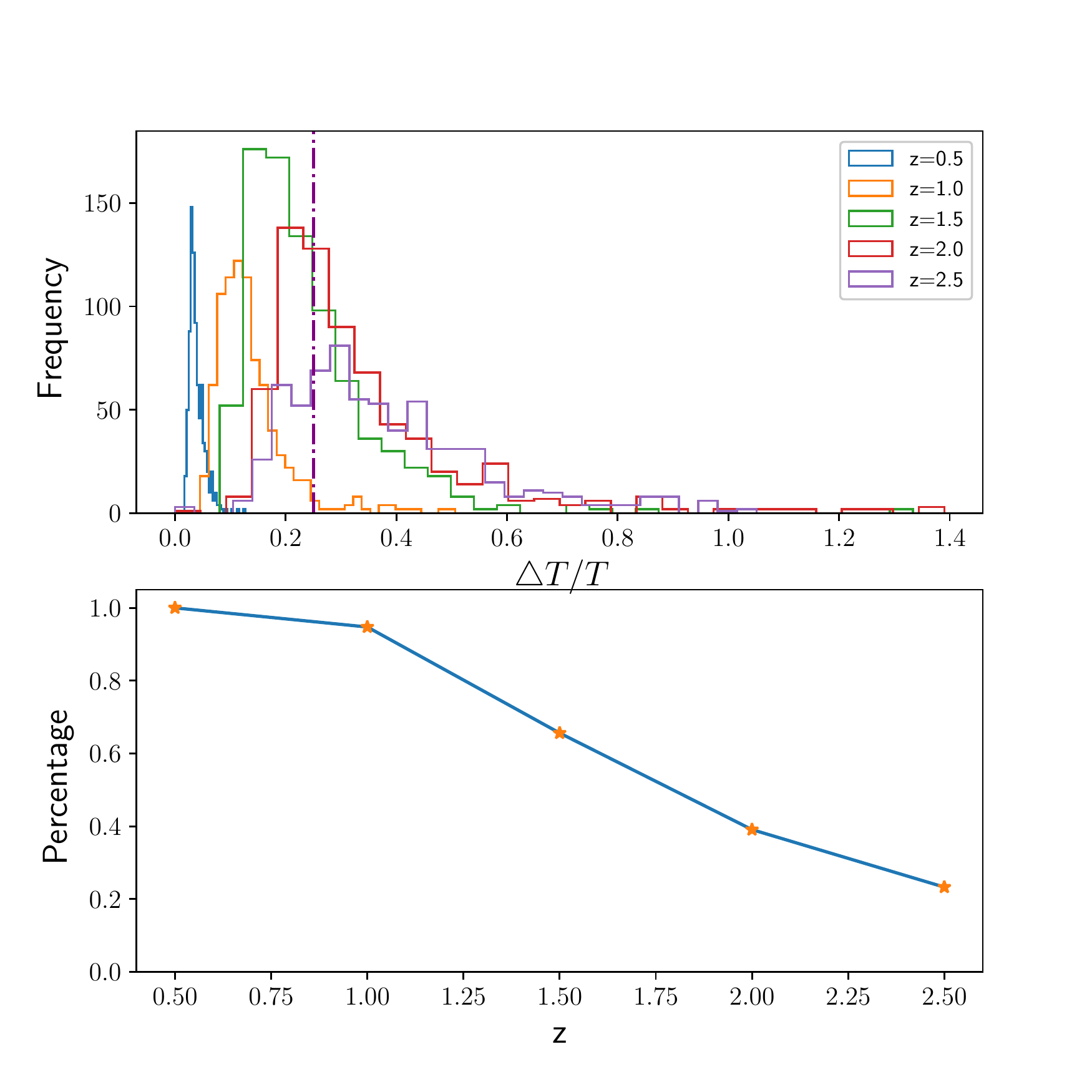}
\par\end{centering}
\caption{\label{fig:Fraction_T_precision} Top: Histogram of $\Delta T/T$ for each redshift bin. The purple dotted line represents the 25\% precision. Bottom: Estimated area under the histograms in the above plot at  $\Delta T/T=25\%$ cut. This area represents the fraction of detected groups at a given redshift that can achieve a temperature precision measurement better than 25\%.}
\end{figure}

As we saw in Section~\ref{sec:hmf}, Athena can discover about 20 galaxy groups at $z\geq2.0$ with $M_{500}\geq5\times 10^{13}\,\mathrm{M_{\odot}}$ over its four year mission. In this section, we estimate how many among those 20 detected groups can achieve precise temperature measurements with $\Delta T/T\leq25\%$. We first calculate the fraction of galaxy groups with $\Delta T/T\leq25\%$ at redshift bins $z=0.5,~1.0,~1.5,~2.0,~2.5$. This is obtained by estimating the area under the histogram, as shown in the top panel of Fig. \ref{fig:Fraction_T_precision}. The bottom panel shows the fraction curve as a function of redshift in the same figure. This fraction curve can then be multiplied by the expected number of detected groups to give the required number. As a result, we found among the 20 galaxy groups ($z\geq2.0$, shown in the third panel of Fig.  \ref{fig:Mass_function_plot} ), about eight galaxy groups can achieve the precise temperature with $\Delta T/T\leq25\%$, which is displayed both in the fourth panel of Fig.  \ref{fig:Mass_function_plot} and in the third panel in Table \ref{tab:no_groups_table_summary}. If $\Delta T/T\leq20\%$ is required, then only about five galaxy groups can achieve the goal.

\section{Discussion\label{Discussion}}

Many factors can determine whether a high-redshift galaxy group can be detected in X-rays. These factors include exposure time, telescope design, and the physical features of the high-redshift galaxy group (e.g., presence of a central AGN, the flatness of the surface brightness profile, and different scaling relations). In the following section, we discuss the effects of these factors in more detail.

\begin{enumerate}
\item Exposure time: It is clear that the longer the exposure is, the better the efficiency of  Athena in discovering the distant groups. Our baseline model assumes an exposure time of 80 ks, which is in agreement with the current WFI's MOP of 103 pointings with an exposure time of 84 ks. Based on these 80 ks simulated results, Athena would have a $\geq80\%$ chance of detecting galaxy groups at $z\leq2.5$, and have a $\geq90\%$ detection efficiency for 130 ks exposure time.  This reduces to $\sim50\%$ for 50 ks exposure time, which demonstrates that already moderately deep exposures are sufficient for a high detection probability. Therefore, the total number of galaxy groups to be discovered by Athena could potentially increase significantly if -- for a fixed total observing time -- the WFI survey covered a larger area with less exposure per pointing.
\item Specific physical characteristics of high-redshift galaxy groups: 
\begin{itemize}
\item Presence of a central AGN: Adding bright X-ray AGN(s) to the galaxy group can have two competing effects. On the one hand, the X-ray emission from the AGN can boost the surface brightness of the galaxy cluster, which can enhance the detection probability. On the other hand, the peak X-ray emission from the AGN can stand out more easily as a point-like source as opposed to the emission from galaxy groups; this can decrease the probability of detecting the source as extended, which we require here. From our simulation, we found the presence of a central AGN will not significantly affect the overall Athena detection of galaxy groups, except at higher redshifts ($z=3.5,~4.0$) where a slightly lower detection probability is seen. This slight detection probability drop at higher redshift indicates the increasing effect of AGN emissions, which causes the extended emission to be overlooked. However, we stress that this effect is small because the predictions of the two models are within the error bars, as shown in Fig. \ref{fig:probability_1d} panel 4. On the other hand, at those high redshifts, the probability of the presence of AGNs within the groups may increase such that possibly several AGNs may contaminate the total emission. Therefore, we suggest that this be studied further in the future.
\item Core radius of surface brightness profile: Changing the surface brightness profile of galaxy groups can also have two competing effects. A larger core radius $r_{c}$ will make groups appear more extended, which can increase the detection probability; but at the same time the X-ray emission is spread out in a very extended region, which makes it more difficult for a galaxy group to stand out from the X-ray background, and thus lower the detection probability. In our results, we found that increasing the core radius to 30\% of $r_{500}$ can increase the detection probability at higher redshift $z\geq2.5$ (first effect). However, this detection probability at $z\geq2.5$ dropped for the core radius if extended to 45\% of $r_{500}$ (second effect). This demonstrates that the $r_{c}$ can be a crucial factor influencing the detection of galaxy groups, especially at higher redshifts. Besides, we note that while the spherically symmetric $\beta$-model we used in the simulations is fixed at $\beta=2/3$, the real groups at $z\geq2.0$ may have a different slope on average because those objects might still be at the formation stage, and different beta values are possible. Therefore, once a significant number of such groups are detected by Athena, their surface brightness profile distributions should be studied in detail. This may additionally help constrain feedback mechanisms because, for example, stronger AGN feedback may result in gas being pushed out to larger radii, modifying the outer surface brightness slope.
\end{itemize}
\item Different scaling relation: 
\begin{itemize}
\item The $L_{\mathrm{X}}\text{--}T$ relation: The recovered $L_{\mathrm{X}}\text{--}T$ relation by Athena can be used to differentiate the different feedback models quantitatively. Among the TR18 models, the TR18 {\nr} model can be disfavored based on the galaxy groups detected at lower redshift ($z\leq1.5$). However, the lower redshift galaxy groups are insufficient to differentiate between TR18 {\csf} (input) and TR18  {\agn}. The high-redshift galaxy groups $z=2.0$ are required to discriminate the two similar models. We stress that little is known regarding the real heating mechanism in action at higher redshifts, and different simulations can lead to different predictions of the scaling relation \citep[e.g.,][]{Le-Brun:2017aa}. TR18 provided the scaling relations for different heating mechanisms up to $z=2.0$, which allowed us to perform the first test on the capability of Athena. This first test already indicates the importance of observing high-redshift galaxy groups, and our results demonstrate the power of the Athena/WFI instrument.
\item Detection probability: The different feedback mechanisms can lead to different predicted numbers of detected groups, as shown in Table~\ref{tab:LT_num_table_Baseline}. Based on observations, one can infer indirectly what physical models are preferred by comparing the observed numbers to theoretical predictions. For example, we assume that the remaining cosmological uncertainties on the expected numbers of groups at high redshift are small in the 2030s because the results from eROSITA and Euclid will be available then \citep[e.g.,][]{Sartoris:2016aa,pillepich2018forecasts}, and the detection of an unexpectedly large or small number of groups based on their X-ray emission with Athena would hint at strong effects on the ICM distributions.
\end{itemize}
\item Athena telescope and instrument parameters: 
\begin{itemize}
\item Optical filter: Removal of the filter can increase the  possibility of discovering the galaxy groups -- about as much as adding four more outer mirror rows. This is due to the increased effective area at low photon energies. Therefore, we suggest further study of whether the survey observations could be carried out with the filter in the open position; that is, whether fields can be selected that do not have many optically or UV-bright sources.
\item 19 rows: Athena with 19 mirror rows during the observation can increase the probability of detecting high-redshift galaxy groups significantly. Assuming the mass constraints of the Ariane 6 launch vehicle allow for this, we therefore suggest for consideration the addition of rows to the currently planned 15 rows. Obviously, this would not only allow us to achieve the goal of discovering distant groups with less exposure time but indeed enable the solution of many more science questions through the freed exposure time; 
\end{itemize}
\item Athena survey area: We assumed an Athena WFI extragalactic survey to be performed during the four year nominal mission duration, resulting in a total area of 48 $\deg^{2}$. Together with the baseline detection probability shown in Fig. \ref{fig:probability_1d}, we found that Athena can detect $\sim20$ galaxy groups as extended sources at redshift $z\geq2.0$ and with masses  $M_{500}\geq 5\times10^{13}\,\mathrm{M_{\odot}}$. This is a conservative estimation in the sense that one may expect Athena to operate longer. Moreover, if during extended mission operations a dedicated survey was performed with an optimized observing strategy, much larger numbers and higher redshift groups could be detected. This should be investigated further. Also, we suggest making certain that Athena is technically able to perform scans over large areas of the sky very efficiently ($\gtrsim 1000 \deg^{2}$; e.g., with the existing X-ray satellite \textit{XMM-Newton} this does not seem possible).
\end{enumerate}


\section{Conclusions}
In this work, we investigate the capabilities of Athena to detect early galaxy groups ($0.5 < z < 4$) using the advanced SIXTE simulator. The simulations take into account the main instrumental features of Athena: dithering mode, vignetting, and PSF degradation with off-axis angle. The simulations contain high-redshift galaxy groups with realistic surface brightness profiles and central AGN contamination, the general AGN population as well as all other X-ray and particle backgrounds. The wavelet-based detection algorithm combined with the \noun{SExtractor} software is used for the source detection, and a maximum likelihood fitting method is used for the source classification.

The result of our adopted baseline physical group evolution model is that high-redshift galaxy groups with a mass of $M_{500}=5\times10^{13}\,\mathrm{M_{\odot}}$ out to $z=2.0$ will be detected with a high probability ($>90$\%) as extended sources by Athena in the expected typical 80 ks WFI observations. Even at $z\sim2.5$ it is still high with ($\sim 80$\%) and then gradually drops to ($\sim 30$\%) at $z\sim4$. The extrapolation of group properties to such high redshifts never before probed is naturally uncertain but increasing the exposure time to 130 ks increases the detection probability even to $> 50\%$. During the first four years of the Athena mission and assuming the currently envisaged strawman WFI survey design,
more than 10,000 groups and clusters with $M_{500}\geq1\times10^{13}\,\mathrm{M_{\odot}}$ will be discovered at $z>0.5$.
Moreover, $20$ galaxy groups with $M_{500}\geq5\times10^{13}\,\mathrm{M_{\odot}}$ at $z\geq2.0$ will be detected. Importantly, Athena will directly provide precise temperature measurements with $\Delta T/T\leq25\%$ for eight out of those 20 galaxy groups, corresponding to 40\%. This result shows that galaxy groups at high redshift can be detected easily as extended sources and their thermo-dynamical properties can be characterized well by Athena provided that the mission covers sufficient survey sky area to capture those rare systems. In fact, the assumed covered survey area (48 $\deg^{2}$) is the primary limiting factor; so, assuming Athena  operates well beyond the four year nominal lifetime, potentially much larger numbers of high-redshift groups can be discovered.

Since little is known about the thermo-dynamical state of high-redshift galaxy groups, different physically motivated group evolution models are simulated and analyzed. In addition, the specific design of the Athena telescope is not yet completely fixed, and different Athena instrumental setups are also simulated and tested. We summarize those results in the following. Moreover, in the Discussion, we offer several suggestions for further studies.

\begin{enumerate}
\item The absence of a central AGN in galaxy groups can slightly increase the detection probability at $z\geq2.5$ as compared to our baseline model, with about a 5 \%\ increase at $z=2.5$ and a 23 \%\ increase at $z=4.0$.
\item Less-peaked group surface brightness profiles, for example, generated through stronger AGN feedback, can influence the detection probability at higher redshifts. The effect might in principle go both ways, increasing  or decreasing the detection probability. In our simulations, we find a 66 \%\ increase in the detection probability at $z=4.0$ for $r_{\rm c}=0.30\times r_{500}$ and a $15$ \%\ increase for $r_{\rm c}=0.45\times r_{500}$. This demonstrates that the detection probability at higher redshifts is sensitive to the core radius $r_{\rm c}$, with our baseline choice being conservative.
\item The extrapolation of galaxy group properties to high redshift results in unknown systematic uncertainty; these are after all the systems Athena will study in detail for the first time. We try to test for the influence by applying different scaling relation models based on recent numerical simulations. The expected number of galaxy groups at $z\geq2.0$ with $M_{500}\geq5\times 10^{13}\,\mathrm{M_{\odot}}$ changes as follows. The Truong et al.\ (TR18) {\nr} scaling relation results in the highest number of expected detections (20), the TR18 {\csf} has 19 detections and the TR18 {\agn} has the smallest expectation, only about 13 detections. The number of actually detected high-redshift groups will, therefore, already provide an indication of the feedback mechanisms operating at high redshifts. 
\item Moreover, Athena is shown to be able to quantitatively differentiate the various heating mechanisms by examining the observed scaling relation (i.e., the $L_{\mathrm{X}}\text{--}T$ relation). Assuming the TR18 {\csf} model, at the lower redshift regime $z\leq1.5$, the TR18 {\nr} can be easily ruled out; while the two heating models, TR18 {\csf} and TR18 {\agn}, can (only) be disentangled by using the galaxy groups at higher redshifts $z\sim 2.0$. 
\item Distinct telescope setups with both 19 mirror rows and the removal of optical and UV filters result in an overall increase in detection probability at $z\geq2.0$. Those results demonstrate that galaxy groups at high redshift have a significantly higher chance of being detected with those two distinct setups. Interestingly, the effects of these two setups are quite similar quantitatively, which means that, for the particular science case here, removal of the filter (which comes for free) results in the same improvement as adding four full outer mirror rows. 
\end{enumerate}

In summary, the capacity of Athena for detecting and characterizing high-redshift groups and, therefore, for constraining feedback mechanisms at high redshifts, is quite promising due to its high sensitivity, excellent angular resolution, wide FoV, and good spectral energy resolution. Our state-of-the-art source detection and characterization algorithms are efficient in detecting faint extended sources, and the employed maximum-likelihood fitting algorithms work well for source classification. Future advances in this area, possibly including machine learning techniques and simultaneous multi-dimensional imaging and spectral analyses, can only enhance  science exploitation opportunities of Athena further.


\begin{acknowledgements}

We would like to thank the anonymous referee for suggestions that significantly improved the manuscript. The authors wish to thank Weiwei Xu, Nhur Truong, Thomas Dauser and J{\"o}rn Wilms for useful help and discussions during the development of this paper. CZ was supported for this research through a stipend from the International Max Planck Research School (IMPRS) for Astronomy and Astrophysics at the Universities of Bonn and Cologne.

\end{acknowledgements}


\bibliographystyle{aa}
\bibliography{Athena.bib} 

\begin{thebibliography}{68}
\expandafter\ifx\csname natexlab\endcsname\relax\def\natexlab#1{#1}\fi

\bibitem[{{Arnaud}(1996)}]{Ar96}
{Arnaud}, K.~A. 1996, in Astronomical Society of the Pacific Conference Series,
  Vol. 101, Astronomical Data Analysis Software and Systems V, ed. G.~H.
  {Jacoby} \& J.~{Barnes}, 17

\bibitem[{{Barbera} {et~al.}(2015){Barbera}, {Branduardi-Raymont}, {Collura},
  {Comastri}, {Eder}, {Kamisi{\'n}ski}, {Lo Cicero}, {Meidinger}, {Mineo},
  {Molendi}, {Parodi}, {Pilch}, {Piro}, {Rataj}, {Rauw}, {Sciortino},
  {Sciortino}, \& {Wawer}}]{Barbera:2015aa}
{Barbera}, M., {Branduardi-Raymont}, G., {Collura}, A., {et~al.} 2015, in
  \procspie, Vol. 9601, UV, X-Ray, and Gamma-Ray Space Instrumentation for
  Astronomy XIX, 960109

\bibitem[{{Barret} {et~al.}(2013){Barret}, {den Herder}, {Piro}, {Ravera}, {Den
  Hartog}, {Macculi}, {Barcons}, {Page}, {Paltani}, {Rauw}, {Wilms},
  {Ceballos}, {Duband}, {Gottardi}, {Lotti}, {de Plaa}, {Pointecouteau},
  {Schmid}, {Akamatsu}, {Bagliani}, {Bandler}, {Barbera}, {Bastia}, {Biasotti},
  {Branco}, {Camon}, {Cara}, {Cobo}, {Colasanti}, {Costa-Kramer}, {Corcione},
  {Doriese}, {Duval}, {Fabrega}, {Gatti}, {de Gerone}, {Guttridge}, {Kelley},
  {Kilbourne}, {van der Kuur}, {Mineo}, {Mitsuda}, {Natalucci}, {Ohashi},
  {Peille}, {Perinati}, {Pigot}, {Pizzigoni}, {Pobes}, {Porter}, {Renotte},
  {Sauvageot}, {Sciortino}, {Torrioli}, {Valenziano}, {Willingale}, {de Vries},
  \& {van Weers}}]{Barret:2013aa}
{Barret}, D., {den Herder}, J.~W., {Piro}, L., {et~al.} 2013, ArXiv e-prints
  [\eprint[arXiv]{1308.6784}]

\bibitem[{{Benson} {et~al.}(2014){Benson}, {Ade}, {Ahmed}, {Allen}, {Arnold},
  {Austermann}, {Bender}, {Bleem}, {Carlstrom}, {Chang}, {Cho}, {Cliche},
  {Crawford}, {Cukierman}, {de Haan}, {Dobbs}, {Dutcher}, {Everett}, {Gilbert},
  {Halverson}, {Hanson}, {Harrington}, {Hattori}, {Henning}, {Hilton},
  {Holder}, {Holzapfel}, {Irwin}, {Keisler}, {Knox}, {Kubik}, {Kuo}, {Lee},
  {Leitch}, {Li}, {McDonald}, {Meyer}, {Montgomery}, {Myers}, {Natoli},
  {Nguyen}, {Novosad}, {Padin}, {Pan}, {Pearson}, {Reichardt}, {Ruhl},
  {Saliwanchik}, {Simard}, {Smecher}, {Sayre}, {Shirokoff}, {Stark}, {Story},
  {Suzuki}, {Thompson}, {Tucker}, {Vanderlinde}, {Vieira}, {Vikhlinin}, {Wang},
  {Yefremenko}, \& {Yoon}}]{Benson:2014aa}
{Benson}, B.~A., {Ade}, P.~A.~R., {Ahmed}, Z., {et~al.} 2014, in \procspie,
  Vol. 9153, Millimeter, Submillimeter, and Far-Infrared Detectors and
  Instrumentation for Astronomy VII, 91531P

\bibitem[{{Bertin} \& {Arnouts}(1996)}]{Bertin:1996aa}
{Bertin}, E. \& {Arnouts}, S. 1996, \aaps, 117, 393

\bibitem[{{Biffi} {et~al.}(2018){Biffi}, {Dolag}, \& {Merloni}}]{Biffi:2018aa}
{Biffi}, V., {Dolag}, K., \& {Merloni}, A. 2018, ArXiv e-prints
  [\eprint[arXiv]{1804.01096}]

\bibitem[{{Borgani} \& {Kravtsov}(2011)}]{Borgani:2011aa}
{Borgani}, S. \& {Kravtsov}, A. 2011, Advanced Science Letters, 4, 204

\bibitem[{{Borm} {et~al.}(2014){Borm}, {Reiprich}, {Mohammed}, \&
  {Lovisari}}]{Borm:2014aa}
{Borm}, K., {Reiprich}, T.~H., {Mohammed}, I., \& {Lovisari}, L. 2014, \aap,
  567, A65

\bibitem[{{Cash}(1979)}]{Cash:1979aa}
{Cash}, W. 1979, \apj, 228, 939

\bibitem[{{Cavagnolo} {et~al.}(2009){Cavagnolo}, {Donahue}, {Voit}, \&
  {Sun}}]{Cavagnolo:2009aa}
{Cavagnolo}, K.~W., {Donahue}, M., {Voit}, G.~M., \& {Sun}, M. 2009, \apjs,
  182, 12

\bibitem[{{Cavaliere} \& {Fusco-Femiano}(1978)}]{Cavaliere:1978aa}
{Cavaliere}, A. \& {Fusco-Femiano}, R. 1978, \aap, 70, 677

\bibitem[{{Clerc} {et~al.}(2012){Clerc}, {Pierre}, {Pacaud}, \&
  {Sadibekova}}]{Clerc12}
{Clerc}, N., {Pierre}, M., {Pacaud}, F., \& {Sadibekova}, T. 2012, \mnras, 423,
  3545

\bibitem[{{Croston} {et~al.}(2013){Croston}, {Sanders}, {Heinz}, {Hardcastle},
  {Zhuravleva}, {B{\^\i}rzan}, {Bower}, {Br{\"u}ggen}, {Churazov}, {Edge},
  {Ettori}, {Fabian}, {Finoguenov}, {Kaastra}, {Gaspari}, {Gitti}, {Nulsen},
  {McNamara}, {Pointecouteau}, {Ponman}, {Pratt}, {Rafferty}, {Reiprich},
  {Sijacki}, {Worrall}, {Kraft}, {McCarthy}, \& {Wise}}]{Croston:2013aa}
{Croston}, J.~H., {Sanders}, J.~S., {Heinz}, S., {et~al.} 2013, arXiv e-prints
  [\eprint[arXiv]{1306.2323}]

\bibitem[{Croton {et~al.}(2006)Croton, Springel, White, De~Lucia, Frenk, Gao,
  Jenkins, Kauffmann, Navarro, \& Yoshida}]{croton2006many}
Croton, D.~J., Springel, V., White, S.~D., {et~al.} 2006, Monthly Notices of
  the Royal Astronomical Society, 365, 11

\bibitem[{{Dauser} {et~al.}(2019){Dauser}, {Falkner}, {Lorenz}, {Kirsch},
  {Peille}, {Cucchetti}, {Schmid}, {Brand}, {Oertel}, {Smith}, \&
  {Wilms}}]{Dauser:2019aa}
{Dauser}, T., {Falkner}, S., {Lorenz}, M., {et~al.} 2019, arXiv e-prints
  [\eprint[arXiv]{1908.00781}]

\bibitem[{{De Bernardis} {et~al.}(2016){De Bernardis}, {Stevens},
  {Hasselfield}, {Alonso}, {Bond}, {Calabrese}, {Choi}, {Crowley}, {Devlin},
  {Dunkley}, {Gallardo}, {Henderson}, {Hilton}, {Hlozek}, {Ho}, {Huffenberger},
  {Koopman}, {Kosowsky}, {Louis}, {Madhavacheril}, {McMahon}, {N{\ae}ss},
  {Nati}, {Newburgh}, {Niemack}, {Page}, {Salatino}, {Schillaci}, {Schmitt},
  {Sehgal}, {Sievers}, {Simon}, {Spergel}, {Staggs}, {van Engelen},
  {Vavagiakis}, \& {Wollack}}]{De-Bernardis:2016aa}
{De Bernardis}, F., {Stevens}, J.~R., {Hasselfield}, M., {et~al.} 2016, in
  \procspie, Vol. 9910, Observatory Operations: Strategies, Processes, and
  Systems VI, 991014

\bibitem[{Despali {et~al.}(2016)Despali, Giocoli, Angulo, Tormen, Sheth, Baso,
  \& Moscardini}]{despali2016universality}
Despali, G., Giocoli, C., Angulo, R.~E., {et~al.} 2016, Monthly Notices of the
  Royal Astronomical Society, 456, 2486

\bibitem[{{Diemer}(2018)}]{Diemer:2018aa}
{Diemer}, B. 2018, \apjs, 239, 35

\bibitem[{{Eckert} {et~al.}(2012){Eckert}, {Vazza}, {Ettori}, {Molendi},
  {Nagai}, {Lau}, {Roncarelli}, {Rossetti}, {Snowden}, \&
  {Gastaldello}}]{Eckert12}
{Eckert}, D., {Vazza}, F., {Ettori}, S., {et~al.} 2012, \aap, 541, A57

\bibitem[{{Erfanianfar} {et~al.}(2013){Erfanianfar}, {Finoguenov}, {Tanaka},
  {Lerchster}, {Nandra}, {Laird}, {Connelly}, {Bielby}, {Mirkazemi}, {Faber},
  {Kocevski}, {Cooper}, {Newman}, {Jeltema}, {Coil}, {Brimioulle}, {Davis},
  {McCracken}, {Willmer}, {Gerke}, {Cappelluti}, \& {Gwyn}}]{Er13}
{Erfanianfar}, G., {Finoguenov}, A., {Tanaka}, M., {et~al.} 2013, \apj, 765,
  117

\bibitem[{{Ettori} {et~al.}(2013){Ettori}, {Pratt}, {de Plaa}, {Eckert},
  {Nevalainen}, {Battistelli}, {Borgani}, {Croston}, {Finoguenov}, {Kaastra},
  {Gaspari}, {Gastaldello}, {Gitti}, {Molendi}, {Pointecouteau}, {Ponman},
  {Reiprich}, {Roncarelli}, {Rossetti}, {Sanders}, {Sun}, {Trinchieri},
  {Vazza}, {Arnaud}, {B{\"o}ringher}, {Brighenti}, {Dahle}, {De Grandi},
  {Mohr}, {Moretti}, \& {Schindler}}]{Ettori:2013ac}
{Ettori}, S., {Pratt}, G.~W., {de Plaa}, J., {et~al.} 2013, arXiv e-prints
  [\eprint[arXiv]{1306.2322}]

\bibitem[{Fabjan {et~al.}(2010)Fabjan, Borgani, Tornatore, Saro, Murante, \&
  Dolag}]{fabjan2010simulating}
Fabjan, D., Borgani, S., Tornatore, L., {et~al.} 2010, Monthly Notices of the
  Royal Astronomical Society, 401, 1670

\bibitem[{{Faccioli} {et~al.}(2018){Faccioli}, {Pacaud}, {Sauvageot}, {Pierre},
  {Chiappetti}, {Clerc}, {Gastaud}, {Koulouridis}, {Le Brun}, \&
  {Valotti}}]{Faccioli:2018aa}
{Faccioli}, L., {Pacaud}, F., {Sauvageot}, J.-L., {et~al.} 2018, \aap, 620, A9

\bibitem[{{Gilli} {et~al.}(2007){Gilli}, {Comastri}, \&
  {Hasinger}}]{Gilli:2007aa}
{Gilli}, R., {Comastri}, A., \& {Hasinger}, G. 2007, \aap, 463, 79

\bibitem[{{Giodini} {et~al.}(2013){Giodini}, {Lovisari}, {Pointecouteau},
  {Ettori}, {Reiprich}, \& {Hoekstra}}]{Giodini:2013ab}
{Giodini}, S., {Lovisari}, L., {Pointecouteau}, E., {et~al.} 2013, \ssr, 177,
  247

\bibitem[{{Gobat} {et~al.}(2011{\natexlab{a}}){Gobat}, {Daddi}, {Onodera},
  {Finoguenov}, {Renzini}, {Arimoto}, {Bouwens}, {Brusa}, {Chary}, {Cimatti},
  {Dickinson}, {Kong}, \& {Mignoli}}]{Go11}
{Gobat}, R., {Daddi}, E., {Onodera}, M., {et~al.} 2011{\natexlab{a}}, \aap,
  526, A133

\bibitem[{{Gobat} {et~al.}(2011{\natexlab{b}}){Gobat}, {Daddi}, {Onodera},
  {Finoguenov}, {Renzini}, {Arimoto}, {Bouwens}, {Brusa}, {Chary}, {Cimatti},
  {Dickinson}, {Kong}, \& {Mignoli}}]{Gobat:2011aa}
{Gobat}, R., {Daddi}, E., {Onodera}, M., {et~al.} 2011{\natexlab{b}}, \aap,
  526, A133

\bibitem[{Henderson {et~al.}(2016)Henderson, Allison, Austermann, Baildon,
  Battaglia, Beall, Becker, De~Bernardis, Bond, Calabrese,
  {et~al.}}]{henderson2016advanced}
Henderson, S., Allison, R., Austermann, J., {et~al.} 2016, Journal of Low
  Temperature Physics, 184, 772

\bibitem[{{Le Brun} {et~al.}(2017){Le Brun}, {McCarthy}, {Schaye}, \&
  {Ponman}}]{Le-Brun:2017aa}
{Le Brun}, A.~M.~C., {McCarthy}, I.~G., {Schaye}, J., \& {Ponman}, T.~J. 2017,
  \mnras, 466, 4442

\bibitem[{{Lehmer} {et~al.}(2012){Lehmer}, {Xue}, {Brandt}, {Alexander},
  {Bauer}, {Brusa}, {Comastri}, {Gilli}, {Hornschemeier}, {Luo}, {Paolillo},
  {Ptak}, {Shemmer}, {Schneider}, {Tozzi}, \& {Vignali}}]{Lehmer:2012aa}
{Lehmer}, B.~D., {Xue}, Y.~Q., {Brandt}, W.~N., {et~al.} 2012, \apj, 752, 46

\bibitem[{Lotti {et~al.}(2017)Lotti, Mineo, Jacquey, Molendi, D'andrea,
  Macculi, \& Piro}]{lotti2017particle}
Lotti, S., Mineo, T., Jacquey, C., {et~al.} 2017, Experimental Astronomy, 44,
  371

\bibitem[{{Lovisari} {et~al.}(2015){Lovisari}, {Reiprich}, \&
  {Schellenberger}}]{Lovisari:2015aa}
{Lovisari}, L., {Reiprich}, T.~H., \& {Schellenberger}, G. 2015, \aap, 573,
  A118

\bibitem[{{Mantz} {et~al.}(2019){Mantz}, {Allen}, {Battaglia}, {Benson},
  {Canning}, {Ettori}, {Evrard}, {von der Linden}, \&
  {McDonald}}]{Mantz:2019aa}
{Mantz}, A.~B., {Allen}, S.~W., {Battaglia}, N., {et~al.} 2019, arXiv e-prints
  [\eprint[arXiv]{1903.05606}]

\bibitem[{McCarthy {et~al.}(2011)McCarthy, Schaye, Bower, Ponman, Booth,
  Vecchia, \& Springel}]{mccarthy2011gas}
McCarthy, I.~G., Schaye, J., Bower, R.~G., {et~al.} 2011, Monthly Notices of
  the Royal Astronomical Society, 412, 1965

\bibitem[{{Meidinger} {et~al.}(2017){Meidinger}, {Eder}, {Eraerds}, {Nandra},
  {Pietschner}, {Plattner}, {Rau}, \& {Strecker}}]{Meidinger:2017aa}
{Meidinger}, N., {Eder}, J., {Eraerds}, T., {et~al.} 2017, ArXiv e-prints
  [\eprint[arXiv]{1702.01079}]

\bibitem[{{Merloni} {et~al.}(2012){Merloni}, {Predehl}, {Becker},
  {B{\"o}hringer}, {Boller}, {Brunner}, {Brusa}, {Dennerl}, {Freyberg},
  {Friedrich}, {Georgakakis}, {Haberl}, {Hasinger}, {Meidinger}, {Mohr},
  {Nandra}, {Rau}, {Reiprich}, {Robrade}, {Salvato}, {Santangelo}, {Sasaki},
  {Schwope}, {Wilms}, \& {German eROSITA Consortium}}]{Mer12}
{Merloni}, A., {Predehl}, P., {Becker}, W., {et~al.} 2012, ArXiv e-prints
  [\eprint[arXiv]{1209.3114}]

\bibitem[{{Moretti} {et~al.}(2003){Moretti}, {Campana}, {Lazzati}, \&
  {Tagliaferri}}]{Moretti:2003aa}
{Moretti}, A., {Campana}, S., {Lazzati}, D., \& {Tagliaferri}, G. 2003, \apj,
  588, 696

\bibitem[{{Nandra} {et~al.}(2013){Nandra}, {Barret}, {Barcons}, {Fabian}, {den
  Herder}, {Piro}, {Watson}, {Adami}, {Aird}, {Afonso}, \&
  et~al.}]{Nandra:2013aa}
{Nandra}, K., {Barret}, D., {Barcons}, X., {et~al.} 2013, ArXiv e-prints
  [\eprint[arXiv]{1306.2307}]

\bibitem[{{Pacaud} {et~al.}(2006){Pacaud}, {Pierre}, {Refregier}, {Gueguen},
  {Starck}, {Valtchanov}, {Read}, {Altieri}, {Chiappetti}, {Gandhi}, {Garcet},
  {Gosset}, {Ponman}, \& {Surdej}}]{Pacaud:2006aa}
{Pacaud}, F., {Pierre}, M., {Refregier}, A., {et~al.} 2006, \mnras, 372, 578

\bibitem[{{Pillepich} {et~al.}(2012){Pillepich}, {Porciani}, \&
  {Reiprich}}]{Pi12}
{Pillepich}, A., {Porciani}, C., \& {Reiprich}, T.~H. 2012, \mnras, 422, 44

\bibitem[{Pillepich {et~al.}(2018)Pillepich, Reiprich, Porciani, Borm, \&
  Merloni}]{pillepich2018forecasts}
Pillepich, A., Reiprich, T.~H., Porciani, C., Borm, K., \& Merloni, A. 2018,
  Monthly Notices of the Royal Astronomical Society, 481, 613

\bibitem[{{Planck Collaboration} {et~al.}(2018){Planck Collaboration},
  {Aghanim}, {Akrami}, {Ashdown}, {Aumont}, {Baccigalupi}, {Ballardini},
  {Banday}, {Barreiro}, {Bartolo}, {Basak}, {Battye}, {Benabed}, {Bernard},
  {Bersanelli}, {Bielewicz}, {Bock}, {Bond}, {Borrill}, {Bouchet}, {Boulanger},
  {Bucher}, {Burigana}, {Butler}, {Calabrese}, {Cardoso}, {Carron},
  {Challinor}, {Chiang}, {Chluba}, {Colombo}, {Combet}, {Contreras}, {Crill},
  {Cuttaia}, {de Bernardis}, {de Zotti}, {Delabrouille}, {Delouis}, {Di
  Valentino}, {Diego}, {Dor{\'e}}, {Douspis}, {Ducout}, {Dupac}, {Dusini},
  {Efstathiou}, {Elsner}, {En{\ss}lin}, {Eriksen}, {Fantaye}, {Farhang},
  {Fergusson}, {Fernandez-Cobos}, {Finelli}, {Forastieri}, {Frailis},
  {Franceschi}, {Frolov}, {Galeotta}, {Galli}, {Ganga}, {G{\'e}nova-Santos},
  {Gerbino}, {Ghosh}, {Gonz{\'a}lez-Nuevo}, {G{\'o}rski}, {Gratton},
  {Gruppuso}, {Gudmundsson}, {Hamann}, {Hand ley}, {Herranz}, {Hivon}, {Huang},
  {Jaffe}, {Jones}, {Karakci}, {Keih{\"a}nen}, {Keskitalo}, {Kiiveri}, {Kim},
  {Kisner}, {Knox}, {Krachmalnicoff}, {Kunz}, {Kurki-Suonio}, {Lagache},
  {Lamarre}, {Lasenby}, {Lattanzi}, {Lawrence}, {Le Jeune}, {Lemos},
  {Lesgourgues}, {Levrier}, {Lewis}, {Liguori}, {Lilje}, {Lilley}, {Lindholm},
  {L{\'o}pez-Caniego}, {Lubin}, {Ma}, {Mac{\'\i}as-P{\'e}rez}, {Maggio},
  {Maino}, {Mandolesi}, {Mangilli}, {Marcos-Caballero}, {Maris}, {Martin},
  {Martinelli}, {Mart{\'\i}nez-Gonz{\'a}lez}, {Matarrese}, {Mauri}, {McEwen},
  {Meinhold}, {Melchiorri}, {Mennella}, {Migliaccio}, {Millea}, {Mitra},
  {Miville-Desch{\^e}nes}, {Molinari}, {Montier}, {Morgante}, {Moss}, {Natoli},
  {N{\o}rgaard-Nielsen}, {Pagano}, {Paoletti}, {Partridge}, {Patanchon},
  {Peiris}, {Perrotta}, {Pettorino}, {Piacentini}, {Polastri}, {Polenta},
  {Puget}, {Rachen}, {Reinecke}, {Remazeilles}, {Renzi}, {Rocha}, {Rosset},
  {Roudier}, {Rubi{\~n}o-Mart{\'\i}n}, {Ruiz-Granados}, {Salvati}, {Sandri},
  {Savelainen}, {Scott}, {Shellard}, {Sirignano}, {Sirri}, {Spencer},
  {Sunyaev}, {Suur-Uski}, {Tauber}, {Tavagnacco}, {Tenti}, {Toffolatti},
  {Tomasi}, {Trombetti}, {Valenziano}, {Valiviita}, {Van Tent}, {Vibert},
  {Vielva}, {Villa}, {Vittorio}, {Wand elt}, {Wehus}, {White}, {White},
  {Zacchei}, \& {Zonca}}]{2018arXiv180706209P}
{Planck Collaboration}, {Aghanim}, N., {Akrami}, Y., {et~al.} 2018, arXiv
  e-prints, arXiv:1807.06209

\bibitem[{{Pointecouteau} {et~al.}(2013){Pointecouteau}, {Reiprich}, {Adami},
  {Arnaud}, {Biffi}, {Borgani}, {Borm}, {Bourdin}, {Brueggen}, {Bulbul},
  {Clerc}, {Croston}, {Dolag}, {Ettori}, {Finoguenov}, {Kaastra}, {Lovisari},
  {Maughan}, {Mazzotta}, {Pacaud}, {de Plaa}, {Pratt}, {Ramos-Ceja}, {Rasia},
  {Sanders}, {Zhang}, {Allen}, {Boehringer}, {Brunetti}, {Elbaz}, {Fassbender},
  {Hoekstra}, {Hildebrandt}, {Lamer}, {Marrone}, {Mohr}, {Molendi},
  {Nevalainen}, {Ohashi}, {Ota}, {Pierre}, {Romer}, {Schindler}, {Schrabback},
  {Schwope}, {Smith}, {Springel}, \& {von der Linden}}]{Athena132}
{Pointecouteau}, E., {Reiprich}, T.~H., {Adami}, C., {et~al.} 2013, ArXiv
  e-prints [\eprint[arXiv]{1306.2319}]

\bibitem[{{Pratt} {et~al.}(2019){Pratt}, {Arnaud}, {Biviano}, {Eckert},
  {Ettori}, {Nagai}, {Okabe}, \& {Reiprich}}]{Pratt:2019aa}
{Pratt}, G.~W., {Arnaud}, M., {Biviano}, A., {et~al.} 2019, \ssr, 215, 25

\bibitem[{{Pratt} {et~al.}(2010){Pratt}, {Arnaud}, {Piffaretti},
  {B{\"o}hringer}, {Ponman}, {Croston}, {Voit}, {Borgani}, \&
  {Bower}}]{Pratt:2010aa}
{Pratt}, G.~W., {Arnaud}, M., {Piffaretti}, R., {et~al.} 2010, \aap, 511, A85

\bibitem[{Predehl {et~al.}(2010)Predehl, Andritschke, B{\"o}hringer, Bornemann,
  Br{\"a}uninger, Brunner, Brusa, Burkert, Burwitz, Cappelluti, Churazov,
  Dennerl, Eder, Elbs, Freyberg, Friedrich, F{\"u}rmetz, Gaida, H{\"a}lker,
  Hartner, Hasinger, Hermann, Huber, Kendziorra, von Kienlin, Kink,
  Kreykenbohm, Lamer, Lapchov, Lehmann, Meidinger, Mican, Mohr, M{\"u}hlegger,
  M{\"u}ller, Nandra, Pavlinsky, Pfeffermann, Reiprich, Robrade, Roh{\'e},
  Santangelo, Sch{\"a}chner, Schanz, Schmid, Schmitt, Schreib, Schrey, Schwope,
  Steinmetz, Str{\"u}der, Sunyaev, Tenzer, Tiedemann, Vongehr, \&
  Wilms}]{10.1117/12.856577}
Predehl, P., Andritschke, R., B{\"o}hringer, H., {et~al.} 2010, in Space
  Telescopes and Instrumentation 2010: Ultraviolet to Gamma Ray, ed. M.~Arnaud,
  S.~S. Murray, \& T.~Takahashi, Vol. 7732, International Society for Optics
  and Photonics (SPIE), 210 -- 219

\bibitem[{{Rau}(2013)}]{Rau13}
{Rau}, A. 2013, {Athena+ WFI Preparation of background files.
  MPE-WFI-TN-2013-001}, (Internal report)

\bibitem[{{Rau} {et~al.}(2013){Rau}, {Meidinger}, {Nandra}, {Porro}, {Barret},
  {Santangelo}, {Schmid}, {Struder}, {Tenzer}, {Wilms}, {Amoros},
  {Andritschke}, {Aschauer}, {Bahr}, {Gunther}, {Furmetz}, {Ott}, {Perinati},
  {Rambaud}, {Reiffers}, {Treis}, {von Kienlin}, \&
  {Weidenspointner}}]{Rau:2013aa}
{Rau}, A., {Meidinger}, N., {Nandra}, K., {et~al.} 2013, ArXiv e-prints
  [\eprint[arXiv]{1308.6785}]

\bibitem[{Rau {et~al.}(2016)Rau, Nandra, Aird, Comastri, Dauser, Merloni,
  Pratt, Reiprich, Fabian, Georgakakis, {et~al.}}]{rau2016athena}
Rau, A., Nandra, K., Aird, J., {et~al.} 2016, in Space Telescopes and
  Instrumentation 2016: Ultraviolet to Gamma Ray, Vol. 9905, International
  Society for Optics and Photonics, 99052B

\bibitem[{{Reichert} {et~al.}(2011){Reichert}, {B{\"o}hringer}, {Fassbender},
  \& {M{\"u}hlegger}}]{Reichert:2011aa}
{Reichert}, A., {B{\"o}hringer}, H., {Fassbender}, R., \& {M{\"u}hlegger}, M.
  2011, \aap, 535, A4

\bibitem[{{Sartoris} {et~al.}(2016){Sartoris}, {Biviano}, {Fedeli}, {Bartlett},
  {Borgani}, {Costanzi}, {Giocoli}, {Moscardini}, {Weller}, {Ascaso},
  {Bardelli}, {Maurogordato}, \& {Viana}}]{Sartoris:2016aa}
{Sartoris}, B., {Biviano}, A., {Fedeli}, C., {et~al.} 2016, \mnras, 459, 1764

\bibitem[{{Schellenberger} \& {Reiprich}(2017)}]{Schellenberger:2017ab}
{Schellenberger}, G. \& {Reiprich}, T.~H. 2017, \mnras, 471, 1370

\bibitem[{Short {et~al.}(2010)Short, Thomas, Young, Pearce, Jenkins, \&
  Muanwong}]{short2010evolution}
Short, C., Thomas, P., Young, O., {et~al.} 2010, Monthly Notices of the Royal
  Astronomical Society, 408, 2213

\bibitem[{{Stacey} {et~al.}(2018){Stacey}, {Aravena}, {Basu}, {Battaglia},
  {Beringue}, {Bertoldi}, {Bond}, {Breysse}, {Bustos}, {Chapman}, {Chung},
  {Cothard}, {Erler}, {Fich}, {Foreman}, {Gallardo}, {Giovanelli}, {Graf},
  {Haynes}, {Herrera-Camus}, {Herter}, {Hlo{\v z}ek}, {Johnstone}, {Keating},
  {Magnelli}, {Meerburg}, {Meyers}, {Murray}, {Niemack}, {Nikola}, {Nolta},
  {Parshley}, {Riechers}, {Schilke}, {Scott}, {Stein}, {Stevens}, {Stutzki},
  {Vavagiakis}, \& {Viero}}]{Stacey:2018aa}
{Stacey}, G.~J., {Aravena}, M., {Basu}, K., {et~al.} 2018, in Society of
  Photo-Optical Instrumentation Engineers (SPIE) Conference Series, Vol. 10700,
  Ground-based and Airborne Telescopes VII, 107001M

\bibitem[{Starck {et~al.}(1998)Starck, Murtagh, \&
  Bijaoui}]{starck_murtagh_bijaoui_1998}
Starck, J.-L., Murtagh, F.~D., \& Bijaoui, A. 1998, Image Processing and Data
  Analysis: The Multiscale Approach (Cambridge University Press)

\bibitem[{Starck \& Pierre(1998)}]{starck1998structure}
Starck, J.-L. \& Pierre, M. 1998, Astronomy and Astrophysics Supplement Series,
  128, 397

\bibitem[{Steidel {et~al.}(2010)Steidel, Erb, Shapley, Pettini, Reddy,
  Bogosavljevi{\'c}, Rudie, \& Rakic}]{steidel2010structure}
Steidel, C.~C., Erb, D.~K., Shapley, A.~E., {et~al.} 2010, The Astrophysical
  Journal, 717, 289

\bibitem[{Sun {et~al.}(2009)Sun, Voit, Donahue, Jones, Forman, \&
  Vikhlinin}]{sun2009chandra}
Sun, M., Voit, G., Donahue, M., {et~al.} 2009, The Astrophysical Journal, 693,
  1142

\bibitem[{{Tinker} {et~al.}(2008){Tinker}, {Kravtsov}, {Klypin}, {Abazajian},
  {Warren}, {Yepes}, {Gottl{\"o}ber}, \& {Holz}}]{Tinker:2008aa}
{Tinker}, J., {Kravtsov}, A.~V., {Klypin}, A., {et~al.} 2008, \apj, 688, 709

\bibitem[{{Truong} {et~al.}(2018){Truong}, {Rasia}, {Mazzotta}, {Planelles},
  {Biffi}, {Fabjan}, {Beck}, {Borgani}, {Dolag}, {Gaspari}, {Granato},
  {Murante}, {Ragone-Figueroa}, \& {Steinborn}}]{Truong:2018aa}
{Truong}, N., {Rasia}, E., {Mazzotta}, P., {et~al.} 2018, \mnras, 474, 4089

\bibitem[{Valtchanov {et~al.}(2001)Valtchanov, Pierre, \&
  Gastaud}]{valtchanov2001comparison}
Valtchanov, I., Pierre, M., \& Gastaud, R. 2001, Astronomy \& Astrophysics,
  370, 689

\bibitem[{{Vikhlinin} {et~al.}(2009){Vikhlinin}, {Kravtsov}, {Burenin},
  {Ebeling}, {Forman}, {Hornstrup}, {Jones}, {Murray}, {Nagai}, {Quintana}, \&
  {Voevodkin}}]{Vikhlinin09}
{Vikhlinin}, A., {Kravtsov}, A.~V., {Burenin}, R.~A., {et~al.} 2009, \apj, 692,
  1060

\bibitem[{{Voit}(2005)}]{Voit:2005ab}
{Voit}, G.~M. 2005, Reviews of Modern Physics, 77, 207

\bibitem[{von Kienlin {et~al.}(2018)von Kienlin, Eraerds, Bulbul, Fioretti,
  Gastaldello, Grant, Hall, Holland, Keelan, Meidinger,
  {et~al.}}]{von2018evaluation}
von Kienlin, A., Eraerds, T., Bulbul, E., {et~al.} 2018, in Space Telescopes
  and Instrumentation 2018: Ultraviolet to Gamma Ray, Vol. 10699, International
  Society for Optics and Photonics, 106991I

\bibitem[{Watson {et~al.}(2013)Watson, Iliev, D'Aloisio, Knebe, Shapiro, \&
  Yepes}]{watson2013halo}
Watson, W.~A., Iliev, I.~T., D'Aloisio, A., {et~al.} 2013, Monthly Notices of
  the Royal Astronomical Society, 433, 1230

\bibitem[{{Willingale} {et~al.}(2013){Willingale}, {Pareschi}, {Christensen},
  \& {den Herder}}]{Willingale:2013aa}
{Willingale}, R., {Pareschi}, G., {Christensen}, F., \& {den Herder}, J.-W.
  2013, ArXiv e-prints [\eprint[arXiv]{1307.1709}]

\bibitem[{Wilms {et~al.}(2014)Wilms, Brand, Barret, Beuchert, den Herder,
  Kreykenbohm, Lotti, Meidinger, Nandra, Peille, {et~al.}}]{wilms2014athena}
Wilms, J., Brand, T., Barret, D., {et~al.} 2014, in Space Telescopes and
  Instrumentation 2014: Ultraviolet to Gamma Ray, Vol. 9144, International
  Society for Optics and Photonics, 91445X

\bibitem[{{Xu} {et~al.}(2018){Xu}, {Ramos-Ceja}, {Pacaud}, {Reiprich}, \&
  {Erben}}]{Xu:2018aa}
{Xu}, W., {Ramos-Ceja}, M.~E., {Pacaud}, F., {Reiprich}, T.~H., \& {Erben}, T.
  2018, \aap, 619, A162

\end{thebibliography}


\begin{appendix}

\section{Observed $\log N-\log S$ relation\label{LogN-LogS}}

\begin{figure}[!ht]
\begin{centering}
\includegraphics[scale=0.55]{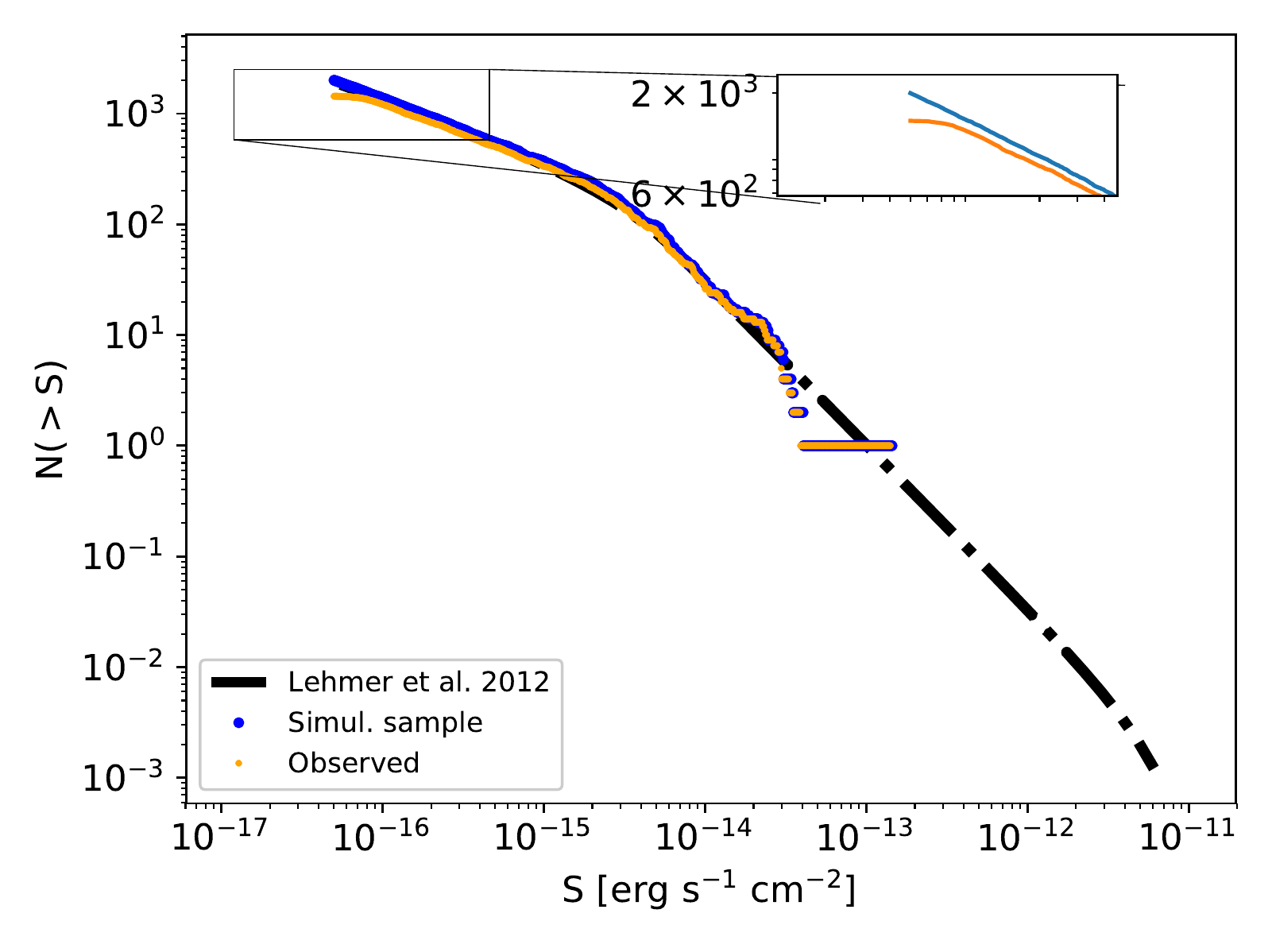}
\par\end{centering}
\caption{\label{fig: LogN-LogS_obs} Simulated AGN $\log N-\log S$ relation. The black curve is the AGN model in the $0.5-2.0$~keV energy band from \citet{Lehmer:2012aa}; the blue dotted curve represents one realization of an AGN population used for the simulation, while the orange dots are the detected AGNs. The zoom-in box shows the $\log N-\log S$ relation at the faint end, which indicates Athena cannot detect any sources fainter than ($7\times10^{-17}\ \text{ergs/s/cm}^{2}$) in an 80ks exposure time observation.}
\end{figure}

We adopted the \cite{Lehmer:2012aa} AGN model in the $0.5-2.0$~keV energy band for our $\log N-\log S$ relation. Figure \ref{fig: LogN-LogS_obs} shows one example of an observed $\log N-\log S$ relation for the Athena simulation with an exposure time of 80 ks. The zoom-in box shows the simulated sample with the lowest flux $5\times10^{-17}\ \text{ergs/s/cm}^{2}$, which is faint enough for our simulations because Athena will not detect any fainter sources ($<7\times10^{-17}\ \text{ergs/s/cm}^{2}$) in our 80 ks simulation. This justifies the lowest flux of $5\times10^{-17}\ \text{ergs/s/cm}^{2}$ that we used to simulate the AGN population.

\section{Detection probability 2D interpolation}

\begin{figure}
\begin{centering}
\includegraphics[trim=30 30 70 30,clip,width=\columnwidth]{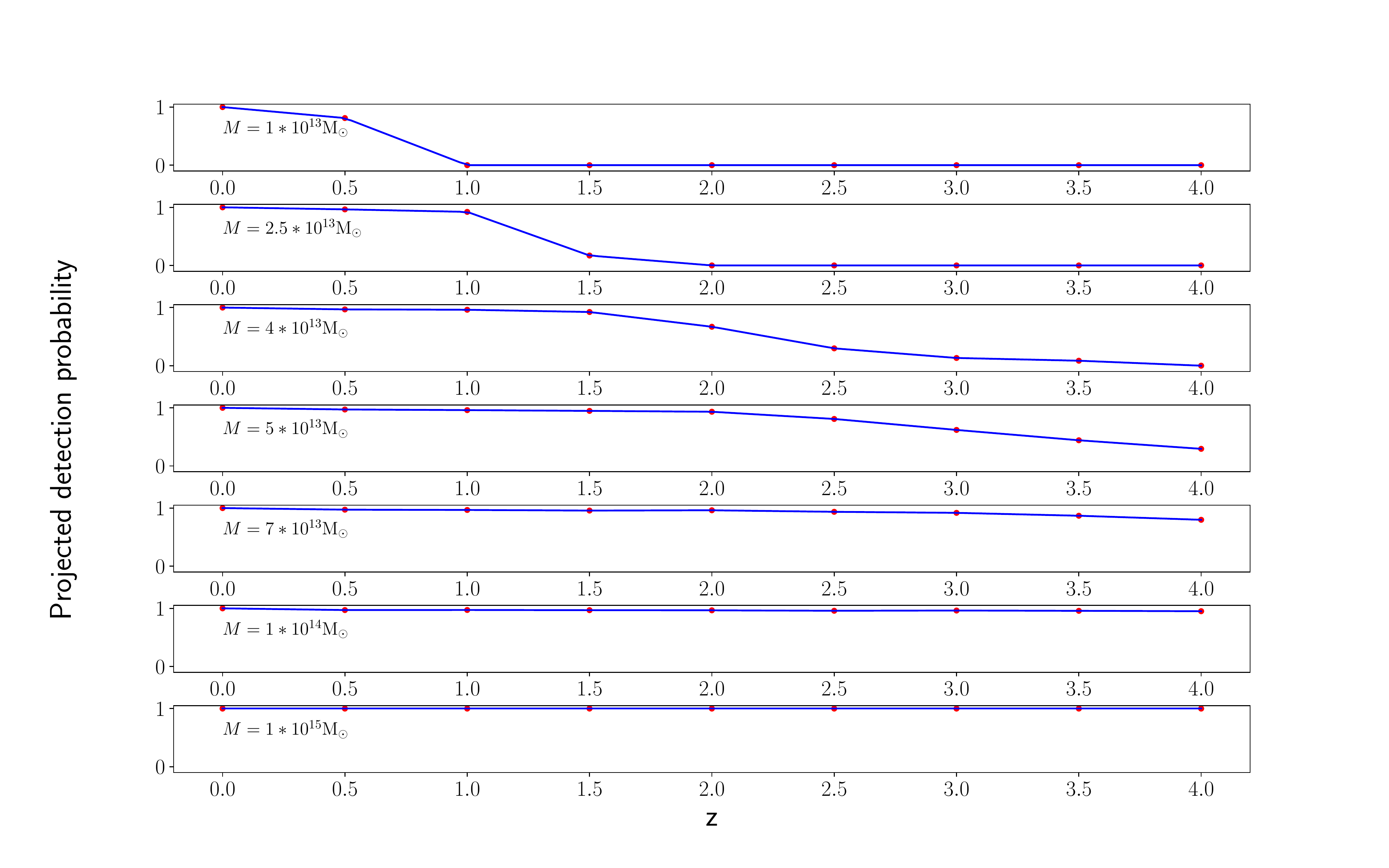}\\
\includegraphics[trim=30 30 70 30,clip,width=\columnwidth]{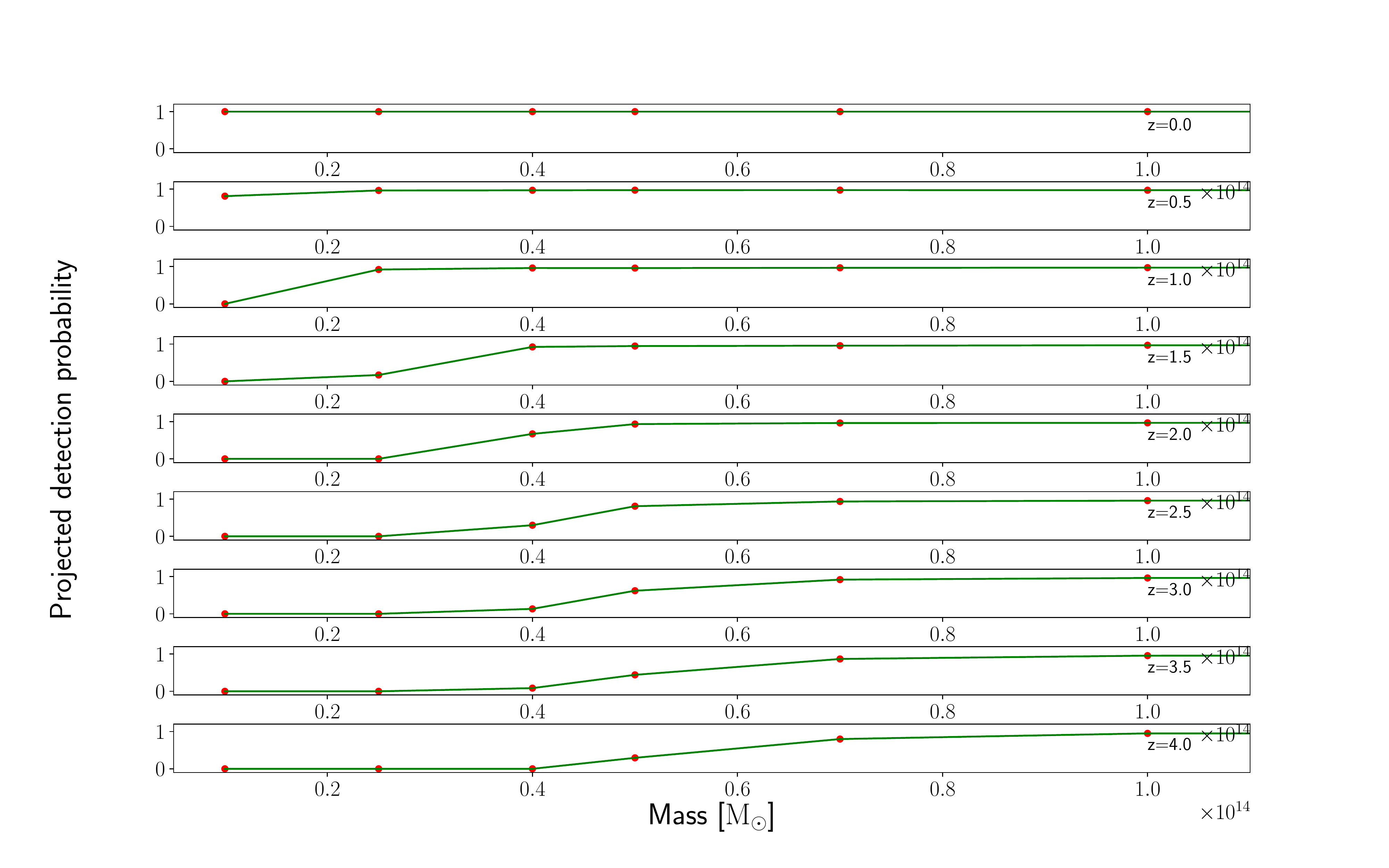}
\par\end{centering}
\caption{\label{fig: Det_2D_interpolation} Two-dimensional detection probability interpolation grid. Top: Detection probability as a function of redshift $z$, for masses of $M_{500}=1\times10^{13},2.5\times10^{13},4\times10^{13},5\times10^{13},7\times10^{13},1\times10^{14},1\times10^{15}\,\mathrm{M}_{\odot}$. Bottom: Detection probability as a function of mass, for redshifts  $z=0,~0.5,~1.0,~1.5,~2.0,~2.5,~3.0,~3.5,~4.0$. }
\end{figure}

The projected detection probability $P_{(z,M)}$ in Eq.~(\ref{eq:MF}) is a function of mass and redshift. Therefore, we need to make a 2D grid for the interpolation between detection probability, redshift $z,$ and mass $M$. The interpolation of mass and detection probability is shown in the top panel of Fig. \ref{fig: Det_2D_interpolation}. The interpolation of redshift and detection probability is shown at the bottom in the same figure. 

For the mass interpolation, we manually add 100\% detection probabilities of galaxy clusters for the masses of $10^{15}\,\mathrm{M}_{\odot}$ and larger in order to extend the integration mass limit. Similarly, for the redshift interpolation, we add 100\% detection probability at redshift $z=0$. 

\section{Photon background parameter\label{background_parameter}}

The cosmic photon background consists of the integrated emission from unresolved extra-galactic point sources and the diffuse Galactic foreground. The parameter values are given in Table~\ref{tab:Background-Xspec-model}. The fraction of resolved sources was assumed to be 60\%. The functional form of the background model is given in Eq.~(\ref{eq:bkg_model}). 

\begin{table*}
\begin{center}
\begin{tabular}{cccc}
Model & Parameter & Value & Unit\tabularnewline
\hline
\hline
apec & $kT$ & $9.9\times10^{-2}$ & keV\tabularnewline
apec & $abundance$ & 1.0 & \tabularnewline
apec & $redshift$ & 0 & \tabularnewline
apec & $norm$ & $1.7\times10^{-6}$ & $\ensuremath{(10^{-14}/(4\pi(D_{A}(1+z))^{2}))\int n_{e}n_{H}dV}$\tabularnewline
wabs & $n_{\mathrm{H}}$ & 0.018 & $\ensuremath{10^{22}\text{cm}^{-2}}$\tabularnewline
apec & $kT$ & 0.225 & keV\tabularnewline
apec & $abundance$ & 1.0 & \tabularnewline
apec & $redshift$ & 0 & \tabularnewline
apec & $norm$ & $7.3\times10^{-7}$ & \tabularnewline
powerlaw & $PhoIndex$ & 1.45 & \tabularnewline
powerlaw & $norm$ & $4.0\times10^{-7}$ & $\text{pho}/\text{keV}/\text{cm}^{2}/s\ @\ 1\ \text{keV}$\tabularnewline
 &  &  & \tabularnewline
\end{tabular}
\par\end{center}
\caption{\label{tab:Background-Xspec-model}\noun{Xspec} model parameters for
the photon background. Normalizations to 1 $\mathrm{arcmin}^{2}$. The angular size distance to the source (cm) is denoted by $D_{A}$  , and $n_{\text{e}}$ and $n_{\text{H}}$ are the electron and hydrogen densities ($\mathrm{cm^{-3}}$).}
\end{table*}

\section{Truong scaling relation tables\label{scaling_relation_TR18}}

The three scaling relation tables used in our simulations are summarized in Table~\ref{tab:Truong_SC_App}. We note that the {\agn} table was provided in \cite{Truong:2018aa}, and the {\csf} and {\agn} were obtained through a private request to the Authors. For the details of the generic fit expression for the parameters  $\log_{10}C$, $\beta$ , $\gamma,$ and $\sigma$, we refer the reader to the Eq. (11) in their paper. Our renormalization procedure in luminosity for those scaling relations is described in Sect.~\ref{subs:TR18_relation}.

\begin{table*}[t]
\begin{centering}
\begin{tabular}{c|ccccc}
\hline 
\hline 
\multicolumn{1}{c}{Type} & Parameter & $\log_{10}C$ & $\beta$ & $\gamma$ & $\sigma$\tabularnewline
\hline 
\multicolumn{1}{c|}{Truong NR}  & $\ensuremath{M-T_{\mathrm{sl}}}$ & $14.440\pm0.008$ & $1.403\pm0.023$ & $-0.943\pm0.042$ & $0.111\pm0.003$\tabularnewline
 & $\ensuremath{L_{\mathrm{X}}-T_{\mathrm{sl}}}$ & $0.925\pm0.010$ & $1.932\pm0.029$ & $1.264\pm0.054$ & $0.143\pm0.004$\tabularnewline
\hline 
Truong CSF  & $\ensuremath{M-T_{\mathrm{sl}}}$ & $14.260\pm0.006$ & $1.579\pm0.020$ & $-0.721\pm0.035$ & $0.092\pm0.003$\tabularnewline
 & $\ensuremath{L_{\mathrm{X}}-T_{\mathrm{sl}}}$ & $0.273\pm0.012$ & $2.538\pm0.037$ & $2.094\pm0.064$ & $0.166\pm0.005$\tabularnewline
\hline 
Truong AGN & $\ensuremath{M-T_{\mathrm{sl}}}$ & $14.297\pm0.005$ & $1.661\pm0.016$ & $-0.847\pm0.027$ & $0.069\pm0.002$\tabularnewline
 & $\ensuremath{L_{\mathrm{X}}-T_{\mathrm{sl}}}$ & $0.552\pm0.011$ & $2.877\pm0.038$ & $1.161\pm0.063$ & $0.162\pm0.005$\tabularnewline
\end{tabular}
\par\end{centering}
\caption{Three scaling relation models from \cite{Truong:2018aa}.
The parameters are the best-fitting parameters from the Bayesian fit
to the NR, CSF, and AGN data in the redshift range $[0-2.0]$. \label{tab:Truong_SC_App}}

\end{table*}


\section{Spectra for different off-axis angles at redshift 2\label{spec_off_axis_fix_z}}

\begin{figure*}
\begin{centering}
\includegraphics[trim=20 40 80 80,clip,width=0.85\textwidth]{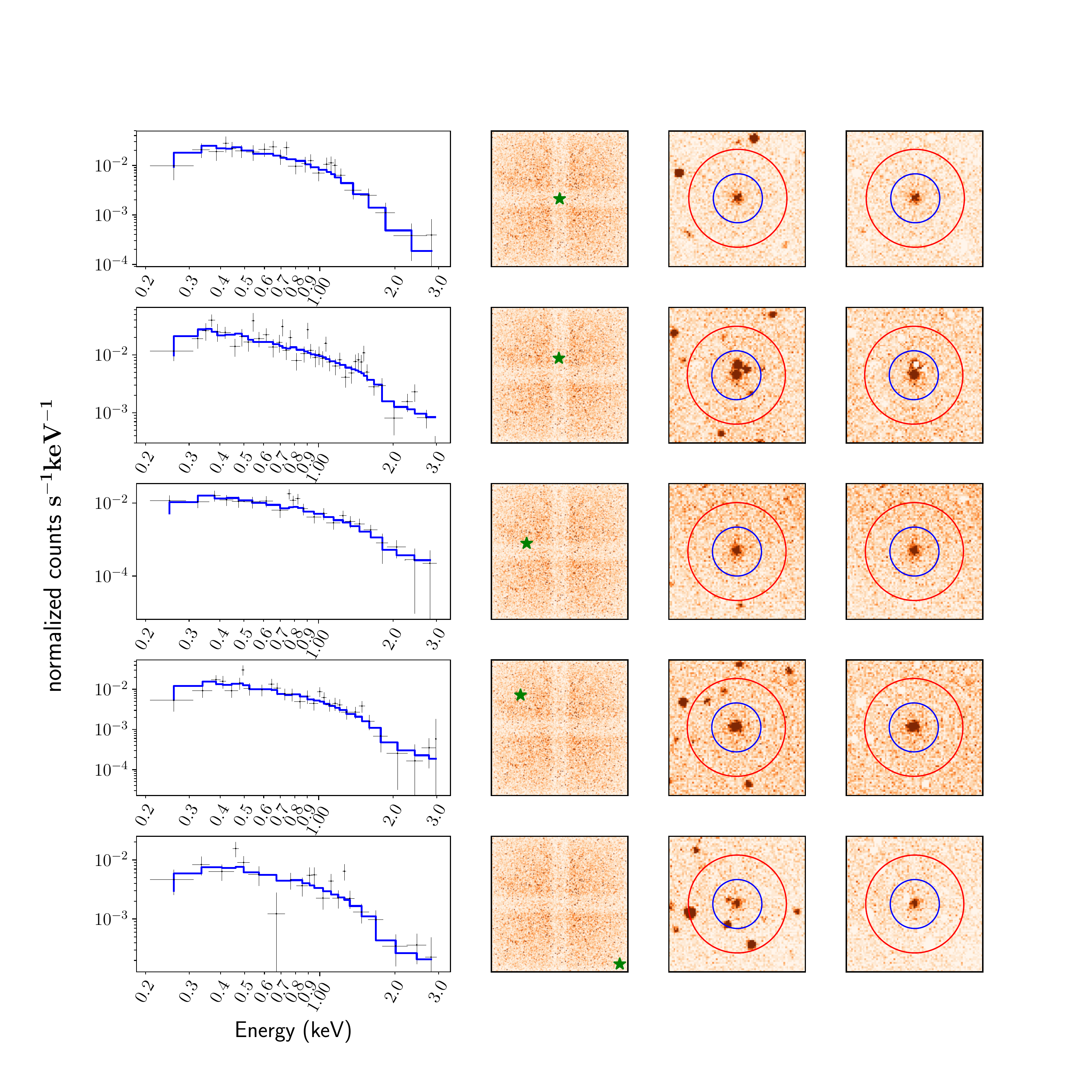}
\par\end{centering}
\caption{\label{fig:spectra_z_diff_off_axis}Simulated background-subtracted WFI spectrum of a group with mass $M_{500}=5\times10^{13}\,\mathrm{M_{\odot}}$ at off-axis angles $\theta=0,~5,~10,~15,~25$~arcmin at redshift 2.0 (from top to bottom), and an exposure time of  80 ks. First panel: Galaxy group spectra with the background subtracted. Second panel: Extracted location of galaxy group in Athena FoV. Third panel: Zoom-in view of the galaxy groups (box size $6\arcmin\times6\arcmin$). The blue circle represents $r_{500}$; the red circle is $2.0\times r_{500}$; the annulus is used for background subtraction. Fourth panel: Detected AGNs are excluded with a PSF corresponding to 90\%  encircled energy fraction (99\% for the bright ones outside $r_{500}$).}
\end{figure*}

Similar to the Fig. \ref{fig:spec_IM_diff_z} in Section~\ref{subs:specral_fitting_plot}, the spectra of galaxy groups at off-axis angles $\theta=0,~5,~10,~15,~25$~arcmin are displayed in Fig. \ref{fig:spectra_z_diff_off_axis}. The extracted spectra are similar at those off-axis angles, because they are obtained from the baseline simulation with a mass value of $M_{500}=5\times10^{13}\,\mathrm{M_{\odot}}$ in 80~ks at redshift $z=2.0$, meaning the groups all have the same $r_{500}=32.5\arcsec$ (indicated as blue circles) and flux $1.71\times10^{-15}\ \text{ergs/s/cm}^{2}$.

\section{ The  {\tt extent}-{\tt extension-likelihood} plane \label{Ext_ML_plot}}

\begin{figure*}
\begin{centering}
\includegraphics[trim=20 50 80 120,clip,width=\textwidth]{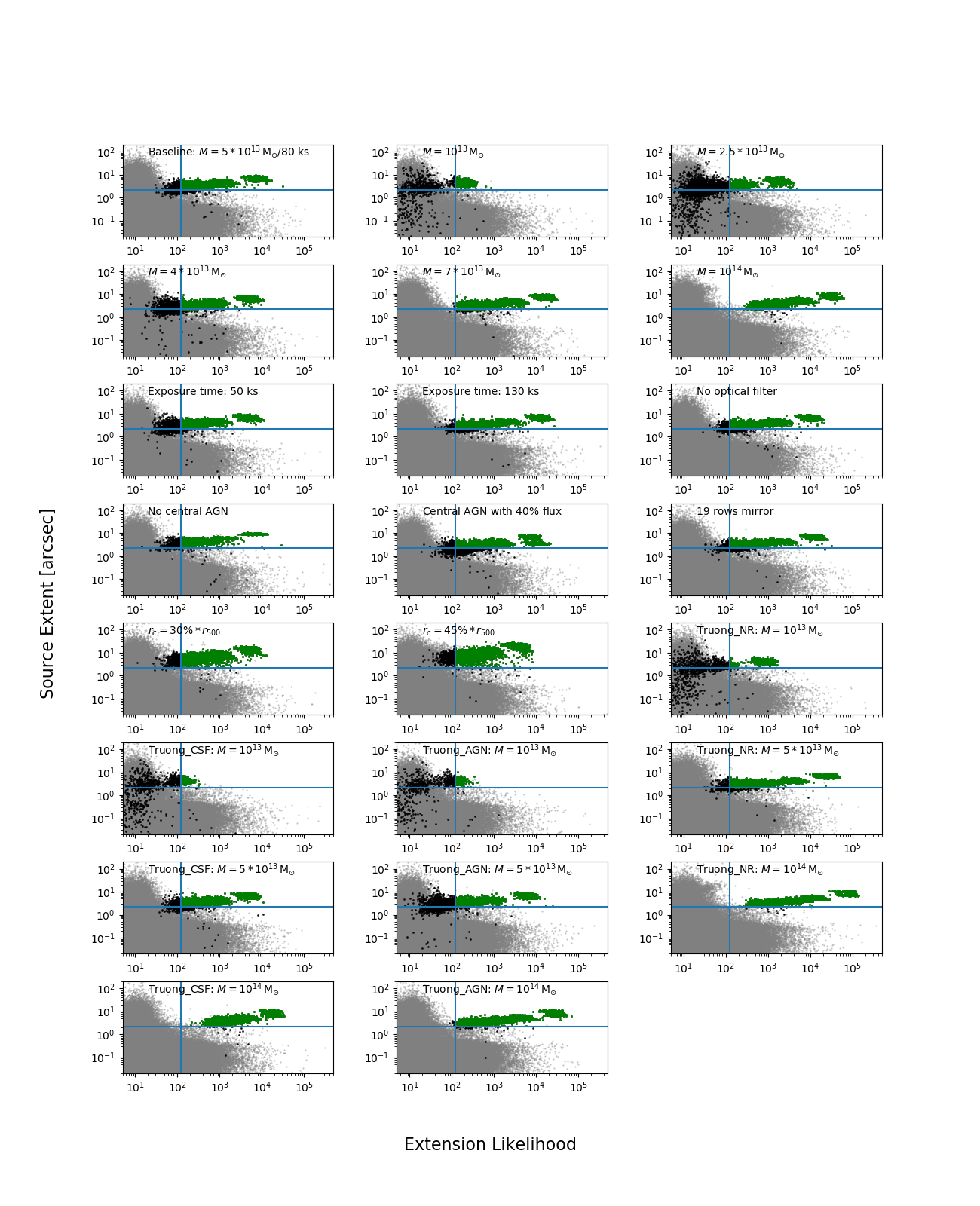}
\par\end{centering}
\caption{\label{fig: Extension_Li_EXT} The {\tt extent}-{\tt extension-likelihood} plane for all simulations. The gray data points are the detected AGNs, while black data points are input galaxy groups, and the green data points are the detected galaxy groups. The selection criteria for extended-like sources with a $1/\mathrm{deg}^{2}$ false detection rate is shown as an intersection of the blue lines at $\mathtt{extension-likelihood}>104$ and $\mathtt{extent}>2.3$. }
\end{figure*}

Figure \ref{fig: Extension_Li_EXT} shows the {\tt extent}-{\tt extension-likelihood} plane for all simulated scenarios in our work. There are 24 simulations in total. Each set of simulations consists of eight redshift bins at $z=0,~0.5,~1.0,~1.5,~2.0,~2.5,~3.0,~3.5,~4.0$, and each redshift bin consists of 20 sets of simulated images. One panel in the figure presents the combination of 20 sets of simulated images with all eight redshifts, giving 160 in total.

\end{appendix}


\end{document}